\documentclass{article}[12pt]
\usepackage{amsfonts}

\newbox\strutbox
\setbox\strutbox=\hbox{\char"B}
\def\strut{\relax\ifmmode\copy\strutbox\else\unhcopy\strutbox\fi}
\def\ialign{\everycr{}\tabskip0pt\halign}
\def\eqalign#1{\null \,\vcenter {\openup\jot \mathsurround 0pt
    \ialign{\strut \hfil$\displaystyle{##}$&$\displaystyle
      {{}##}$\hfil\crcr#1\crcr}}\,}

\def\eqalignno#1{\tabskip 0pt plus 1 fill \halign to\displaywidth
{\hfil$\tabskip0pt\everycr{}\displaystyle{##}$\tabskip 0pt &
$\tabskip0pt\everycr{}\displaystyle {{}##}$\hfil 
\tabskip 0pt plus 1 fill&\llap {$\tabskip0pt\everycr{} ##$}
\tabskip 0pt \crcr #1\crcr }}

\begin{document}
\title{Quantum Amplitudes in Black-Hole Evaporation: Spins 1 and 2}
\author{A.N.St.J.Farley and P.D.D'Eath\protect\footnote
{Department of Applied Mathematics and Theoretical Physics,
Centre for Mathematical Sciences,
University of Cambridge, Wilberforce Road, Cambridge CB3 0WA,
United Kingdom}}

\maketitle

\begin{abstract}
Quantum amplitudes for $s=1$ Maxwell fields 
and for $s=2$ linearised gravitational-wave perturbations 
of a spherically-symmetric Einstein/massless scalar 
background, describing gravitational collapse 
to a black hole, are treated by analogy 
with the previous treatment of $s=0$ scalar-field 
perturbations of gravitational collapse at late times.  
Both the spin-1 and the spin-2 perturbations split 
into parts with odd and even parity.  
Their detailed angular behaviour is analysed, 
as well as their behaviour under infinitesimal coordinate
transformations and their linearised field equations.  
In general, we work in the Regge-Wheeler gauge, 
except that, at a certain point, it becomes necessary 
to make a gauge transformation to an asymptotically-flat 
gauge, such that the metric perturbations 
have the expected fall-off behaviour at large radii.  
In both the $s=1$ and $s=2$ cases, 
we isolate suitable 'coordinate' variables 
which can be taken as boundary data 
on a final space-like hypersurface $\Sigma_{F}{\,}$.
(For simplicity of exposition, 
we take the data on the initial surface
$\Sigma_{I}$ to be exactly spherically-symmetric.)  
The (large) Lorentzian proper-time interval 
between $\Sigma_{I}$ and $\Sigma_{F}{\,}$,
measured at spatial infinity, is denoted by $T{\,}$.  
We then consider the classical boundary-value problem 
and calculate the second-variation classical Lorentzian 
action $S^{(2)}_{\rm class}{\,}$, 
on the assumption that the time interval $T$ 
has been rotated into the complex:  
$T\rightarrow {\mid} T{\mid}\exp(-{\,}i\theta)$, 
for $0<\theta\leq\pi/2{\,}$.  
This complexified classical boundary-value problem 
is expected to be well-posed, in contrast 
to the boundary-value problem 
in the Lorentzian-signature case $(\theta =0){\,}$, 
which is badly posed, since it refers to hyperbolic 
or wave-like field equations.  
Following Feynman, we recover the Lorentzian quantum 
amplitude by taking the limit 
as $\theta\rightarrow 0_{+}$ 
of the semi-classical amplitude
$\exp\bigl(iS^{(2)}_{\rm class}\bigr){\,}$.  
The boundary data for $s=1$ involve 
the (Maxwell) magnetic field, 
while the data for $s=2$ involve the magnetic part 
of the Weyl curvature tensor.  
These relations are also investigated, 
using 2-component spinor language, 
in terms of the Maxwell field strength 
$\phi_{AB}=\phi_{(AB)}$ and the Weyl spinor
$\Psi_{ABCD}=\Psi_{(ABCD)}{\,}$.  
The magnetic boundary conditions are related 
to each other and to the natural $s={{1}\over{2}}$ 
boundary conditions by supersymmetry. 
\end{abstract}

\begin{section}{ Introduction}
This paper describes part of a project concerned 
with the calculation of quantum amplitudes 
(not just probabilities) associated 
with quantum fields, including gravity itself, 
in the case that strong gravitational fields 
may be present.  The most obvious example 
-- the original motivation for this work 
-- concerns quantum radiation associated 
with gravitational collapse to a black hole [1-11].  
But the framework adopted here is more general, 
and certainly does not depend on whether 
there is a classical Lorentzian-signature collapse 
to a black hole.  It includes the case 
of local collapse which is not sufficient to lead 
to (Lorentzian) curvature singularities, 
and also quantum processes in cosmology, where, 
for example, anisotropies in the Cosmic Microwave 
Background Radiation (CMBR) can be computed, 
and depend crucially on the underlying Lagrangian 
for gravity and matter [12].

To exemplify the underlying ideas, we consider 
the case of local collapse 
(whether or not to a black hole).  
Thus, the gravitational field is taken 
to be asymptotically flat.  For simplicity, 
consider Einstein gravity coupled minimally 
to a massless scalar field $\phi{\,}$.  
In classical gravitation, we are used to describing 
this by means of a Cauchy problem, giving evolution 
to the future (say) of an initial space-like 
hypersurface ${\cal S}{\,}$, 
which extends to spatial infinity.  
We write ${\,}g_{\mu\nu}{\;}{\,}(\mu,\nu =0,1,2,3){\,}$ 
for the components of the 4-dimensional metric, 
and then denote by $h_{ij}=g_{ij}{\;}{\,}(i,j=1,2,3)$ 
the components of the intrinsic spatial metric 
on ${\cal S}{\,}$ in the case that ${\cal S}{\,}$ 
is given by the condition $x^{0}{\,}={\rm const}{\,}$.  
Cauchy data would, loosely speaking, 
consist of $h_{ij}{\,},\phi$ and their corresponding 
normal derivatives on ${\cal S}{\,}$.  
By contrast, in quantum theory, 
one typically asks for the amplitude to go 
from an initial configuration 
such as $(h_{ij}{\,},\phi)_{I}$ 
on an initial hypersurface $\Sigma_{I}{\,}$, 
to a final configuration $(h_{ij},\phi)_{F}$ 
on a final hypersurface $\Sigma_{F}{\,}$.  
The problem of finding the quantum amplitude 
should (naively) be completely posed, 
once one has also specified 
the (Lorentzian) proper-time interval between 
the surfaces $\Sigma_{I}$ and $\Sigma_{F}{\,}$, 
as measured near spatial infinity.

Much of the 'non-intuitive' nature 
of quantum mechanics can be traced 
to the 'boundary-value' nature of such a quantum 
amplitude [13], as compared with the familiar 
classical initial-value problem.  
A crucial mathematical aspect of this difference, 
responsible for a good part of the 'non-intuition', 
is that the {\it classical} version of the problem 
of calculating a quantum amplitude, as posed above, 
would involve solving the classical field equations 
(typically hyperbolic), subject to the given boundary 
data $(h_{ij}{\,},\phi)_{I,F}$ on the hypersurfaces 
$\Sigma_{I}{\,},\Sigma_{F}{\,}$, 
separated near spatial infinity by a Lorentzian time 
interval $T{\,}$.  As is well known, a boundary-value 
problem for a hyperbolic equation 
is typically not well posed.  
For typical boundary data, a classical solution 
will not exist 14,15]; or, if it does exist, 
it will be non-unique.  The straightforward cure 
for this ill, due to Feynman [13], 
is of course to rotate the Lorentzian 
time-interval $T$ into the complex:
${\,}T\rightarrow{\,}{\mid}T{\mid}{\,}\exp(-{\,}i\theta){\,}$, 
with ${\,}0<\theta\leq\pi/2{\;}$.

A simplified classical boundary-value example, 
showing this behaviour, is described in [16].  
When this example is posed originally (and badly) 
in 2-dimensional Minkowski space-time, 
one considers a scalar field $\phi(t{\,},x){\,}$, 
obeying the wave equation 
$$-{\,}{{{\partial}^{2}\phi}\over{{\partial}{t}^{2}}}{\,}
+{\,}{{{\partial}^{2}{\phi}}\over{{\partial}x^{2}}}{\;}{\,}
={\;}{\,}0{\quad};
{\qquad}{\quad}0<{t}<T{\;}{\,},
{\quad}-\infty <x< +{\,}\infty
{\quad},\eqno(1.1)$$
on the assumption that ${\,}\phi{\,}$ decays rapidly as 
${\,}{\mid}x{\mid}{\,}{\rightarrow}{\,}{\infty}{\;}$.  
A simple choice of Dirichlet boundary data is to take
$$\phi(t{\,}=0{\,},{\,}x){\;}{\,}={\;}{\,}0{\quad},
{\qquad}{\quad}\phi(t{\,}=T{\,},{\,}x){\;}{\,}
={\;}{\,}{\phi}_{1}(x)
{\qquad}.\eqno(1.2)$$

The time-interval $T$ at spatial infinity is then, 
as above, rotated into the lower complex half-plane:
$$T{\,}\rightarrow{\,}{\mid}T{\mid}{\,}
\exp(-{\,}i\theta){\;}{\,};
{\qquad}0{\,}<{\,}{\theta}{\;}{\leq}{\;}\pi/2
{\qquad}.\eqno(1.3)$$
For convenience, we define, for a given fixed 
${\,}{\theta}{\;}{\;}(0<{\,}{\theta}{\;}{\leq}{\;}\pi/2){\,}$, 
the 'rotated-time' coordinate
$$y{\;}{\,}={\;}{\,}t{\,}\exp(-{\,}i\theta)
{\quad}.\eqno(1.4)$$
In terms of the new coordinates $(y{\,},x){\,}$, 
the wave equation (1.1) reads
$$-{\,}e^{2i\theta}{\;}
{{{\partial}^{2}\phi}\over{\partial y^{2}}}{\,}
+{\,}{{{\partial}^{2}\phi}\over{\partial x^{2}}}{\;}{\,}
={\;}{\,}0
{\qquad},\eqno(1.5)$$ 
and the boundary conditions (1.2) become
$$\phi(y=0{\,},{\,}x){\;}{\,}={\;}{\,}0{\quad},
{\qquad}{\quad}\phi\bigl(y=Te^{i\theta},{\,}x\bigr){\;}{\,}
={\;}{\,}\phi_{1}(x)
{\qquad}.\eqno(1.6)$$
Here, the extreme case ${\,}{\theta}{\,}={\,}{\pi}/2{\,}$ 
corresponds to the Riemannian (Euclidean) sector, 
and to a well-posed real elliptic Dirichlet 
boundary-value problem for the Laplace equation.  
The potential $\phi(y{\,},x)$ 
is thus required to be a complex solution 
of Eqs.(1.5,6), which, for 
$0<{\,}{\theta}{\;}{\,}{\leq}{\;}{\pi}/2{\;}$, 
describe a {\it strongly elliptic} partial differential 
equation in the sense of [17].  
The property of strong ellipticity guarantees existence 
and uniqueness in this linear example.  
As verified in [16], however, the 'classical solution' 
becomes singular in the Lorentzian limit 
${\theta}{\;}{\rightarrow}{\;}0_{+}{\;}$.

In our coupled non-linear gravitational/scalar-field 
example, the extreme case 
${\,}\theta{\,}={\,}\pi/2{\,}$ would correspond 
to a purely Euclidean time-interval ${\mid}T{\mid}{\,}$, 
and classically one would then be solving the field 
equations for Riemannian gravity with a scalar field 
$\phi{\,}$.  Since these field equations 
are 'elliptic {\it modulo} gauge' 
-- see [17] 
-- one would expect to have a well-posed classical 
boundary-value problem, with existence and uniqueness.  
The intermediate case ${\,}0<\theta<\pi/2{\,}$ 
requires the interval $T$ and any classical solution 
to involve the complex numbers non-trivially.  
If the problem turns out to be strongly elliptic, 
up to gauge, then the complex case 
${\,}0<\theta<\pi/2{\,}$ 
would again be expected to have the good existence 
and uniqueness properties of the real elliptic case.

In practice, in the black-hole evaporation problem 
or (say) in cosmological examples, 
one typically treats the case in which both 
the gravitational and the scalar initial data 
are close to spherical symmetry.  
Hence, as a leading approximation, 
one begins by studying the spherically-symmetric 
Einstein/scalar system.  
This was treated in [18] for Lorentzian signature 
and is outlined in [19] for Riemannian signature.  
In the Riemannian case, the metric is taken 
(without loss of generality) in the form
$$ds^2{\quad}
={\quad}e^{b}{\,}d{\tau}^{2}{\,}+{\,}e^{a}{\,}dr^{2}{\,} 
+{\,}r^{2}{\,}(d{\theta}^{2}+{\,}{\sin}^{2}{\theta}{\;}d{\phi}^2)
{\quad},\eqno(1.7)$$
where
$$b{\;}{\,}={\;}{\,}b(\tau,r){\;}{\;},
{\qquad}{\qquad}a{\;}{\,}={\;}{\,}a(\tau,r)
{\qquad},\eqno(1.8)$$
and the scalar field is taken of the form 
${\,}\phi(\tau,r){\,}$ [19].  
The scalar field equation reads:
$$\ddot\phi{\,}+{\,}e^{{b-a}}{\,}{\phi}^{\prime\prime}{\,}
+{\,}{{1}\over{2}}{\,}\bigl(\dot a -\dot b\bigr){\,}\dot{\phi}{\,} 
+{\,}r^{-1}{\,}e^{{b-a}}{\,}\bigl(1+e^{a}\bigr){\,}\phi'{\quad}
={\quad}0
{\quad}{\;},\eqno(1.9)$$
where $(\dot{\;}{\,})$ denotes 
${\partial}(~)/{\partial}{\tau}{\,}$ 
and $(~)^{\prime}$ denotes ${\partial}(~)/{\partial}r{\,}$.   
Together with Eq.(1.9), 
a slightly redundant set of gravitational field 
equations is given by:
$$\eqalignno{a^\prime{\quad}&
={\,}-4{\pi}r{\,}\bigl(e^{a-b}{\,}{\dot\phi}^{2}
-{\phi^{\prime}}^2{\,}\bigr)
+r^{-1}\bigl(1-e^{a}\bigr)
{\quad},&(1.10)\cr
b^{\prime}{\quad}&
={\,}-4{\pi}r{\,}\bigl(e^{a-b}{\,}{\dot\phi}^{2}
-{\phi^{\prime}}^2{\,}\bigr)
-r^{-1}\bigl(1-e^{a}\bigr)
{\quad},&(1.11)\cr
\dot{a}{\quad}&
={\quad}8{\pi}r{\,}\dot{\phi}{\,}\phi^{\prime}
{\quad},&(1.12)\cr}$$
$${\ddot a}+e^{b-a}{\,}b^{\prime\prime}
+{{1}\over{2}}\bigl({\dot a}-{\dot b}\bigr){\dot a} 
-r^{-1}e^{b-a}\bigl(1-e^{a}\bigr)
\bigl(b^{\prime}+2r^{-1}\bigr)
=8\pi{\,}\bigl({\dot\phi}^{2}
+e^{b-a}{\,}{\phi^\prime}^{2}\bigr).
\eqno(1.13)$$ 

The metric and the classical field equations 
in Lorentzian signature [18] can be derived 
from the above by the formal replacement
$$t{\quad}={\quad}\tau{\,}e^{-i\theta}
{\quad},\eqno(1.14)$$ 
where ${\,}{\theta}{\,}={\,}\pi/2{\,}$ 
is independent of 4-dimensional position.  
Similarly for complex metrics with suitable behaviour 
at infinity, with ${\,}0{\,}<{\,}\theta{\,}\leq{\,}\pi/2{\;}$.

Even in the spherically-symmetric case, 
very little is known rigorously about existence 
and uniqueness for the Riemannian (or complex) 
boundary-value problem.  
For this case, numerical investigation of the weak-field 
Riemannian boundary-value problem was begun in [19], 
and has recently been extended towards the strong-field 
region [20].  For weak scalar boundary data, global 
quantities such as the mass $M$ and Euclidean action $I$ 
appear to scale quadratically, 
in accordance with analytic weak-field estimates [20].  
In the limit of strong-field scalar boundary data, 
it may be that a typical pattern will emerge 
numerically for the general 'shape' 
of the classical Riemannian gravitational 
and scalar fields.  
In that case, it might be possible to find analytic 
approximations for the strong-field limit 
(quite different from those valid in the weak-field case), 
which could provide further analytical insight 
into the solutions of the coupled Riemannian 
Einstein/scalar boundary-value problem.  
In particular, it would be extremely valuable 
to have strong-field approximations 
which were valid into the complex region, with 
${\,}0<{\,}\theta{\,}<{\,}\pi/2{\,}$.  
One might conjecture that, 
as one approaches the Lorentzian limit 
${\,}\theta{\,}\rightarrow{\,}0_{+}{\,}$, 
for very strong spherically-symmetric boundary data, 
the solutions (albeit complex) correspond 
to classical Einstein/scalar solutions which form 
a singularity, surrounded by a black hole.  

In the case of quantum amplitudes for Lagrangians 
with Einstein gravity coupled to matter, 
one can consider {\it anisotropic} boundary data 
posed in the 'field language' of this paper, 
by taking (for the present Einstein/scalar case) 
non-spherically-symmetric boundary data 
$(h_{ij}{\,},\phi)_{I,F}{\,}$.  
Then, at least in the asymptotically-flat case 
with time-interval $T$ at spatial infinity, 
one is inevitably led to consider 
the complexified boundary-value problem, with
${\,}T\rightarrow{\mid}T{\mid}{\,}\exp(-{\,}i\theta){\,}$, 
but with unchanged data $(h_{ij}{\,},\phi)_{I,F}$ 
on the other boundaries.  This corresponds 
(by a slight re-definition of $\theta{\,}$) 
to the procedure adopted in our model 2-dimensional 
boundary-value problem of Eqs.(1.1-6).  
Even for fairly small ${\,}\theta{\,}$, 
solution of this boundary-value problem is expected 
to smooth out variations or oscillations 
of the boundary data, when one moves 
into the interior by a few multiples 
of the relevant wavelength.  
If the problem is genuinely strongly elliptic, 
up to gauge, then one will be able to extend 
the classical solution analytically into the complex.

Strictly, in order that quantum amplitudes 
should be meaningful for any Einstein-gravity/matter 
Lagrangian under consideration, one should work 
only with theories invariant under local supersymmetry 
-- that is, with supergravity models or supergravity 
coupled to supermatter [21,22].  
Thus, for example, the bosonic Einstein/massless-scalar 
model above should be replaced by the simplest 
locally-supersymmetric theory which contains it [21].  
This $N=1$ supergravity/supermatter model contains 
a {\it complex} scalar field $\phi{\,}$, 
with a massless spin-${{1}\over{2}}$ partner; 
the graviton acquires a spin-${{3}\over{2}}$ 
gravitino partner.  
Generally, for 'Riemannian' boundary data, 
one expects that the resulting 'Euclidean' 
quantum amplitude has the semi-classical form
$${\rm Amp}{\quad}
\sim{\quad}(A_{0}+{\hbar}A_{1}+{\hbar}^{2}A_{2}
+{\ldots}{\;}){\;}{\,}\exp\bigl(-{\,}I_{\rm class}/{\hbar}\bigr)
{\quad},\eqno(1.15)$$
asymptotically in the limit that 
${\,}\bigl(I_{\rm class}/\hbar\bigr)\rightarrow 0{\;}$.  
Here, $I_{\rm class}$ is the classical 
'Euclidean action' of a Riemannian solution 
of the coupled Einstein and bosonic-matter 
classical field equations, 
subject to suitable boundary conditions.  
In the complex r\'egime of this paper, 
we shall use the expressions $I$ 
and $-{\,}iS$ interchangeably, 
where $S$ denotes the 'Lorentzian action'.  
For simplicity, we assume that there is a unique 
classical solution, up to gauge, coordinate 
and local supersymmetry transformations.  
But it is quite feasible, in certain theories 
and for certain boundary data, 
to have instead (say) a complex-conjugate pair 
of classical solutions [23].  
The classical action $I_{\rm class}$ 
and loop terms 
$A_{0}{\,},A_{1}{\,},A_{2}{\,},\ldots{\;}$ 
depend in principle on the boundary data.  
In the case of supermatter coupled to $N=1$ 
supergravity, each of 
${\,}I_{\rm class}{\,},A_{0}{\,},A_{1}{\,},A_{2}{\,},\ldots{\;}$ 
will also obey differential constraints connected 
with the local coordinate and local supersymmetry 
invariance of the theory, 
and with any other local invariances 
such as gauge invariance (if appropriate) [15,24].

In particular, in the locally-supersymmetric case, 
the semi-classical expansion (1.15) 
may become extremely simple [15,25,26].  
For example, for $N=1$ supergravity, 
for purely bosonic (Einstein) boundary data, 
one has [15]:
$${\rm Amp}{\quad}
\sim{\quad}A_{0}{\,}
\exp\bigl(-{\,}I_{\rm class}/{\hbar}\bigr)
{\quad},\eqno(1.16)$$
where, further, the one-loop factor $A_{0}$ 
is in fact a constant.  When the boundary data 
are allowed to include both bosonic 
and fermionic parts (suitably posed), 
one expects that a classical solution 
of the coupled bosonic/fermionic field equations 
will still exist.  In this case, the expression 
$I_{\rm class}$ in Eq.(1.16) denotes the full classical 
action, including both bosonic and fermionic 
contributions.  The fermionic contributions (naturally) 
also depend on the boundary data, and, 
as is standard in the holomorphic
representation used here for fermions [27-29], 
live in a Grassmann algebra over the complex numbers.  
Related properties hold for $N=1$ supergravity coupled 
to gauge-invariant supermatter [22,25].  
Whether or not the supermatter is also invariant 
under a gauge group, there will be analogous 
consequences for the semi-classical expansion (1.15) 
of the quantum amplitude.  
In particular, one expects finite loop terms 
$A_{0}{\,},A_{1}{\,},A_{2}{\,},{\ldots}{\;}$ [15,25,26],
but typically not the maximal simplicity of the pure 
supergravity amplitude (1.16).

In the case (1.16) of pure supergravity, 
the classical action is all that is needed 
for the quantum computation.  
A corresponding situation arises 
with ultra-high-energy collisions, 
whether between black holes [30], 
in particle scattering [31], 
or in string theory [32].

To fix one's physical intuition, one can assume that, 
near the initial surface ${\Sigma}_{I}{\,}$, 
the gravitational and scalar fields are approximately 
spherically symmetric and vary extremely slowly 
with time, corresponding to diffuse bosonic matter 
near ${\Sigma}_{I}{\,}$.
The final hypersurface ${\Sigma}_{F}{\,}$ should 
preferably be taken at a sufficiently late time $T$ 
that all the quantum radiation due to the evaporation 
of the black hole will by then have been emitted.  
The final data $(h_{ij}{\,},\phi)_{F}$ for gravity 
and the scalar field, together, if a Maxwell field 
is included, with the spatial components $(A_{i})_{F}$ 
of the vector potential, are taken to have small
anisotropic parts 
-- this corresponds, in 'particle language', 
to a choice of final particle state.  
The resulting 'weak-field' quantum amplitude, 
to be found below, can be described in terms 
of products of zero- or one-particle 
harmonic-oscillator eigenstates.  
One could, of course, also consider final data 
which deviate strongly from spherical symmetry.  
Their quantum amplitudes will still be roughly 
proportional to the (complex) quantity 
${\,}{\lim}_{\theta{\rightarrow}0_{+}}
\exp(-{\,}I_{\rm class}){\,}$.  
For the weak perturbations (above) 
away from spherical symmetry, $I_{\rm class}$ 
is nearly quadratic, but for strong perturbations, 
$I_{\rm class}$ will be very non-linear, 
and the probabilities of such final configurations 
will be microscopic.

In considering the classical boundary-value problem, 
even though spin-1 and fermionic fields 
may also be involved, we shall for simplicity 
first consider the fields $g_{\mu\nu}$ 
and $\phi{\,}$.  
As above, we consider the Riemannian 
(or complex-rotated Riemannian) classical 
boundary-value problem, given boundary data 
$(h_{ij}{\,},\phi)_{I}$ and 
$(h_{ij}{\,},\phi)_{F}{\,}$, 
where the initial and final boundary 
hypersurfaces ${\Sigma}_{I}$ 
and ${\Sigma}_{F}$ are separated at spatial infinity 
by a complex (Riemannian-time) interval of the form
${\,}{\mid}T{\mid}{\,}\exp(-{\,}i\theta){\,}$.  
Since all fields are to be regarded as perturbations 
of a 'background' spherically-symmetric 
configuration, we are assuming that the classical 
solutions 
${\,}(g_{\mu\nu}{\,},\phi)$ 
of the coupled Einstein/scalar field equations 
may be decomposed into a 'background' 
spherically-symmetric part 
$({\gamma}_{\mu\nu}{\,},\Phi)$, 
together with a 'small' perturbative part.  
The linearised perturbative fields, 
whether spin-0 scalar [11,33,34], 
spin-1 or spin-2 (this paper), 
spin-${{1}\over{2}}$ [35] 
or spin-${{3}\over{2}}$ (in progress [36]), 
can be expanded out in the appropriate spin-weighted 
spherical harmonics [37-39].

Typically, the perturbative scalar-field configuration 
(say) ${\phi}^{(1)}_{F}{\,}$, 
given on the late-time surface ${\Sigma}_{F}{\,}$, 
will involve an enormous number of modes, 
both angular and radial, but with a minute coefficient 
for each mode.  
That is, the given ${\phi}^{(1)}_{F}$ 
may contain extremely detailed angular structure, 
and also be spread over a considerable radius 
from the centre of spherical symmetry 
of the background
$({\gamma}_{\mu\nu}{\,},\Phi)$, 
again with detailed radial structure.  
Similar comments should apply to the perturbative 
part $h_{ijF}$ of the spatial gravitational field 
on ${\Sigma}_{F}{\,}$, 
and to the final spatial spin-1 potential $A_{iF}{\,}$, 
if appropriate.  
As a result, 
in the (complexified) nearly-Lorentzian r\'egime, 
one will have (classically) radiation of various 
spins, typically with wavelengths much shorter 
than the characteristic length- or time-scale 
corresponding to the Schwarzschild mass $M{\,}$; 
this radiation will propagate approximately 
by geometrical optics.  In turn, the effective 
energy-momentum tensor $T_{\mu\nu}$ 
due to this radiation will, on the average, 
be nearly spherically symmetric, 
and will indeed have the form appropriate 
to a radially-outgoing null fluid [40,41].  
The classical 'space-time' metric resulting 
from such a null-fluid effective $T_{\mu\nu}$ 
is precisely of the Vaidya type [41].  
This resembles the Schwarzschild geometry, 
except that the r\^ole of the Schwarzschild 
mass $M$ is taken by a mass function 
$m(t{\,},r){\,}$, 
which varies extremely slowly with respect 
both to $t$ and to $r$ in the space-time region 
containing outgoing radiation.  
The perturbative fields, 
such as ${\,}{\phi}^{(1)}{\,}$, 
propagate adiabatically in this classical solution.

In Sec.2 we shall describe, in tensor language, 
the boundary data on the final surface $\Sigma_{F}$ 
which are natural for the $s=1$ Maxwell 
and for the $s=2$ graviton cases.  
These are, respectively, the Maxwell magnetic 
field $B_{i}$ and the magnetic part $H_{ik}$ 
of the Weyl tensor.  In Sec.3 we rephrase the $s=1$ 
and $s=2$ problems in terms of 2-component spinors; 
the Maxwell field strength is determined 
by the symmetric spinor $\phi_{AB}=\phi_{(AB)}$ 
and the Weyl curvature by 
$\Psi_{ABCD}=\Psi_{(ABCD)}{\,}$.  
In both cases, the natural boundary data involve 
a 'projection' on symmetric spinors; 
for $s=1$ and $2{\,}$, the data are related 
by supersymmetry.  
In Sec.4, for the Maxwell field, we begin 
the process of decomposing the classical problem 
in terms of odd- and even-parity harmonics, 
following Regge and Wheeler.  
A similar procedure is carried in Sec.5 
for the odd-parity gravitational perturbations.  
Returning to the Maxwell case in Sec.6, 
the (magnetic) boundary conditions on $\Sigma_{F}$ 
are described, together with the classical action 
$S^{EM}_{\rm class}$ 
as a functional of the final boundary data.  
Secs.7,8 and 9 are concerned with gravitational
perturbations.  In Sec.7, the odd-parity $s=2$ 
problem is treated roughly by analogy 
with the Maxwell problem of Sec.6.  
Even-parity gravitational perturbations are treated 
in Secs.8 and 9.  A preliminary analysis in Sec.8 
leads to the classical action functional 
$S^{(2)}_{\rm class}$ 
in Sec.9.  The Conclusion is in Sec.10. 
\end{section}

\begin{section}{ Boundary data for the Maxwell field 
and for gravity}
The Maxwell contribution to the total Lorentzian 
action $S$ is
$$S^{EM}{\quad}
={\,}-{\,}{{1}\over{16\pi}}
\int_{{\cal M}}d^{4}x{\;}
(-g)^{{1}\over{2}}{\,}F_{\mu\nu}{\,}F^{\mu\nu}
{\quad},\eqno(2.1)$$
where ${\,}F_{\mu\nu}=F_{[\mu\nu]}{\,}$ is the Maxwell 
field strength, while the space-time metric 
${\,}g_{\mu\nu}{\,}$ is assumed here to have Lorentzian 
signature, with 
${\,}g{\,}={\,}{\rm det}(g_{\mu\nu})<0{\,}$.  
The resulting classical Maxwell field equations are
$$\nabla_{\mu}F^{\mu\nu}{\quad}
={\quad}0
{\quad}.\eqno(2.2)$$
The further condition that $F_{\mu\nu}$ be derivable 
from a vector potential $A_{\mu}{\,}$, as
$$F_{\mu\nu}{\quad}
={\quad}\nabla_{\mu}A_{\nu}-\nabla_{\nu}A_{\mu}{\quad}
={\quad}\partial_{\mu}A_{\nu}{\,}
-{\,}\partial_{\nu}A_{\mu}
{\quad},\eqno(2.3)$$  
may equivalently be written in the form of the dual 
field equations
$$\nabla_{\mu}(^{*}F^{\mu\nu}){\quad}
={\quad}0
{\quad},\eqno(2.4)$$
where
$${~}^{*}F_{\mu\nu}{\quad}
={\;}{\,}{{1}\over{2}}{\,}
\epsilon_{\mu\nu}^{~~~\rho\sigma}{\,}F_{\rho\sigma}
\eqno(2.5)$$ 
is the dual field strength [42,43]. 
(The Poincar\'e lemma [44] may be applied, 
since we are working within a manifold ${\cal M}$ 
which may be regarded as a slice of ${\Bbb R}^4{\,}$, 
with boundary $\partial{\cal M}$ consisting 
of two ${\Bbb R}^3$ hypersurfaces.)  
The action (2.1) is invariant under Maxwell 
gauge transformations
$$A_{\mu}{\quad}\rightarrow{\quad}A_{\mu}{\,}
+{\,}\partial_{\mu}\Lambda
{\quad},\eqno(2.6)$$ 
\noindent
in the interior, 
where $\Lambda(x)$ is a function of position.

As in [33,34,45] for scalar (spin-0) perturbations 
of spherically symmetric Einstein/massless-scalar 
gravitational collapse, and as in the treatment 
of spin-2 (graviton) perturbations below, 
we shall need the classical action $S_{\rm class}{\,}$, 
namely the action $S$ evaluated at a classical 
solution of the appropriate (slightly complexified) 
boundary-value problem, as a functional 
of the boundary data.  From this, one obtains 
the semi-classical quantum amplitude, 
proportional to $\exp(iS_{\rm class}){\,}$, 
and hence by a limiting procedure the Lorentzian 
quantum amplitude.  In the present (spin-1) 
Maxwell case, the classical action 
${\,}S^{EM}_{\rm class}{\,}$ resides solely 
on the boundary ${\,}\partial{\cal M}{\,}$, 
which consists of the initial space-like 
hypersurface $\Sigma_{I}$ and final hypersurface 
$\Sigma_{F}{\,}$.  There will be no contribution 
from any large cylinder of radius 
$R_{\infty}\rightarrow\infty{\,}$, 
provided that we impose the physically reasonable 
restriction that, as $r\rightarrow\infty{\,}$, 
the potential $A_{\mu}$ should die off faster 
than $r^{-1}{\,}$, and that the field strength $F_{\mu\nu}$ 
should die off faster than $r^{-2}{\,}$.  
That is, we impose reasonable fall-off conditions 
at large $r$ on field configurations, 
such that the action $S$ should be finite.  
(Compare the usual fall-off conditions for instantons 
in Euclidean Yang-Mills theory [44,48,49].)  
For the above class of Maxwell field configurations, 
the boundary form of the classical Maxwell action is
$$S^{EM}_{\rm class}{\quad} 
={\,}-{\;}{{1}\over{8\pi}}
\int^{\Sigma_{F}}_{\Sigma_{I}}d^{3}x{\;} 
h^{{1}\over{2}}{\,}n_{\mu}{\,}A_{\nu}{\,}F^{\mu\nu}
{\quad}.\eqno(2.7)$$
Here, as in Sec.1, ${\,}h_{ij}=g_{ij}{\;}{\;}(i,j=1,2,3)$ 
gives the intrinsic Riemannian 3-metric on the boundary 
hypersurface $\Sigma_{I}$ or $\Sigma_{F}{\,}$, 
and we write ${\,}h{\,}={\,}{\rm det}(h_{ij})>0{\,}$.  
Further, $n^{\mu}$ denotes the (Lorentzian) 
unit future-directed timelike vector, 
normal to the space-like hypersurface $\Sigma_{I}$ 
or $\Sigma_{F}{\,}$.

Given the $3{\,}+{\,}1$ split of the 4-metric 
$g_{\mu\nu}$ at each boundary, due to the ability 
to project vectors and tensors normally using $n^{\mu}$ 
and tangentially using the projector [48]
$$h_{\mu\nu}{\quad}
={\quad}g_{\mu\nu}{\,}+{\,}n_{\mu}{\,}n_{\nu}
{\quad},\eqno(2.8)$$
at the boundary, one can project the potential $A_{\mu}$ 
and field strength $F_{\mu\nu}$ into 'normal' 
and 'spatial' parts on $\Sigma_{I}$ and $\Sigma_{F}{\,}$.  
In particular, one defines the densitised electric field 
vector on the boundary:
$${\cal E}^{i}{\quad}
={\,}-{\,}h^{{1}\over{2}}E^{i}
{\quad},\eqno(2.9)$$
where
$$E^{\nu}{\quad}
={\quad}n_{\mu}{\,}F^{\mu\nu}
{\quad}\eqno(2.10)$$
obeys ${\,}n_{\nu}E^{\nu}{\,}={\,}0{\,}$.  
Further, in a Hamiltonian formulation [49], 
when one regards the spatial components $A_{i}$ 
of the vector potential as 'coordinates', 
then the canonical momentum $\pi^{i}{\,}$, 
automatically a vector density, is given by
$$\pi^{i}{\quad}
={\,}-{\,}{{{\cal E}^{i}}\over{4\pi}}
{\quad}.\eqno(2.11)$$
Note that the normal component $A_{t}=-{\,}\varphi{\,}$, 
where $\varphi$ is the Maxwell scalar potential, 
is gauge-dependent, but that $\varphi$ 
does not need to be specified on the spacelike 
boundaries $\Sigma_{I}$ and $\Sigma_{F}{\,}$, 
and is indeed allowed to vary freely there 
and throughout the space-time.  Its conjugate momentum 
therefore vanishes.  In the gravitational case, 
analogous properties hold for the lapse function $N$ 
and the shift vector $N^{i}$ [15,49].

As described in [49-51], it is natural in specifying 
a classical boundary-value problem for the Maxwell field, 
with data given on the space-like boundaries $\Sigma_{I}$ 
and $\Sigma_{F}{\,}$ and at spatial infinity 
(with Lorentzian proper-time separation $T{\,}$), 
to fix the spatial magnetic field components, 
described in densitised form by 
$${\cal B}^{i}{\quad}
={\quad}{{1}\over{2}}{\,}\epsilon^{ijk}{\,}F_{jk}
{\quad},\eqno(2.12)$$
on $\Sigma_{I}$ and $\Sigma_{F}{\,}$.  
The ${\cal B}^{i}$ cannot be specified freely 
on the boundary, but are further subject 
to the (linear) restriction
$$\partial_{i}{\cal B}^{i}{\quad}
={\quad}0
{\quad}.\eqno(2.13)$$
These components are gauge-invariant, 
and therefore physically measurable, 
in contrast to those of the spatially-projected vector
potential $A_{i}{\,}$.  We shall regard the space 
of such ${\cal B}^{i}(x){\,}$, on $\Sigma_{I}$ 
or $\Sigma_{F}{\,}$, as the 'coordinates' 
for Maxwell theory.  From the space-time Maxwell 
equations (2.2), one also deduces the constraint
$$\partial_{i}{\cal E}^{i}{\quad} 
={\quad}0
{\quad}.\eqno(2.14)$$

Turning to $s=2$ (graviton) perturbations 
of a spherically-symmetric background, 
we describe the boundary conditions found below 
to be appropriate both for odd- and even-parity 
vacuum $s=2$ perturbations.  
The most suitable $s=2$ boundary data involve 
prescribing the magnetic part of the Weyl 
curvature tensor
${\,}C_{\alpha\beta\gamma\delta}{\,}$ [42,43,52,53] 
on $\Sigma_{I}$ and on $\Sigma_{F}{\,}$.  
For simplicity, as above, we are taking 
the gravitational initial data on $\Sigma_{I}$ 
to be exactly spherically symmetric ('no incoming gravitons').  
Of course, in a large part of the space-time, 
one is nearly {\it in vacuo}, the Ricci tensor then obeying 
${\,}R_{\alpha\beta}{\,}\simeq{\,}0{\;}$, 
whence
${\,}C_{\alpha\beta\gamma\delta}{\,}
\simeq{\,}R_{\alpha\beta\gamma\delta}{\;}$, 
the Riemann tensor.  More generally, the Weyl tensor 
is defined by
$$R_{\alpha\beta\gamma\delta}{\quad}
={\quad}C_{\alpha\beta\gamma\delta}{\,}
+{\,}g_{\alpha [ \gamma}R_{\delta ] \beta}{\,}
-{\,}g_{\beta [ \gamma}R_{\delta ] \alpha}{\,}
-{\,}{{1}\over{3}}{\,}R{\,}g_{\alpha [ \gamma}g_{\delta ]\beta}
{\quad},\eqno(2.15)$$
where ${\,}R{\,}={\,}g^{\alpha\beta}{\,}R_{\alpha\beta}{\,}$ 
gives the Ricci scalar, and where square brackets 
denote anti-symmetrisation.

The algebraic symmetries of the Weyl tensor 
at a point are summarised by
$$\eqalign{C_{\alpha\beta\gamma\delta}{\quad}
&={\quad}C_{[\alpha\beta][\gamma\delta]}{\quad}
={\quad}C_{\gamma\delta\alpha\beta}{\quad},\cr
C_{\alpha[\beta\gamma\delta]}{\quad}&
={\quad}0{\quad},
{\qquad}{\quad}C^{\alpha}_{~~\beta\alpha\delta}{\quad}
={\quad}0
{\quad}.\cr}\eqno(2.16)$$
These imply that $C_{\alpha\beta\gamma\delta}$ 
has 10 algebraically-independent components at each point.  
At a bounding space-like hypersurface, 
such as $\Sigma_{F}{\,}$, one can, 
by analogy with the Maxwell case, apply a $3{\,}+{\,}1$ 
decomposition to the Weyl tensor
${\,}C_{\alpha\beta\gamma\delta}{\,}$, 
which splits into two symmetric trace-free spatial tensors, 
the electric part $E_{ik}$ and the magnetic part $H_{ik}$ 
of the Weyl tensor [52,53].  Thus, the 10 space-time 
components of ${\,}C_{\alpha\beta\gamma\delta}{\,}$ 
have been decomposed into the 5 spatial components 
of $E_{ik}$ and 5 more of $H_{ik}{\,}$.  
(Correspondingly, in Maxwell theory above, 
the 6 non-trivial components of the field strength 
$F_{\mu\nu}$ became the 3 of $E_{i}$ plus the 3 of $B_{i}{\;}$.)

For convenience of exposition, consider an 'adapted' 
coordinate system 
${\,}(x^{0}{\,},x^{1}{\,},x^{2}{\,},x^{3}){\,}$ 
in a neighbourhood of ${\,}\Sigma_{F}{\,}$, 
such that $\Sigma_{F}$ lies at ${\,}x^{0}{\,}={\,}0{\,}$, 
and such that ${\,}n^{0}{\,}=1$ at all points 
of $\Sigma_{F}{\,}$.  The spatial 3-metric is, as usual, 
denoted by $h_{ij}{\,}$, and we again write 
${\,}h{\,}={\,}{\rm det}(h_{ij}){\,}$.  
The electric part of the Weyl tensor is defined 
in 4-dimensional language to be
$$E_{\alpha\gamma}{\quad} 
={\quad}C_{\alpha\beta\gamma\delta}{\,}n^{\beta}{\,}n^{\delta}
{\quad}.\eqno(2.17)$$
In an adapted coordinate system, 
this corresponds to the 'spatial' equation
$$E_{ik}{\quad}
={\quad}C_{i0k0}
{\quad}.\eqno(2.18)$$
The magnetic part of the Weyl tensor is defined to be
$$H_{\alpha\gamma}{\quad} 
={\;}{\,}{{1}\over{2}}{\,}\eta_{\alpha\beta}^{~~~\rho\sigma}{\,}
C_{\rho\sigma\gamma\delta}{\,}n^{\rho}{\,}n^{\delta}
{\quad},\eqno(2.19)$$
where
$$\eta_{\alpha\beta\gamma\delta}{\quad}
={\quad}\eta_{[\alpha\beta\gamma\delta]}{\quad}
={\quad}(-g)^{{1}\over{2}}{\,}\epsilon_{\alpha\beta\gamma\delta}
\eqno(2.20)$$
is the alternating tensor, with 
${\,}g{\,}={\,}{\rm det}(g_{\mu\nu}){\,}$, 
and where 
${\,}\epsilon_{\alpha\beta\gamma\delta}{\;} 
={\;}\epsilon_{[\alpha\beta\gamma\delta]}{\,}$ 
is the alternating symbol, normalised such that 
${\,}\epsilon_{0123}{\,}={\,}1{\,}$.  
In an adapted coordinate system, one finds that
$$H_{ik}{\quad} 
={\,}-{\,}{{1}\over{2}}{\,}h^{-1/2}{\,}h_{in}{\,}
\epsilon^{n\ell m}{\,}C_{\ell m k 0}
{\quad}.\eqno(2.21)$$
Both $E_{ik}$ and $H_{ik}{\,}$, so defined, 
are the components of 3-dimensional (spatial) tensors, 
obeying the algebraic restrictions (symmetric, traceless):
$$\eqalignno{E_{ik}{\quad}&
={\quad}E_{(ik)}{\quad},
{\qquad}{\quad}h^{ik}{\,}E_{ik}{\quad}
={\quad}0
{\quad};&(2.22)\cr
H_{ik}{\quad}&
={\quad}H_{(ik)}{\quad},
{\qquad}{\quad}h^{ik}{\,}H_{ik}{\quad}
={\quad}0
{\quad},&(2.23)\cr}$$
where round brackets denote symmetrisation.  
By analogy with the vacuum Maxwell case 
for $E_{i}{\,},{\,}B_{i}{\,}$ above, here $E_{ik}$ 
and $H_{ik}$ also obey differential constraints 
on the bounding 3-surface.  From the Bianchi 
identities [42,43], one has ({\it in vacuo})
$${~}^{3}\nabla_{k}E^{ik}{\quad}
={\quad}0{\quad},
{\qquad}{\quad}^{3}\nabla_{k}H^{ik}{\quad}
={\quad}0
{\quad},\eqno(2.24)$$ 
where ${~}^{3}\nabla_{k}$ denotes the intrinsic 
3-dimensional covariant derivative, 
which preserves the 3-metric $h_{ij}{\;}$.

The classical Einstein-Hilbert action functional 
for gravity, with magnetic data $H_{ik}$ specified 
on the boundaries, will be discussed in Secs.5,7-9 below, 
for the case of weak anisotropic perturbations.
\end{section}

\begin{section}{Boundary conditions 
in two-component spinor language}
As mentioned in the Introduction, a more unified view 
of the boundary conditions for perturbed data, 
as specified on the initial and final space-like 
hypersurfaces $\Sigma_{I}$ and $\Sigma_{F}{\,}$, 
can be gained from their description in terms 
of 2-component spinors [15,42,43].  
In this Section, we again begin with $s=1$ 
Maxwell perturbations.

Consider (in Lorentzian signature) a real Maxwell 
field-strength tensor $F_{\mu\nu}{\,}$, 
obeying the Maxwell equations (2.2,4).  
In the theory of 2-component spinors, 
a space-time index $\mu$ is related to a pair 
of spinor indices
$AA'{\;}{\;}({\;}A=0{\,},1{\;};{\;}A'=0'{\,},1')$ 
through the (hermitian) spinor-valued $1$-forms 
$e^{AA'}_{~~~~\mu}{\,}$, defined by [15,42,43]:
$$e^{AA'}_{~~~~\mu}{\quad}
={\quad}e^{a}_{~\mu}{\;}\sigma_{a}^{~AA'}
{\quad}.\eqno(3.1)$$
Here, $e^{a}_{~\mu}$ denotes a (pseudo-)orthonormal 
basis of 1-forms ${\;}(a=0{\,},1{\,},2{\,},3){\,}$, 
while $\sigma_{a}^{~AA'}$ denotes 
the Infeld-van der Waerden translation symbols [15,42,43].  
At this point, one has to make a definite decision 
between $SL(2,{\Bbb C})$ spinors, which are the most appropriate 
for studying real Lorentzian-signature geometry, 
and $SO(4)$ spinors, in which Riemannian geometry 
is most simply described [54].  We follow the Lorentzian 
spinor conventions of [15].  This does not prevent one 
from describing complex or Riemannian geometry 
-- one simply allows the 'tetrad' $e^{a}_{~\mu}$ of 1-forms
above to become suitably complex.

Knowledge of $F_{\mu\nu}$ at a point is equivalent 
to knowledge of 
$$F_{AA'BB'}{\quad}
={\quad}F_{\mu\nu}{\;}
e_{AA'}^{~~~~~\mu}{\;}e_{BB'}^{~~~~~~\nu}
{\quad}\eqno(3.2)$$
at that point.  Here, $F_{AA'BB'}$ is hermitian, 
for real $F_{\mu\nu}{\,}$.  
Further, the space-time antisymmetry 
${\,}F_{\mu\nu}{\,}={\,}F_{[\mu\nu]}{\,}$ 
implies that the decomposition 
$$F_{AA'BB'}{\quad}
={\quad}\epsilon_{AB}{\;}\tilde\phi_{A'B'}{\,}
+{\,}\epsilon_{A'B'}{\;}\phi_{AB}
{\quad}\eqno(3.3)$$
holds, where ${\,}\epsilon_{AB}{\,}$ 
and ${\,}\epsilon_{A'B'}{\,}$ are the unprimed and primed 
alternating spinors [42,43], and where
$$\phi_{AB}{\quad}
={\quad}{{1}\over{2}}{\,}F_{AA'B}^{~~~~~~A'}{\quad}
={\quad}\phi_{BA}
\eqno(3.4)$$
is a symmetric spinor.  In the present Lorentzian case 
with real Maxwell field, $\tilde\phi_{A'B'}$ is the spinor
hermitian-conjugate to $\phi_{AB}{\,}$.  
(In the Riemannian context, Eq.(3.3) would give 
the splitting of the Maxwell field strength 
into its self-dual and anti-self-dual parts [26,44].)  
Knowledge of the 3 complex components of $\phi_{AB}$ 
at a point is equivalent to knowledge of the 6 real 
components of $F_{\mu\nu}$ at that point; 
also, the $\phi_{AB}$ are, in principle, 
physically measurable, just as the $F_{\mu\nu}$ are.

In terms of the dual field strength ${~}^{*}F_{\mu\nu}$ 
of Eq.(2.5), one finds that
$$F_{\mu\nu}{\,}+{\,}i{\;}^{*}F_{\mu\nu}{\quad}
={\quad}2{\,}\phi_{AB}{\;}\epsilon_{A'B'}{\,}
e^{AA'}_{~~~~\mu}{\,}e^{BB'}_{~~~~~\nu}
{\quad},\eqno(3.5)$$
together with the conjugate equation.  
The vacuum Maxwell field equations (2.2,4) 
can then be combined to give
$$\nabla^{AA'}\phi_{AB}{\quad}
={\quad}0{\quad},
{\qquad}\nabla^{AA'}\tilde\phi_{A'B'}{\quad}
={\quad}0
{\quad},\eqno(3.6)$$
where 
${\,}\nabla^{AA'}{\,}={\;}e^{AA'\mu}{\,}\nabla_{\mu}{\;}$.

Here, we are again interested in the decomposition 
of the Maxwell field strength with respect 
to a space-like bounding hypersurface and its associated 
unit (future-directed) normal vector $n^{\mu}{\,}$.  
Define the normal spinor
$$n^{AA'}{\quad}
={\quad}n^{\mu}{\,}e^{AA'}_{~~~~\mu}
{\quad}.\eqno(3.7)$$
Then the (purely spatial) electric and magnetic field 
vectors $E_{k}$ and $B_{k}$ can be expressed through
$$\eqalignno{E_{k}+i{\,}B_{k}
&{\quad}
={\quad}2{\,}\phi_{AB}{\;}n^{A}_{~~B'}{\,}
e^{BB'}_{~~~~~k}
{\quad},&(3.8)\cr
E_{k}-i{\,}B_{k}
&{\quad}
={\quad}2{\,}\bar\phi_{A'B'}{\;}n^{~~A'}_{B}{\,}
e^{BB'}_{~~~~k}
{\quad}.&(3.9)\cr}$$ 
In 4-vector language, the corresponding co-vector 
fields $E_{\mu}$ and $B_{\mu}$ are defined by
$$E_{\mu}{\quad}
={\quad}n^{\nu}{\,}F_{\nu\mu}{\quad},
{\qquad}B_{\mu}{\quad}
={\quad}n^{\nu}{\;}^{*}F_{\nu\mu}
{\quad},\eqno(3.10)$$
obeying
$$n^{\mu}{\,}E_{\mu}{\quad} 
={\quad}0{\quad}
={\quad}n^{\mu}{\,}B_{\mu}
{\quad}.\eqno(3.11)$$
Next, for ${\,}\epsilon{\;}={\,}\pm 1{\;}$, define
$$\Psi^{AB}_{\epsilon}{\quad} 
={\quad}2{\,}\epsilon{\,}n^{A}_{~~A'}{\,}n^{B}_{~~B'}{\,}
\tilde\phi^{A'B'}+{\,}\phi^{AB}
{\quad}.\eqno(3.12)$$ 
Here, ${\,}\Psi^{AB}_{\epsilon}{\,}$ is symmetric 
on $A$ and $B{\,}$; 
this spinor may be re-expressed in terms of $E_{k}$ 
and $B_{k}{\,}$, on making use of the symmetry 
of ${\,}n_{B}^{~~B'}{\,}e^{BA'k}{\,}$ on its free spinor 
indices $B'A'$ [15].  Here, we define
$$e^{BA'k}{\quad} 
={\quad}h^{k\ell}{\;}e^{BA'}_{~~~~~\ell}
{\quad},\eqno(3.13)$$
where $h^{k\ell}$ is the inverse spatial metric.  
The above symmetry property then reads
$$n_{B}^{~~B'}{\,}e^{BA'k}{\quad}
={\quad}n_{B}^{~~A'}{\,}e^{BB'k}
{\quad}.\eqno(3.14)$$
From Eq.(3.14), we find the decomposition
$$\Psi^{AB}_{\epsilon}{\quad} 
={\quad}n^{B}_{~~B'}{\,}e^{AB'k}{\,}
\Bigl[(\epsilon -1)E_{k} 
-i{\,}(\epsilon +1){\,}B_{k}\Bigr]
{\quad}.\eqno(3.15)$$
In particular, our boundary condition in Sec.6 below 
of fixing the magnetic field (a spatial co-vector field) 
on each of the initial and final space-like hypersurfaces 
$\Sigma_{I}$ and $\Sigma_{F}$ is equivalent to fixing 
the spinorial expression
$$\Psi^{AB}_{+}{\quad} 
={\;}-{\,}2{\,}i{\,}n^{B}_{~~B'}{\,}e^{AB'k}{\,}B_{k}
{\quad}\eqno(3.16)$$
on each boundary.  Note that, even though we regard $B_{k}$ 
as having 3 real components, the left-hand side, 
being symmetric on $(AB){\,}$, 
appears to have 3 complex components.  
In fact, $\Psi^{AB}_{+}$ as defined through Eq.(3.16) 
obeys a further hermiticity requirement, appropriate 
for spinors in 3 Riemannian dimensions 
(that is, on the hypersurfaces $\Sigma_{I}$ 
and $\Sigma_{F}$) [15,42,43], so re-balancing matters.

For comparison with much of the work done on black holes 
and their perturbations, one needs the Newman-Penrose 
formalism [55] 
-- an essentially spinorial description of the geometry.  
Here, considering only unprimed spinors at present, 
a normalised dyad $(o^{A}{\,},{\,}\iota^{A})$ 
at a point is defined to be a basis 
for the 2-complex-dimensional vector space 
of spinors ${\,}\omega^{A}$ at that point, 
normalised according to
$$o_{A}{\,}\iota^{A}{\quad}
={\quad}1{\quad}
={\,}-{\,}\iota_{A}{\,}o^{A}
{\quad}.\eqno(3.17)$$ 
The unprimed field strength 
${\,}\phi_{AB}{\,}={\,}\phi_{(AB)}{\,}$ 
can be projected onto the dyad, 
to give the 3 Newman-Penrose quantities
$$\phi_{0}{\;}
={\;}\phi_{AB}{\;}o^{A}{\,}o^{B}{\quad},
{\quad}\phi_{1}{\;}
={\;}{{1}\over{2}}{\,}\phi_{AB}{\;}o^{A}{\,}\iota^{B}{\quad},
{\quad}\phi_{2}{\;}
={\;}\phi_{AB}{\;}\iota^{A}{\,}\iota^{B}
{\quad},\eqno(3.18)$$ 
each of which is a complex scalar field (function).  
Using the Newman-Penrose formalism to describe 
perturbations in the background of a rotating Kerr 
black-hole geometry, Teukolsky [56] derived decoupled 
separable equations for the quantities 
${\,}\phi_{0}{\;}{\;}(s=1){\,}$ 
and 
${\,}r^{2}{\,}\phi_{2}{\;}{\;}(s=-1){\,}$ 
(for further review, see [39,57].)  
In our non-rotating case, with spherically-symmetric 
background, the Newman-Penrose quantity of most interest 
to us, following the work of this paper, is $\phi_{1}{\,}$.  
In the language of [55,58], $\phi_{1}$ has spin 
and conformal weight zero.  This is best described 
in the Kinnersley null tetrad [59] for the Schwarzschild 
or Kerr geometry, in our coordinate system.  
A null tetrad [55] 
${\,}\ell^{\mu}{\,},{\,}n^{\mu}{\,},
{\,}m^{\mu}{\,},{\,}\bar m^{\mu}{\,}$ 
of vectors at a point is a set obeying, 
in an obvious notation, 
${\,}{\ell}.{\ell}=n.n=m.m={\bar m}.{\bar m}=0{\;}{\,};
{\;}{\ell}.n=-{\,}m.{\bar m}=1{\;}{\,};
{\;}{\ell}.m={\ell}.{\bar m}=n.m=n.{\bar m}=0{\;}$.
Knowledge of such a null tetrad is equivalent to knowledge 
of the corresponding normalised spinor dyad 
$(o^{A}{\,},{\,}\iota^{A}){\,}$, 
through the relations
$$\eqalign{l^{\mu}{\quad}
&\leftrightarrow{\quad}o^{A}{\,}o^{A'}{\quad}, 
{\qquad}{\quad}n^{\mu}{\quad}
\leftrightarrow{\quad}\iota^{A}{\,}\iota^{A'}{\quad},\cr
m^{\mu}{\quad}
&\leftrightarrow{\quad}o^{A}{\,}\iota^{A'}{\quad},
{\qquad}{\quad}\bar m^{\mu}{\quad}
\leftrightarrow{\quad}\iota^{A}{\,}o^{A'}
{\quad}.\cr}\eqno(3.19)$$
In terms of the Regge-Wheeler variables to be used 
for the decomposition of the linearised Maxwell field 
strength $F^{(1)}_{\mu\nu}$ given in Secs.4,6, one has
$$\phi_{1}{\quad}
={\;}{\,}{{1}\over{2r^{2}}}\sum_{\ell m}
\Bigl(\psi^{(e)}_{1\ell m}
+i{\,}\psi^{(o)}_{1\ell m}\Bigr){\,}Y_{\ell m}(\Omega)
{\quad},\eqno(3.20)$$
where the $Y_{\ell m}(\Omega)$ are the normalised spherical 
harmonics of [60].  As a result, 
one finds that ${\,}r^{2}{\,}\phi_{1}{\,}$
obeys the wave equation (4.47) below.  
With regard to the boundary conditions on $\Sigma_{I}$ 
and $\Sigma_{F}{\,}$, when the variables 
$\psi^{(e)}_{1\ell m}$ 
and 
$\psi^{(o)}_{1\ell m}$ 
are being used, the correct boundary data (Sec.6) 
will involve specifying both ${\,}\psi^{(o)}_{1\ell m}{\,}$ 
and 
${\,}\partial_{t}\psi^{(e)}_{1\ell m}{\,}$ 
on $\Sigma_{I}$ and $\Sigma_{F}{\,}$.

Turning again to the gravitational field, 
in 2-component spinor language [42,43], 
one has the decomposition of the Weyl tensor: 
$$C_{\alpha\beta\gamma\delta}{\quad}
\leftrightarrow{\quad}\epsilon_{AB}{\,}\epsilon_{CD}{\;}
\tilde\Psi_{A'B'C'D'}{\,}
+{\,}\epsilon_{A'B'}{\,}\epsilon_{C'D'}{\;}\Psi_{ABCD}
{\quad},\eqno(3.21)$$
where6
$$\Psi_{ABCD}{\quad}
={\quad}\Psi_{(ABCD)}
\eqno(3.22)$$
is the totally symmetric (complex) Weyl spinor, 
and (in Lorentzian signature) $\tilde\Psi_{A'B'C'D'}{\,}$ 
is its hermitian conjugate.  
The dual of the Weyl tensor is defined as
$$^{*}C_{\alpha\beta\gamma\delta}{\;}{\,}
={\;}{{1}\over{2}}{\,}\eta_{\alpha\beta\rho\sigma}{\,}
C^{\rho\sigma}_{~~~\gamma\delta}
{\quad}.\eqno(3.23)$$
One finds that:
$$\Bigl(C_{\alpha\beta\gamma\delta}{\,} 
+{\,}i{\;}^{*}C_{\alpha\beta\gamma\delta}\Bigr){\quad}
\leftrightarrow{\quad}2{\,}\epsilon_{A'B'}{\,}
\epsilon_{C'D'}{\,}\Psi_{ABCD}
{\quad},\eqno(3.24)$$
together (in Lorentzian signature) 
with the hermitian-conjugate equation. 
(If instead we had used the Euclidean definition 
of spinors, then Eq.(3.21) would describe the splitting 
of the Weyl tensor into self-dual 
and anti-self-dual parts [44,58].)  
The (vacuum) Bianchi identities read [42,43]:
$$\nabla^{AA'}{\,}\Psi_{ABCD}{\quad}
={\quad}0{\quad},
{\qquad}\nabla^{AA'}{\,}\tilde\Psi_{A'B'C'D'}{\quad}
={\quad}0
{\quad}.\eqno(3.25)$$
On the bounding surface $\Sigma_{F}$ (say), 
one finds analogously that 
$$E_{k\ell}{\,}+i{\,}H_{k\ell}{\quad} 
={\quad}2{\,}\Psi_{ABCD}{\,}
\bigl(n^{A}_{~~B'}{\,}e^{BB'}_{~~~~k}\bigr)
\bigl(n^{C}_{~~D'}{\,}e^{DD'}_{~~~~~\ell}\bigr)
\eqno(3.26)$$
and its hermitian conjugate.  
Thence, the magnetic tensor $H_{k\ell}$ is given by
$$H_{k\ell}{\quad} 
={\;}{\,}\biggl(-{\,}i{\,}\Psi_{ABCD}{\,}
\Bigl(n^{A}_{~~B'}{\,}e^{BB'}_{~~~~k}\Bigr)
\Bigl(n^{C}_{~~D'}{\,}e^{DD'}_{~~~~~\ell}\Bigr)\biggr)
+\Bigl[{\rm h.c.}\Bigr]
{\quad},\eqno(3.27)$$
with a corresponding equation for $E_{k\ell}{\,}$.  
These two equations can straightforwardly be inverted 
to give an expression analogous to $\Psi^{AB}_{+}{\,}$ 
for the $s=1$ Maxwell case of Eq.(3.16).  
This analogous expression,
${\,}{\Psi}^{ABCD}_{+}{\,}={\,}{\Psi}^{(ABCD)}_{+}{\,}$, 
is again totally symmetric on its indices, 
and is given for ${\,}{\epsilon}{\,}=\pm{\,}1{\,}$ 
by the generalisation of Eq.(3.12), as
$${\Psi}^{ABCD}_{\epsilon}{\quad}
={\quad}4{\,}{\epsilon}{\,}n^{A}_{~~A'}{\,}n^{B}_{~~B'}{\,}
n^{C}_{~~C'}{\,}n^{D}_{~~D'}{\;}{\tilde \Psi^{A'B'C'D'}}
+{\;}{\Psi}^{ABCD}
{\quad}.\eqno(3.28)$$
Thus, again for the magnetic case ${\,}{\epsilon}=+{\,}1{\;}$, 
${\;}{\Psi}^{ABCD}_{+}{\,}$ provides a spinorial version 
of the magnetic part $H_{k\ell}{\,}$ of the Weyl tensor, 
to be fixed on the final boundary $\Sigma_{F}{\,}$ 
in our $s=2{\,}$ treatment of the gravitational
boundary-value problem, perturbed about spherically-symmetric 
collapse.  Of course, the perturbative boundary data 
$H_{k\ell}{\,}$ must further be chosen 
such that the divergence condition (2.24) holds: 
$^{3}\nabla_{k}H^{ik}{\,}={\,}0{\,}$ 
on ${\,}\Sigma_{F}{\,}$, just as, from Eq.(2.13) 
in the Maxwell case, the condition
${\,}^{3}\nabla_{k}B^{k}{\,}={\,}0{\,}$ must hold.

Two of the five complex components of ${\,}\Psi_{ABCD}{\,}$ 
are contained in the Newman-Penrose quantities [42,43,55]
$$\Psi_{0}{\quad} 
={\quad}\Psi_{ABCD}{\;}
o^{A}{\,}o^{B}{\,}o^{C}{\,}o^{D}{\quad},
{\qquad}\Psi_{4}{\quad}
={\quad}\Psi_{ABCD}{\;}
\iota^{A}{\,}\iota^{B}{\,}\iota^{C}{\,}\iota^{D}
{\quad},\eqno(3.29)$$
where ${\,}(o^{A}{\,},{\,}\iota^{A})$ is a normalised 
spinor dyad [43,43,55].  Again taking the Kinnersley 
null tetrad [39,59], it was further shown 
by Teukolsky [56] that $\Psi_{0}$ and ${\,}\Psi_{4}$ 
each obey decoupled separable wave equations.  
Following work by Chrzanowski [61], 
it was confirmed by Wald [62] that, given a solution 
$\Psi_{0}$ or $\Psi_{4}$ of the Teukolsky equation 
for a (nearly-) Kerr background, 
all the vacuum metric perturbations can be reconstructed 
in a certain gauge through a series of simple linear 
operations on $\Psi_{0}$ (or on $\Psi_{4}$) [39,61,62].  
Once the linearised metric perturbations are known, 
then, of course, one can compute other Newman-Penrose 
quantities at linearised order, such as
$$\Psi_{2}{\quad}
={\quad}\Psi_{ABCD}{\,}o^{A}{\,}o^{B}{\,}\iota^{C}{\,}\iota^{D}
{\quad}.\eqno(3.30)$$
Note here that, for an unperturbed Schwarzschild 
background (say), only the middle Newman-Penrose quantity 
${\Psi}_{2}{\,}$, out of the set 
${\Psi}_{i}{\;}{\,}(i=0{\,},\ldots ,4{\,}){\,}$, 
is non-zero, with
${\,}{\Psi}_{2}{\,}=-{\,}M/{r^3}{\,}$ 
in Schwarzschild coordinates [57].

The analogous process is implicit in the (Maxwell) 
discussion above, leading to Eq.(3.20):  
For $s=1$ perturbations of the Kerr metric, 
the corresponding linearised Maxwell vector potential 
$A_{\mu}$ (in a particular gauge) can be reconstructed 
by simple steps from the Newman-Penrose quantites 
${\,}\phi_{0}$ or ${\,}\phi_{2}{\,}$ [39,61,62].
Hence, the middle Newman-Penrose quantity ${\,}\phi_{1}{\,}$ 
can also be found, leading to Eq.(3.20).

In the more complicated $s=2$ case, 
although we have not yet finished detailed calculations 
on this point, it does now look reasonable to expect that, 
for gravitational perturbations 
about a spherically-symmetric gravitational collapse, 
there should exist a relation analogous 
to the ${\,}s=1{\,}$ Eq.(3.20).  In this case, 
for a Schwarzschild background, this relation would involve 
expanding out the first-order perturbations 
of the middle Newman-Penrose quantity ${\Psi}_{2}{\,}$ 
in terms of the $s=2$ Regge-Wheeler description 
of Secs.5,7-9 below.
\end{section}

\begin{section}{Regge-Wheeler formalism 
-- Maxwell case}
A more unified treatment of the angular harmonics 
appearing in the separation process 
for $s=1$ (Maxwell) and $s=2$ (graviton) perturbations 
of a spherically-symmetric background can be given 
in terms of vector and tensor harmonics [63].  
In [11], we expanded the $s=0$ (massless-scalar) 
perturbations in terms of scalar spherical harmonics 
$Y_{\ell m}(\theta{\,},\phi){\,}$, which have even parity.  
Vector and tensor spherical harmonics, however, 
can have odd as well as even parity.

Any vector field in a spherically-symmetric background, 
such as the classical $s=1$ (photon) solutions appearing 
in the background gravitational-collapse geometry , 
can be expanded in terms of vector spherical harmonics 
on the unit 2-sphere [37,38].  
Correspondingly, angular vector and tensor indices 
are raised and lowered with the metric 
$\hat\gamma_{ab}{\,}$, given by 
$$\hat\gamma_{\theta\theta}{\;}{\,}
={\;}{\,}1{\quad},
{\qquad}\hat\gamma_{\phi\phi}{\;}{\,}
={\;}{\,}\sin^{2}\theta{\quad},
{\qquad}\hat\gamma_{\theta\phi}{\;}{\,}
={\;}{\,}\hat\gamma_{\phi\theta}{\;}{\,}
={\quad}0
{\quad}.\eqno(4.1)$$
The even-parity vector harmonics [37,38] 
have angular components
$$(\Psi_{\ell m})_a{\quad}
={\quad}\bigl(\partial_{a}Y_{\ell m}\bigr)
{\quad},\eqno(4.2)$$ 
where ${\,}a=(\theta{\,},\phi){\,}$.  
The odd-parity vector harmonics are
$$(\Phi_{\ell m})_{a}{\quad} 
={\quad}\epsilon_{a}^{~~b}{\,}
\bigl(\partial_{b}Y_{\ell m}\bigr)
{\quad}.\eqno(4.3)$$
Here, $\epsilon_{a}^{~~b}$ denotes the tensor 
with respect to angular indices 
$(a=\theta{\,},\phi{\;};{\;}b=\theta{\,},\phi){\,}$, 
such that the lowered version 
${\,}\epsilon_{ab}=-{\,}\epsilon_{ba}{\,}$ 
is anti-symmetric, with
${\,}\epsilon_{01}
=\bigl({\hat\gamma}\bigr)^{{1}\over{2}}
=\sin\theta{\;}$,
where
${\,}{\hat\gamma}={\rm det}\bigl(\hat \gamma_{ab}\bigr){\,}$.  
Thus,
$$\epsilon_{\theta}^{~~\phi}{\;}{\,}
={\;}{\,}{{-{\,}1}\over{\sin\theta}}{\quad},
{\qquad}\epsilon_{\phi}^{~~\theta}{\;}{\,}
={\;}{\,}\sin\theta{\quad},
{\qquad}\epsilon_{\phi}^{~~\phi}{\;}{\,}
={\;}{\,}\epsilon{_\theta}^{~\theta}{\;}{\,} 
={\;}{\,}0
{\quad}.\eqno(4.4)$$
The forms of the angular harmonics appearing 
in the $s=1$ photon calculations below can be deduced 
from these vector-spherical-harmonic expressions.

Analogously, any rank-2 tensor field 
such as a linearised (graviton) $s=2$ classical solution, 
to be treated in Secs.5,7-9, below, 
can be expanded in terms of tensor spherical harmonics.  
The even-parity tensor harmonics are
$$(\Psi_{\ell m})_{ab}{\quad}
={\quad}Y_{\ell m\mid ab}{\quad}, 
{\qquad}(\Phi_{\ell m})_{ab}{\quad}
={\quad}\hat\gamma_{ab}{\;}Y_{\ell m}
{\quad},\eqno(4.5)$$
where a bar $\mid$ denotes a covariant derivative 
with respect to the metric $\hat\gamma_{ab}{\,}$.  
The odd-parity tensor harmonics are
$$(\chi_{\ell m})_{ab}{\quad}
={\quad}{{1}\over{2}}
\biggl[\epsilon_{a}^{~~c}\Bigl(\Psi_{\ell m}\Bigr)_{cb} 
+{\,}\epsilon_{b}^{~~c}
\Bigl(\Psi_{\ell m}\Bigr)_{ca}\biggr]
{\quad}.\eqno(4.6)$$

From 1957, Regge and Wheeler [63] developed 
the formalism for treating both spin-1 and spin-2 
classical perturbations of the Schwarzschild solution 
and of other spherically-symmetric solutions, 
corresponding to Maxwell and gravitational (graviton) 
perturbations.  Here, in the Regge-Wheeler (RW) formalism, 
for Maxwell theory we decompose the real linearised field 
strength $F^{(1)}_{\mu\nu}$ and linearised vector 
potential $A^{(1)}_{\mu}$ into tensor and vector 
spherical harmonics, respectively [37,38].  
We are assuming that the background (unperturbed) 
classical solution consists, as above, 
of a spherically-symmetric gravitational 
and massless-scalar field 
$(\gamma_{\mu\nu}{\,},{\,}\Phi)$, 
with no background Maxwell field:  
$A^{(0)}_{\mu}=0{\,},{\;}F^{(0)}_{\mu\nu}=0{\,}$.

For each spin $s=0{\,},1{\,},2{\,}$, 
the corresponding perturbation modes split 
into those with even parity and those with odd parity.  
Under the parity inversion: 
$\theta\rightarrow(\pi-\theta){\,},
{\;}\phi\rightarrow(\pi+\phi){\,}$, 
we define the even perturbations as those with parity 
$\pi=(-1)^{\ell}{\,}$, while the odd perturbations 
have parity $\pi=(-1)^{\ell +1}{\,}$.  
For Maxwell theory ${\,}(s=1){\;}$, the $\ell=0$ mode 
corresponds to a static perturbation, 
in which a small amount of electric charge 
is added to the black hole; 
in particular, a Schwarzschild solution 
will be 'displaced' infinitesimally along the family 
of Reissner-Nordstr\"om solutions.  
For radiative modes with ${\,}\ell =1{\,}$ (dipole) 
and higher, we set
$$\eqalignno{F^{(1)}_{\mu\nu}(x){\quad}&
={\quad}\sum^{\infty}_{\ell=1}\sum^{\ell}_{m=-\ell} 
{\,}\biggl[\Bigl(F^{(o)}_{\mu\nu}\Bigr)_{\ell m}
+\Bigl(F^{(e)}_{\mu\nu}\Bigr)_{\ell m}\biggr]
{\quad},&(4.7)\cr
A^{(1)}_{\mu}(x){\quad}&
={\quad}\sum^{\infty}_{\ell=1}\sum^{\ell}_{m=-\ell}
{\,}\biggl[\Bigl(A^{(o)}_{\mu}\Bigr)_{\ell m}
+\Bigl(A^{(e)}_{\mu}\Bigr)_{\ell m}\biggr]
{\quad}.&(4.8)\cr}$$
On substituting this decomposition into the boundary 
expression (2.7) for the classical Maxwell action 
${\,}S^{EM}_{\rm class}{\;}$, we find
$$\eqalign{S^{EM}_{\rm class}
=&{\,}-{\,}{{1}\over{8\pi}}\sum_{\ell m\ell' m'}
{\int}d\Omega\int^{R_{\infty}}_{0}dr{\,}r^{2}{\,}
e^{(a-b)/2}{\,}\gamma^{ij}\Bigl(A^{(o)}_{j}\Bigr)_{\ell m}
\Bigl(F^{(o)}_{ti}\Bigr)^{*}_{\ell' m'}
\biggl\bracevert^{\Sigma_{F}}_{\Sigma_{I}}\cr
&-{\,}{{1}\over{8\pi}}\sum_{\ell m\ell' m'}{\int}d\Omega
\int^{R\infty}_{0}dr{\,}r^{2}{\,}e^{(a-b)/2}{\,}\gamma^{ij}
\Bigl(A^{(e)}_{j}\Bigr)_{\ell m}
\Bigl(F^{(e)}_{ti} \Bigr)^{*}_{\ell' m'}
\biggr\bracevert^{\Sigma_{F}}_{\Sigma_{I}}
{\,},\cr}\eqno(4.9)$$  
where $\Omega$ denotes the angular coordinates 
$\theta{\,},\phi{\,}$, and we work at present 
with the Lorentzian form of the spherically-symmetric 
background metric 
(for which the Riemannian form was given in Eq.(1.7), 
by suitable choice of coordinates), writing:
$$ds^{2}{\;}{\,}
={\,}-{\,}e^{b(t,r)}{\,}dt^{2}+{\,}e^{a(t,r)}{\,}dr^{2}
+{\,}r^{2}{\,}(d\theta^{2}+{\,}\sin^{2}\theta{\;}d\phi^{2})
{\;}.\eqno(4.10)$$
For later reference, we introduce a standard expression 
for such a spherically-symmetric space-time, 
defining the 'mass function' $m(t{\,},r)$ by
$$e^{-a}{\;}{\,}
={\;}{\,}1{\,}-{\,}{{2m(t{\,},r)}\over{r}}
{\quad},\eqno(4.11)$$
within the Vaidya-like region of the space-time, 
in which the black hole is evaporating.  
Clearly, the odd and even contributions decouple
in the action of Eq.(4.9).

For the subsequent detailed treatment 
of the angular harmonics involved, we follow Zerilli's 
decomposition [64] of $F^{(1)}_{\mu\nu}$ and
$A^{(1)}_{\mu}{\,}$.  Thus, we set
$$F^{(1)}_{\mu\nu}(x){\;}{\,}
={\;}{\,}\sum_{\ell m}\Bigl(F^{(1)}_{\mu\nu}\Bigr)_{\ell m}
{\quad},\eqno(4.12)$$ 
$$\Bigl(F^{(1)}_{\mu\nu}\Bigr)_{\ell m}{\;}{\,}
={\;}{\,}\pmatrix{0& {f_{1}}& {f_{2}}& {f_{3}}\cr
{-f_{1}}& 0& {f_{4}}&{f_{5}}&\cr
{-f_{2}}& {-f_{4}}& 0& {f_{6}}\cr
{-f_{3}}& {-f_{5}}& {-f_{6}}& 0\cr}
{\quad}.\eqno(4.13)$$
For a given choice of $(\ell{\,},m){\,}$, we take
$$\eqalignno{f_{1}{\quad}&
={\;}{\,}\Bigl(\hat F^{(e)}_{tr}\Bigr)_{\ell m}
Y_{\ell m}(\Omega)
{\quad},&(4.14)\cr
f_{2}{\quad}&
={\;}{\,}\bigl(\sin\theta\bigr)^{-1}
\Bigl(\hat F^{(o)}_{t\theta}\Bigr)_{\ell m}
\bigl(\partial_{\phi}Y_{\ell m}\bigr)
+\Bigl(\hat F^{(e)}_{t\theta}\Bigr)_{\ell m}
\bigl(\partial_{\theta}Y_{\ell m}\bigr)
{\quad},&(4.15)\cr
f_{3}{\quad}&
={\;}-{\,}\Bigl(\hat F^{(o)}_{t\theta}\Bigr)_{\ell m}
\bigl(\sin\theta\bigr)\bigl(\partial_{\theta}Y_{\ell m}\bigr)
+{\;}\Bigl(\hat F^{(e)}_{t\theta}\Bigr)_{\ell m}
\bigl(\partial_{\phi}Y_{\ell m}\bigr)
{\quad},&(4.16)\cr
f_{4}{\quad}&
={\;}{\,}\bigl(\sin\theta\bigr)^{-1}
\Bigl(\hat F^{(o)}_{r\theta}\Bigr)_{\ell m}
\bigl(\partial_{\phi}Y_{\ell m}\bigr) 
+\Bigl(\hat F^{(e)}_{r\theta}\Bigr)_{\ell m}
\bigl(\partial_{\theta}Y_{\ell m}\bigr)
{\quad},&(4.17)\cr
f_{5}{\quad}&
={\;}-{\,}\Bigl(\hat F^{(o)}_{r\theta}\Bigr)_{\ell m}
\bigl(\sin\theta\bigr)\bigl(\partial_{\theta}Y_{\ell m}\bigr)
+\Bigl(\hat F^{(e)}_{r\theta}\Bigr)_{\ell m}
\bigl(\partial_{\phi}Y_{\ell m}\bigr)
{\quad},&(4.18)\cr
f_{6}{\quad}&
={\;}{\,}\Bigl(\hat F^{(o)}_{\theta\phi}\Bigr)_{\ell m}
\bigl(\sin\theta\bigr){\;}Y_{\ell m}(\Omega)
{\quad}.&(4.19)\cr}$$
Here, the ${\,}Y_{\ell m}(\Omega){\,}$ are scalar spherical 
harmonics [60], and a caret indicates that the quantity 
is a function of $t$ and $r$ only.

Again, following [63], for the vector potential we set
$$\eqalignno{\Bigl(A^{(o)}_{\mu}\Bigr)_{\ell m}(x)&
=\biggl(0,0,
{{a_{2\ell m}(t,r)\bigl(\partial_{\phi}Y_{\ell m}\bigr)}
\over{\ell(\ell +1)\sin\theta}},
{{-{\,}a_{2\ell m}(t,r)\bigl(\sin\theta\bigr)
\bigl(\partial_{\theta}Y_{\ell m}\bigr)}\over{\ell(\ell +1)}}\biggr)
{\,},&(4.20)\cr
\Bigl(A^{(e)}_{\mu}\Bigr)_{\ell m}(x)&
=\biggl(-{\,}a_{0\ell m}(t,r){\,}Y_{\ell m}(\Omega){\,},
{\,}a_{1\ell m}(t,r){\;}Y_{\ell m}(\Omega){\,},{\,}0{\,},{\,}0\biggr)
{\;}.&(4.21)\cr}$$
Now Eq.(4.9) can be expanded out in the form:
$$\eqalign{S^{EM}_{\rm class}{\;}&
={\,}-{\,}{{1}\over{8\pi}}\sum_{\ell m\ell' m'}
\int d\Omega\int^{R\infty}_{0}dr{\;}
e^{(a-b)/2}{\quad}\times\cr
&\times{\;}{\,}\Biggl[\Bigl(A_{\theta}^{(o)}\Bigr)_{\ell m}
\Bigl(F^{(o)}_{t\theta}\Bigr)^{*}_{\ell' m'} 
+{{\Bigl(A^{(o)}_{\phi}\Bigr)_{\ell m}
\Bigl(F^{(o)}_{t\phi}\Bigr)^{*}_{\ell' m'}}
\over{\sin^{2}\theta}}\Biggr]
\biggl\bracevert^{\Sigma_{F}}_{\Sigma_{I}}\cr
&-{\,}{{1}\over{8\pi}}\sum_{\ell m \ell' m'}\int 
d\Omega\int^{R\infty}_{0}dr{\;}r^{2}{\,}
e^{(a-b)/2}\Bigl(A^{(e)}_{r}\Bigr)_{\ell m}
\Bigl(F^{(e)}_{t,r}\Bigr)^{*}_{\ell' m'}
\biggl\bracevert^{\Sigma_{F}}_{\Sigma_{I}}
{\;}.\cr}\eqno(4.22)$$

Of course, the components of the field strength are given 
in terms of those of the vector potential by Eq.(2.3); 
for example,
${\,}F^{(1)}_{t\theta}
={\,}\partial_{t}A^{(1)}_{\theta}
-{\,}\partial_{\theta}A^{(1)}_{t}{\,}$.
This gives the relations
$$\eqalignno{\Bigl(\hat F^{(o)}_{t\theta}\Bigr)_{\ell m}{\;}{\,}&
={\quad}{{\bigl(\partial_{t}a_{2\ell m}\bigr)}
\over{\ell(\ell +1)}}
{\quad},&(4.23)\cr
\Bigl(\hat F^{(e)}_{t\theta}\Bigr)_{\ell m}{\;}{\,}&
={\quad}a_{0\ell m}
{\quad},&(4.24)\cr
\Bigl(\hat F^{(o)}_{r\theta}\Bigr)_{\ell m}{\;}{\,}&
={\quad}{{\bigl(\partial_{r}a_{2\ell m}\bigr)
\over{\ell(\ell+1)}}}
{\quad},&(4.25)\cr
\Bigl(\hat F^{(e)}_{r\theta}\Bigr)_{\ell m}{\;}{\,}&
={\quad}a_{1\ell m}
{\quad},&(4.26)\cr
\Bigl(\hat F^{(e)}_{tr}\Bigr)_{\ell m}{\;}{\,}&
={\quad}\bigl(\partial_{r}a_{0\ell m}\bigr){\,}
-{\,}\bigl(\partial_{t}a_{1\ell m}\bigr)
{\quad},&(4.27)\cr
\Bigl(\hat F^{(o)}_{\theta\phi}\Bigr)_{\ell m}{\;}{\,}&
={\quad}a_{2\ell m}
{\quad}.&(4.28)\cr}$$
The action (4.22) then simplifies to give
$$\eqalign{S^{EM}_{\rm class}{\,} 
={\,}&-{\,}{{1}\over{8\pi}}\sum_{\ell m}{\,}\ell(\ell +1)
\int^{R\infty}_{0}dr{\,}e^{(a-b)/2}{\,}f^{(o)}_{\ell m}
\Bigl(\partial_{t}f^{(o)*}_{\ell m}\Bigr)
\biggl\bracevert^{\Sigma_{F}}_{\Sigma_{I}}\cr
&-{\,}{{1}\over{8\pi}}\sum_{\ell m}\int^{R\infty}_{0}
dr{\,}r^{2}{\,}e^{-(a+b)/2}{\,}a_{1\ell m}
\Bigl(\bigl(\partial_{t}a^{*}_{1\ell m}\bigr)
-\bigl(\partial_{r}a^{*}_{0\ell m}\bigr)\Bigr)
\biggl\bracevert^{\Sigma_{F}}_{\Sigma_{I}}
{\;},\cr}\eqno(4.29)$$
where
$$f^{(o)}_{\ell m}{\quad}
={\quad}{{a_{2\ell m}}\over{\ell(\ell +1)}}
\eqno(4.30)$$
determines the odd-parity Maxwell tensor 
{\it via} Eqs.(4.23,25,28).

The form of the classical action (4.29) 
can be further simplified by using the Maxwell field 
equations (2.2,4).  This will lead finally 
to the form (4.45) below, 
in which ${\,}S^{EM}_{\rm class}{\,}$ 
is expressed explicitly in terms of boundary data, 
as needed in the subsequent calculation 
of the quantum amplitude 
(see [11,16] for the spin-0 analogue).

The linearised Maxwell equations can be written as [49]:
$$F^{(1)\mu\nu}_{~~~~~~;\nu}{\quad}
={\quad}{(-{\,}\gamma)}^{-{{1}\over{2}}}{\,}\partial_{\nu}
\Bigl[{(-{\,}\gamma)}^{{1}\over{2}}{\,}F^{(1)\mu\nu}\Bigr]{\quad}
={\quad}0
{\quad}.\eqno(4.31)$$
The ${\,}\mu{\,}={\,}t{\,},r{\,}$ equations give
$$\eqalignno{\partial_{r}
\biggl[r^{2}\Bigl(\hat F^{(e)}_{tr}\Bigr)_{\ell m}\biggr]
-{\,}\ell(\ell+1){\,}e^{a}
\Bigl(\hat F^{(e)}_{t\theta}\Bigr)_{\ell m}{\;}{\,}&
={\quad}0
{\quad},&(4.32)\cr
e^{a}{\,}\partial_{t}\Bigl(\hat F^{(e)}_{t\theta}\Bigr)_{\ell m} 
-\partial_{r}\biggl[e^{-a}
\Bigl(\hat F^{(e)}_{r\theta}\Bigr)_{\ell m}\biggr]{\;}{\,}&
={\quad}0
{\quad},&(4.33)\cr 
\partial_{r}\biggl[e^{-a}
\Bigl(\hat F^{(o)}_{r\theta}\Bigr)_{\ell m}\biggr]
-e^{a}{\,}\partial_{t}\Bigl(\hat F^{(o)}_{t\theta}\Bigr)_{\ell m} 
-r^{-2}\Bigl(\hat F^{(o)}_{\theta\phi}\Bigr)_{\ell m}
{\;}{\,}&={\quad}0
{\quad},&(4.34)\cr
\partial_{t}\Bigl(\hat F^{(e)}_{tr}\Bigr)_{\ell m} 
-{{e^{-a}{\,}\ell(\ell+1)}\over{r^{2}}}{\,}
\Bigl(\hat F^{(e)}_{r\theta}\Bigr)_{\ell m}{\;}{\,}&
={\quad}0
{\quad}.&(4.35)\cr}$$
The ${\,}\mu =\theta{\,},{\,}\phi{\,}$ components 
give the same equations.  
Note that Eq.(4.32) is just the (source-free) 
constraint equation
${\,}\partial_{i}{\cal E}^{(1)i}{\,}={\,}0{\,}$ 
of Eq.(2.14).

The equations (4.23,25,28) together 
imply the decoupled wave equation for odd perturbations:
$$(\partial_{r^*})^{2}{\,}a_{2\ell m}
-(\partial_{t})^{2}{\,}a_{2\ell m}
-V_{1\ell}(r){\,}a_{2\ell m}{\quad}
={\quad}0
{\quad},\eqno(4.36)$$ 
where
$$V_{1\ell}(r){\;}{\,}
={\;}{\,}{{e^{-a}{\,}\ell(\ell+1)}\over{r^{2}}}{\quad}
>{\quad}0
\eqno(4.37)$$
is the (massless) spin-1 effective potential and where, 
as usual [63], we write 
${\,}\partial_{r^*}{\,}={\,}e^{-a}{\,}\partial_r{\;}$.  
If the geometry were exactly Schwarzschild, 
then the coordinate ${\,}r^{*}{\,}$ so defined 
would be the Regge-Wheeler or 'tortoise' radial 
coordinate, given by [49,63]:
$$r^{*}{\;}{\,}
={\;}{\,}r{\,}+{\,}2M\ln(r-{\,}2M)
{\quad}.\eqno(4.38)$$  
As in [11,16], we assume that the adiabatic approximation 
is valid in a neighbourhood of the initial and final 
surfaces, $\Sigma_{I}$ and $\Sigma_{F}{\,}$.  
In that case, we can, as before, 
effectively work with the field equations 
on a Schwarzschild background, 
except that the Schwarzschild mass $M_{0}$ is replaced 
by the mass function $m(t{\,},r){\,}$, 
as defined in Eq.(4.11), where $m(t{\,},r)$ 
varies extremely slowly with respect both to time and to radius.

Equation (4.33) gives 
${\,}\partial_{t}(\hat F^{(e)}_{t\theta})_{\ell m}{\,}$ 
also in terms of 
${\,}(\hat F^{(e)}_{r\theta})_{\ell m}{\,}$, 
while Eq.(4.35) gives
${\,}\partial_{t}(\hat F^{(e)}_{tr})_{\ell m}{\,}$ 
also in terms of 
${\,}(\hat F^{(e)}_{r\theta})_{\ell m}{\,}$.  
Together, Eqs.(4.27,33,35) imply that
$${(\partial_{r^*})}^{2}f^{(e)}_{\ell m}{\,}
-{\,}({\partial}_{t})^{2}f^{(e)}_{\ell m}{\,}
-{\,}V_{1\ell}{\;}f^{(e)}_{\ell m}{\quad}
={\quad}0
{\quad},\eqno(4.39)$$
where we define
$$f^{(e)}_{\ell m}{\;}{\,}
={\;}{\,}e^{-a}{\,}a_{1\ell m}
{\quad}.\eqno(4.40)$$
Thus, with a suitably defined variable 
${\,}f^{(e)}_{\ell m}{\,}$, 
the even perturbations obey the same decoupled wave 
equation (4.36) as the odd perturbations.

Finally [65], we set
$$\psi^{(e)}_{\ell m}(t{\,},r){\;}{\,}
={\;}{\,}{{r^{2}}\over{\ell(\ell+1)}}
\Bigl(\bigl(\partial_{t}a_{1\ell m}\bigr)
-\bigl(\partial_{r}a_{0\ell m}\bigr)\Bigr)
{\quad},\eqno(4.41)$$
which is clearly gauge-invariant.  
Now ${\,}\psi^{(e)}_{\ell m}{\,}$ 
is related simply to the (even-parity) function 
${\,}f^{(e)}_{\ell m}{\,}$ 
of the previous paragraph: 
Eqs.(4.33,39) imply that
$$\bigl(\partial_{t}\psi^{(e)}_{\ell m}\bigr){\;}{\,}
={\,}-{\,}f^{(e)}_{\ell m}
{\quad}.\eqno(4.42)$$ 

Equations (4.40,41) can now be used to simplify 
the classical Lorentzian action (4.29).  
For ease of comparison with the (second-variation) 
classical spin-2 action, where the pattern is similar, 
we define
$$\eqalignno{\psi^{(o)}_{1\ell m}{\quad}&
={\quad}a_{2\ell m}(t{\,},r)
{\quad},&(4.43)\cr
\psi^{(e)}_{1\ell m}{\quad}& 
={\quad}\ell(\ell+1){\,}\psi^{(e)}_{\ell m}(t{\,},r)
{\quad}.&(4.44)\cr}$$ 
Then, given weak-field Maxwell boundary data, 
specified by the linearised magnetic-field mode 
components ${\,}\{B^{(1)i}_{\ell m}\}{\,}$ 
on each of the boundaries $\Sigma_{I}$ 
and $\Sigma_{F}{\,}$, 
the corresponding classical Maxwell action is
$$\eqalign{&S^{EM}_{\rm class}
\Bigl[\bigl\{B^{(1)i}_{\ell m}\bigr\}\Bigr]\cr 
&={\,}{{1}\over{8\pi}}{\,}\sum_{\ell m}{\,}
{{(\ell -1)!}\over{(\ell +1)!}}\int^{R\infty}_{0}dr{\,}
e^{a}\biggl(\psi^{(e)}_{1\ell m}
\bigl(\partial_{t}\psi^{(e)*}_{1\ell m}\bigr)
-\psi^{(o)}_{1\ell m}
\bigl(\partial_{t}\psi^{(o)*}_{1\ell m}\bigr)\biggr)
\biggl\bracevert^{\Sigma_{F}}_{\Sigma_{I}}
{\,}.\cr}\eqno(4.45)$$
Of course, the limit $R_{\infty}\rightarrow\infty$ 
must be understood in Eq.(4.45).

Note further that, from Eqs.(4.36,39), one has
$$\bigl(\partial_{r*}{\,}\psi^{(e)}_{\ell m}\bigr){\;}{\,} 
={\,}-{\,}a_{0\ell m}
{\quad},\eqno(4.46)$$
whence ${\,}\psi^{(e)}_{1\ell m}{\,}$ 
also obeys the same decoupled wave equation (4.36,39) 
as for ${\,}a_{2\ell m}{\,}$ 
and for ${\,}f^{(e)}_{\ell m}{\,}$, namely,
$$(\partial_{r*})^{2}{\,}\psi^{(e)}_{1\ell m} 
-(\partial_{t})^{2}{\,}\psi^{(e)}_{1\ell m}
-V_{1\ell}{\;}\psi^{(e)}_{1\ell m}{\quad}
={\quad}0
{\quad}.\eqno(4.47)$$ 
The spin-1 radial equation (4.36,39,47) 
in a Schwarzschild background, 
both for odd- and even-parity Maxwell fields, 
was first given in 1962 by Wheeler [66].

This suggests a 'preferred route' for understanding 
the even-parity perturbations 
(which are more complicated than in the odd-parity case,
which only involves the single function 
${\,}a_{2\ell m}(t,r){\,}$, 
obeying the decoupled field equation (4.36)):  
Given suitable boundary conditions, 
one first solves the linear decoupled wave equation 
in two variables $t$ and $r{\,}$, namely, Eq.(4.47), 
for ${\,}\psi^{(e)}_{1\ell m}(t{\,},r){\,}$.  
By differentiation, following Eqs.(4.44,46), 
one then finds ${\,}a_{0\ell m}(t{\,},r){\,}$.  
Then, by integrating Eq.(4.41), one obtains also 
${\,}a_{1\ell m}(t{\,},r){\,}$, and hence, from Eq.(4.40), 
${\,}f^{(e)}_{\ell m}(t{\,},r){\,}$.  
From Eqs.(4.23-28), one now has all the non-zero 
components of the Maxwell field strength $F_{\mu\nu}{\,}$ 
in this even-parity case.
\end{section}

\begin{section}{Regge-Wheeler formalism 
-- odd-parity gravitational perturbations}
Our boundary-value problem, as posed 
in the Introduction and in [11,16], 
involves specifying on the final space-like hypersurface 
${\,}\Sigma_{F}{\,}$ the spatial components 
${\,}h^{(1)}_{ij}(x){\;}{\;}(i,j=1{\,},2{\,},3){\,}$, 
of the real perturbations 
${\,}h^{(1)}_{\mu\nu}(x){\,}$ of the 4-metric 
${\,}(\mu{\,},\nu=0{\,},1{\,},2{\,},3){\,}$.  
We shall construct a basis of tensor spherical harmonics 
with which to expand the angular dependence 
of $h^{(1)}_{ij}{\,}$.  
In general, we make a multipole decomposition 
for real metric perturbations, of the form:
$$h^{(1)}_{ij}(x){\;}{\,}
={\;}{\,}\sum^{\infty}_{\ell =2}
\sum^{\ell}_{m=-\ell}
\Bigl[\bigl(h^{(-)}_{ij}\bigr)_{\ell m}(x) 
+\bigl(h^{(+)}_{ij}\bigr)_{\ell m}(x)\Bigr]
{\quad},\eqno(5.1)$$
where $-$ and $+$ denote odd- and even-parity contributions,
respectively.  We comment below on the limit $\ell =2$ 
in the summation over $\ell{\,}$.

In Sec.4, $s=1$ (Maxwell) perturbations 
of spherically-symmetric gravitational backgrounds 
were treated in the Regge-Wheeler (RW) formalism [63], 
and split naturally into odd and even type, 
according to their behaviour under parity inversion:
$\theta\rightarrow{\;}(\pi-\theta){\,},
{\;}\phi\rightarrow{\;}(\pi +\phi){\,}$.  
The even 'electric-type' perturbations have parity 
${\,}\pi =(-1)^{\ell}{\,}$, while the odd 'magnetic-type' 
perturbations have parity ${\,}\pi =(-1)^{\ell +1}{\,}$.  
The analogous (orthogonal) decomposition also holds 
for the $s=2$ gravitational-wave perturbations.

In the $s=1$ Maxwell case, the lowest $\ell =0$ mode 
does not propagate:  
in the electric case, it corresponds to the addition 
of a small charge to the black hole, 
to turn a Schwarzschild solution 
into a Reissner-Nordstr\"om solution [49] 
with charge $Q{\,}{\ll}{\,}M{\,}$.
Correspondingly, in the $s=2$ case of gravitational 
perturbations, the multipoles with $\ell<2$ 
are non-radiatable.  For example, the even-parity 
gravitational perturbations with $\ell =0$ correspond 
to a small static charge in the mass, 
while the $\ell =0$ odd-parity perturbation 
is identically zero.  For $\ell =1{\,}$, 
the odd-parity (dipole) gravitational perturbations 
must be stationary [39], and even-parity dipole 
perturbations correspond to a coordinate 
displacement of the origin [67] and can be removed 
by a gauge transformation.  
For a general spin $s=0{\,},{\,}1{\,},2{\,}$,
perturbations with ${\,}\ell<{\mid}s{\mid}{\,}$ 
relate to total conserved quantities in the system.  
In the present $s=2$ gravitational-wave case, 
we consider accordingly only the propagating $\ell = 2$ 
(quadrupole) and higher-$\ell$ modes.

In this Section and in Sec.7 below, 
we restrict attention to odd-parity 
gravitational-wave perturbations.  
Following Moncrief [68], we write
$$\Bigl(h^{(-)}_{ij}\Bigr)_{\ell m}(x){\;}{\,} 
={\;}{\,}h^{(-)}_{1\ell m}(t{\,},r)
\Bigl[(e_{1})_{ij}\Bigr]_{\ell m}
+h^{(-)}_{2\ell m}(t{\,},r)\Bigl[(e_{2})_{ij}\Bigr]_{\ell m}
{\quad}.\eqno(5.2)$$
(N.B. : one should not confuse the subscripts 
$1{\,},2$ here with spin subscripts.)  
The non-zero components of the symmetric tensor fields 
$\bigl[(e_{1,2})_{ij}\bigr]_{\ell m}$ 
are defined by
$$\bigl[(e_1)_{r\theta}\bigr]_{\ell m}{\;}{\,}
={\;}-\bigl({\partial}_{\phi}Y_{\ell m}\bigr)/\bigl(\sin\theta\bigr)
{\quad},\eqno(5.3)$$
$$\bigl[(e_1)_{r\phi}\bigr]_{\ell m}{\;}{\,}
={\;}{\,}\bigl(\sin\theta\bigr)\bigl({\partial}_{\theta}Y_{\ell m}\bigr)
{\quad},\eqno(5.4)$$
$$\bigl[(e_2)_{\theta\theta}\bigr]_{\ell m}{\;}{\,}
={\;}{\,}\bigl(\sin\theta\bigr)^{-2}
\Bigl[\bigl(\sin\theta\bigr){\partial}^{2}_{\theta\phi}
-\bigl(\cos\theta\bigr){\,}{\partial}_{\phi}\Bigr]Y_{\ell m}
{\quad},\eqno(5.5)$$
$$\bigl[(e_2)_{\theta\phi}\bigr]_{\ell m}{\;}{\,}
={\;}{\,}{{1}\over{2}}
\Bigl[\Bigl(\sin\theta\bigr)^{-1}{\,}{\partial}^{2}_{\phi}
-\bigl(\cos\theta\bigr){\,}{\partial}_{\theta}
-\bigl(\sin\theta\bigr){\,}{\partial}^{2}_{\theta}\Bigr]Y_{\ell m}
{\quad},\eqno(5.6)$$
$$\bigl[(e_2)_{\phi\phi}\bigr)]_{\ell m}{\;}{\,}
={\;}{\,}\Bigl[\bigl(\cos\theta\bigr){\,}{\partial}_{\theta}
-\bigl(\sin\theta\bigr){\partial}^{2}_{\theta\phi}\Bigr]Y_{\ell m}
{\quad}.\eqno(5.7)$$

This basis is normalised according to:
$$\eqalignno{\int d\Omega{\,}
\bigl[(e_{1})^{ij}\bigr]_{\ell m}{\;}
\bigl[(e_{2})_{ij}\bigr]^{*}_{\ell' m'}{\;}{\,} 
&={\;}{\,}0
{\quad},&(5.8)\cr
\int d\Omega{\,}\bigl[(e_{1})^{ij}\bigr]_{\ell m}{\;}
\bigl[(e_{1})_{ij}\bigr]^{*}_{\ell' m'}{\;}{\,} 
&={\;}{\,}{{2e^{-a}}\over{r^{2}}}{\;}\ell(\ell +1){\,}
\delta_{\ell\ell'}{\,}\delta_{mm'}
{\quad},&(5.9)\cr
\int d\Omega{\,}\bigl[(e_{2})^{ij}\bigr]_{\ell m}{\;}
\bigl[(e_{2})_{ij}\bigr]^{*}_{\ell' m'}{\;}{\,} 
&={\;}{\,}{{\ell(\ell +1)(\ell +2)(\ell -1)}\over{2r^{4}}}{\,}
\delta_{\ell\ell'}{\,}\delta_{mm'}
{\,},&(5.10)\cr}$$
where 
${\int}d\Omega{\;}(~~){\,} 
={\,}{\int}^{2\pi}_{0}d\phi{\;}{\int}^{\pi}_{0}{\,}d\theta{\;}
\sin\theta{\;}(~~){\,}$,
and where these indices are raised and lowered using 
the background 3-metric $\gamma^{ij}{\,},\gamma_{ij}{\,}$.  
Note that both
$\bigl[(e_{1})_{ij}\bigr]_{\ell m}$ 
and
$\bigl[(e_{2})_{ij}\bigr]_{\ell m}$ 
are traceless.

In the standard $3{\,}+{\,}1$ decomposition 
for the gravitational field [49], 
the 4-metric $g_{\mu\nu}$ is decomposed into the spatial 
3-metric ${\,}h_{ij}{\,}={\,}g_{ij}{\,}$ on a hypersurface
$\{x^{0}={\rm const.}\}{\,}$, 
together with the lapse function $N$ and the shift vector 
field $N^i{\,}$, such that the 4-dimensional space-time 
metric has the form [15,49]:
$$ds^{2}{\;}{\,}
={\;}{\,}h_{ij}{\,}(dx^{i}+N^{i}{\,}dt){\,}
(dx^{j}+N^{j}{\,}dt){\,}-{\,}N^{2}dt^{2}
{\quad}.\eqno(5.11)$$
For odd-parity perturbations of the lapse, one has
$$N^{(1)(-)}{\;}{\,}={\;}{\,}0
{\quad},\eqno(5.12)$$
while the odd-parity shift vector takes the form
$$\Bigl[N_{i}^{~(-)}\Bigr]_{\ell m}{\;}{\,} 
={\quad}h^{(-)}_{0\ell m}(t{\,},r){\,}
\biggl[{\,}0{\,},
-{\,}{{1}\over{\bigl(\sin\theta\bigr)}}
\bigl(\partial_{\phi}Y_{\ell m}\bigr){\,},
{\,}\bigl(\sin\theta\bigr)
\bigl(\partial_{\theta}Y_{\ell m}\bigr)\biggr]
{\;}.\eqno(5.13)$$
For a real 4-metric ${\,}g_{\mu\nu}{\,}$, 
both $h^{(1)}_{ij}$ and
$N^{(1)(-)}_{i}$ are real, 
and one has
$$h^{(-)*}_{0,1,2\ell m}{\;}{\,} 
={\;}{\,}(-1)^{m}{\;}h^{(-)}_{0,1,2\ell,-m}
{\quad}.\eqno(5.14)$$

In the Hamiltonian formulation of general relativity, 
the momentum $\pi^{ij}$ conjugate to the 'coordinate' 
$h_{ij}$ is a symmetric spatial tensor density [49].  
As with the 3-metric $h_{ij}$ above [Eq.(5.1)], 
the linearised perturbations of $\pi_{ij}$ 
can be decomposed into multipoles with odd or even parity:
$$\pi^{(1)}_{ij}(x){\;}{\,} 
={\;}{\,}\sum^{\infty}_{\ell =2}\sum^{\ell}_{m=-\ell}
\Bigl[\bigl(\pi^{(-)}_{ij}\bigr)_{\ell m}(x)
+\bigl(\pi^{(+)}_{ij}\bigr)_{\ell m}(x)\Bigr]
{\quad}.\eqno(5.15)$$
Restricting attention at present to the odd modes, one has
$$\bigl(\pi^{(-)}_{ij}\bigr)_{\ell m}{\;}{\,} 
={\;}{\,}(^{(3)}\gamma)^{{1}\over{2}}{\,}\Bigl(p_{1\ell m}(t{\,},r)
\bigl[(e_{1})_{ij}\bigr]_{\ell m} 
+p_{2\ell m}(t{\,},r)
\bigl[(e_{2})_{ij}\bigr]_{\ell m}\Bigr)
{\;},\eqno(5.16)$$
where 
$\bigl[(e_{1})_{ij}\bigr]_{\ell m}$ 
and 
$\bigl[(e_{2})_{ij}\bigr]_{\ell m}$ 
are given above.  One finds that
$$\eqalignno{p_{1\ell m}(t{\,},r){\;}{\,} 
&={\;}{\,}{{1}\over{2N^{(0)}}}{\,}
\Biggl(\Bigl(\partial_{t}h^{(-)}_{1\ell m}\Bigr)
-{\,}r^{2}{\,}\partial_{r}
\biggl({{h^{(-)}_{0\ell m}
\over{r^{2}}}}\biggr)\Biggr)
{\quad},&(5.17)\cr
p_{2\ell m}(t{\,},r){\;}{\,} 
&={\;}{\,}{{1}\over{2N^{(0)}}}{\,}
\biggl(\Bigl(\partial_{t}h^{(-)}_{2\ell m}\Bigr)
+2{\,}h^{(-)}_{0\ell m}\biggr)
{\quad}.&(5.18)\cr}$$

One can typically simplify the form of the perturbations 
by a gauge transformation 
(linearised coordinate transformation) 
in a neighbourhood of the final space-like hypersurface 
$\Sigma_{F}{\,}$.  Suppose that the infinitesimal 
transformation is along a vector field ${\,}\xi^{\mu}{\,}$.  
Then the metric perturbations transform infinitesimally by
$$g_{\mu\nu}{\quad} 
\rightarrow{\quad}g_{\mu\nu}{\,} 
-{\,}\nabla_{\mu}\xi_{\nu}{\,}
-{\,}\nabla_{\nu}\xi_{\mu}
{\quad}.\eqno(5.19)$$ 
For odd perturbations, in the notation of Eq.(4.2), 
consider the infinitesimal vector field 
$\xi^{(-)\mu}{\,}$, 
with components [63] given by
$$\eqalign{\Bigl(\xi^{(-)t}\Bigr)_{\ell m}{\;}{\,} 
&={\quad}0{\quad},
{\qquad}\Bigl(\xi^{(-)r}\Bigr)_{\ell m}{\;}{\,}
={\quad}0{\quad},\cr
\Bigl(\xi^{(-)a}\Bigr)_{\ell m}{\;}{\,} 
&={\quad}{{\Lambda_{\ell m}(t{\,},r)}\over{r^{2}}}{\,}
\bigl(\Phi_{\ell m}\bigr)^{a}
{\quad}.\cr}\eqno(5.20)$$
The resulting 'gauge transformation' is summarised by
$$\eqalignno{h^{(-)'}_{0\ell m}{\quad} 
&={\quad}h^{(-)}_{0\ell m} 
-\bigl(\partial_{t}\Lambda_{\ell m}\bigr)
{\quad},&(5.21)\cr
h^{(-)'}_{1\ell m}{\quad}
&={\quad}h^{(-)}_{1\ell m} 
-\bigl(\partial_{r}\Lambda_{\ell m}\bigr)
+{{2\Lambda_{\ell m}}\over{r}}
{\quad},&(5.22)\cr
h^{(-)'}_{2\ell m}{\quad} 
&={\quad}h^{(-)}_{2\ell m} 
+2\Lambda_{\ell m}
{\quad}.&(5.23)\cr}$$

We have here neglected time-derivatives of the metric 
components: 
we are assuming that an Ansatz for the gauge functions 
$\Lambda_{\ell m}(t{\,},r)$ based on separation 
of variables will be valid, involving frequencies 
which satisfy the adiabatic approximation [11,16] 
described below.  
In the Regge-Wheeler gauge [63], we set 
${\,}h^{(-)RW}_{0\ell m}{\,}
={\;}h^{(-)'}_{0\ell m}{\,}$ 
and 
${\,}h^{(-)RW}_{1\ell m}{\,}={\;}h^{(-)'}_{1\ell m}{\;}$, 
as in Eqs.(5.21,22), but require also
$$h^{(-)RW}_{2\ell m}{\;}{\,} 
={\quad}0{\quad}
={\quad}h^{(-)}_{2\ell m}
+2{\,}\Lambda_{\ell m}
{\quad}.\eqno(5.24)$$ 
For each ${\,}\ell{\,}$, 
one can obtain solutions for arbitrary ${\,}m{\,}$ 
by rotation from the case ${\,}m=0{\,}$.  
Note also that the above equations 
show how the RW perturbations can be uniquely 
recovered from the perturbations in an arbitrary gauge.

Since odd-and even-parity perturbations decouple, 
the odd-parity field equations in the RW gauge 
are obtained by substituting Eq.(2.7), 
together with the equation ${\,}N^{(-)}_{i}={\,}h^{(-)}_{ti}$ 
for the linearised shift vector, 
into the source-free linearised Einstein field 
equations [49].  The (Lorentzian) spherically-symmetric 
background metric is taken, as above, in the form 
of Eq.(4.10), and the mass function $m(t{\,},r)$ 
is defined by Eq.(4.11).
Then the odd-parity linearised field equations, 
taking respectively the 
$(t\phi){\,},{\,}(r\phi){\,}$ 
and 
$(\theta\phi){\,}$ components, 
read:
$$\bigl(\partial_{r}\bigr)^{2}h^{(-)RW}_{0\ell m}
-\partial_{t}\partial_{r}h^{(-)RW}_{1\ell m}
-{{2}\over{r}}{\,}\partial_{t}h^{(-)RW}_{1\ell m}
+F_{1\ell}(t{\,},r){\,}h^{(-)RW}_{0\ell m}{\;}
={\;}0
{\;},\eqno(5.25)$$ 
and 
$$\eqalign{&\bigl(\partial_{t}\bigr)^{2}h^{(-)RW}_{1\ell m}
-\partial_{t}\partial_{r}h^{(-)RW}_{0\ell m}
+{{2}\over{r}}{\,}\partial_{t}h^{(-)RW}_{0\ell m}\cr
&+{{1}\over{2}}\Bigl({\dot a}+{\dot b}\Bigr)
\Bigl(\partial_{r}h^{(-)RW}_{0\ell m}
-\partial_{t}h^{(-)RW}_{1\ell m}\Bigr)\cr
&-\biggl({{1}\over{r}}\Bigl({\dot a}+{\dot b}\Bigr)
+{{1}\over{2}}{\,}b'{\,}\Bigl({\dot a}-{\dot b}\Bigr)\biggr)
h^{(-)RW}_{0\ell m}
-F_{2\ell}(t{\,},r){\,}h^{(-)RW}_{1\ell m}{\;}
={\;}0\cr}
{\;},\eqno(5.26)$$
and
$$\partial_{t}
\Bigl(e^{(a-b)/2}{\,}h^{(-)RW}_{0\ell m}\Bigr)
-\partial_{r}
\Bigl(e^{(b-a)/2}{\,}h^{(-)RW}_{1\ell m}\Bigr){\;}{\,} 
={\;}{\,}0
{\quad}.\eqno(5.27)$$
Here,
$$F_{1\ell}(t{\,},r){\,}
={\,}{{e^{a}}\over{r^{2}}}
\biggl({{4m}\over{r}}+4m'
-\ell(\ell +1)\biggr)
+e^{a-b}\biggl({\ddot a} 
+{{1}\over{2}}{\,}{\dot a}
\Bigl({\dot a}-{\dot b}\Bigr)\biggr)+Z
{\;},\eqno(5.28)$$
and
$$F_{2\ell}(t{\,},r){\;}
={\,}-{\,}{{e^{b}}\over{r^{2}}}{\,}(\ell +2){\,}(\ell -1)
+Z{\,}e^{b-a}+{\,}{\ddot a}{\,}
+{{1}\over{2}}{\,}{\dot a}
\bigl({\dot a}-{\dot b}\bigr)
{\;},\eqno(5.29)$$
with
$$Z{\;}{\,}
={\,}-{\,}{{2e^{a}}\over{r}}
\Biggl(m''+\biggl({{2m'e^{a}}\over{r^{2}}}\biggr) 
\Bigl(m'+rm\Bigr)\Biggr)
{\;}.\eqno(5.30)$$

Further, the Einstein field equations imply
$$m'{\;}{\,}
={\;}{\,}4\pi r^{2}\rho
{\quad}.\eqno(5.31)$$
Here, $\rho$ is the total energy density 
of all the radiative fields;
in the present case $\rho$ has contributions 
from $s=0{\,}$ (massless scalar), $s=2{\,}$ (graviton) 
and, if the Lagrangian contains Maxwell 
or Yang-Mills fields,  also $s=1{\,}$.  
In the adiabatic limit, in which $m(t{\,},r)$ 
varies extremely slowly, Eq.(5.31) provides part 
of the description of the approximate Vaidya metric 
for the region of space-time in which the black hole 
is evaporating; a much fuller treatment of the Vaidya 
region is given in [69].

Now consider the lowest-order perturbative contribution 
to the classical action of our present coupled 
bosonic system, for which the non-zero background part 
consists of the spherically-symmetric gravitational 
and scalar fields, $(\gamma_{\mu\nu}{\,},{\,}\Phi){\,}$.  
It is assumed that the background fields obey the coupled 
Einstein/massless-scalar field equations and contribute 
$S^{(0)}_{\rm class}$ to the classical action.  
Any spin-1 fields present, whether Maxwell or Yang-Mills, 
propagate at lowest order in the background, 
and the spin-1 contribution to the classical action 
begins at second order, as described in Sec.4, 
and adds to the various lowest-order perturbative 
Einstein-Hilbert and scalar contributions.  
To simplify the exposition at this point, 
let us temporarily neglect the spin-1 field.

After detailed calculation [33,34], 
one finds that the classical Lorentzian action 
for an (anisotropic) Einstein/massless-scalar solution, 
subject to perturbed boundary data
$\bigl(h^{(0)}_{ij}+{\,}h^{(1)}_{ij}{\,},
{\,}{\Phi}{\,}+{\,}{\phi}^{(1)}\bigr)$ 
on the initial and final surfaces ${\Sigma}_{I}$ 
and ${\Sigma}_{F}{\,}$, can be written 
as the background contribution $S^{(0)}_{\rm class}$ 
above, plus a quadratic-order contribution, 
plus higher-order terms.  Thus:
$$\eqalignno{S_{\rm class}{\;}{\,}
&={\;}{\,}S^{(0)}_{\rm class}{\,}
+{\,}S^{(2)}_{\rm class}{\,}
+{\,}\ldots{\quad},\cr     
&={\;}{{1}\over{32\pi}}\int_{{\Sigma}_{F}}d^{3}x{\,}
\Bigl({\pi}^{(0)ij}{\,}h^{(0)}_{ij}
+{\pi}^{(1)ij}{\,}h^{(1)_{ij}}\Bigr)+{\,}{\ldots}\cr
&+{\,}{{1}\over{2}}\int_{{\Sigma}_{F}}d^{3}x{\,}
\Bigl({\Phi}{\,}{\Pi_{\phi}}
+{\phi}^{(1)}{\,}{\pi_{\phi}}^{(1)}\Bigr){\,}+{\,}{\ldots}\cr
&-{\,}M{\,}T
{\quad}.&(5.32)\cr}$$ 
Again, for simplicity, it is assumed 
that no perturbations are prescribed on the initial 
boundary ${\Sigma}_{I}{\,}$, but that there are non-zero 
perturbations on the final boundary ${\Sigma}_{F}{\,}$.  
As described after Eq.(5.14), ${\pi}^{ij}$ is defined 
to be the momentum conjugate to the 'coordinate' 
$h_{ij}{\,}$.  Similarly, ${\pi}_{\phi}$ is the momentum 
conjugate to the variable $\phi{\,}$, and is given 
by the normal future-directed derivative 
${\,}{\partial}{\phi}/{\partial}n{\,}$.  
In Eq.(5.32), $T{\,}$, as usual, 
denotes the Lorentzian proper-time interval between 
the initial and final boundaries, 
as measured at spatial infinity.  
Further, $M$ denotes the mass of the space-time; 
for a well-posed (complexified) classical boundary-value 
problem, the mass defined on the initial surface 
must agree with the mass defined on the final surface.

We can now compute the odd-parity contribution 
to the classical gravitational action.  
In an arbitrary gauge, taking only the intrinsic metric 
perturbation $h^{(1)}_{ij}$ to be non-zero 
on ${\Sigma}_{F}{\,}$, but ${\phi}^{(1)}=0{\,}$ there, 
one has the second-variation part of the action
$$S^{(2)}_{\rm class}\Bigl[h^{(1)}_{ij}\Bigr]{\;}{\,} 
={\;}{\,}{{1}\over{32\pi}}
\int_{\Sigma_{F}}d^{3}x{\;}
\pi^{(1)ij}{\;}h^{(1)}_{ij}
{\quad}.\eqno(5.33)$$
On discarding a total divergence, 
the spin-2 classical action can also be written as
$$\eqalign{S^{(2)}_{\rm class}{\;}{\,}
=&{\;}{\,}{{1}\over{64\pi}}
\int_{\Sigma_{F}}d^{3}x{\;}
\sqrt{^{(3)}\gamma}{\;}n^{(0)\mu}{\,}
\Bigl({\bar h}^{(1)\mu\nu}{\,}
\nabla_{\alpha}h^{(1)}_{\mu\nu}{\,}
-2{\,}h^{(1)}_{\alpha\nu}{\;}
\nabla_{\rho}h^{(1)\nu\rho}\Bigr)\cr
&+{\;}{{1}\over{16\pi}}\int_{\Sigma_{F}}{\,}d^{3}x{\;}
\sqrt{^{(3)}\gamma}{\;}
\biggl({{N^{(1)}_{~~~i}}\over{N^{(0)}}}\biggr){\,}
{\bar h}^{(1)ik}_{~~~~~\mid k}
{\quad}.\cr}\eqno(5.34)$$
Since odd- and even-parity perturbations are orthogonal, 
there are no cross-terms in the action.  
In an arbitrary gauge, the odd-parity contribution 
to the classical gravitational action 
can be re-written as 
$$\eqalign{&S^{(2)}_{\rm class}
\Bigl[(h^{(-)}_{ij})_{\ell m}\Bigr]{\,}
={\,}{{1}\over{32\pi}}
\int_{\Sigma_{F}}d^{3}x
\sum_{\ell\ell'mm'}
\Bigl(\pi^{(-)ij}\Bigr)_{\ell m}
\Bigl(h^{(-)}_{ij}\Bigr)^{*}_{\ell' m'}\cr
=&{\,}{{1}\over{32\pi}}\sum_{\ell m}\ell(\ell +1)
\int^{R\infty}_{0}dr{\;}h^{(-)*}_{1\ell m}
\biggl(\Bigl(\partial_{t}h^{(-)}_{1\ell m}\Bigr)
+{{2}\over{r}}{\,}h^{(-)}_{0\ell m}
-\Bigl(\partial_{r}h^{(-)}_{0\ell m}\Bigr)\biggr)
\Biggl\arrowvert_{T}\cr
&+{\,}{{1}\over{128\pi}}\sum_{\ell m}
{{(\ell +2)!}\over{(\ell -2)!}}
{\int}_{0}^{R_{\infty}}dr{\;}e^{a}{\,}
{{h^{(-)*}_{2\ell m}}\over{r^{2}}}
\biggl(\Bigl(\partial_{t}h^{(-)}_{2\ell m}\Bigr)
+2{\,}h^{(-)}_{0\ell m}\biggr)
\Biggl\arrowvert_{T}
{\;},\cr}\eqno(5.35)$$
Here, we have used Eqs.(5.8-10,17,18), 
and have taken the perturbations to vanish initially.  
Note that, if we were to evaluate Eq.(5.35) in the RW 
gauge, for which ${\,}h^{(-)RW}_{2\ell m}={\,}0{\;}{\,}$, 
so that the second integral would vanish, 
and then substitute for the Regge-Wheeler functions 
{\it via} Eqs.(5.21-24), then, in the adiabatic
approximation, we would arrive back at Eq.(5.35) 
up to a boundary term
$$h^{(-)*}_{2\ell m}{\;}P^{(-)}_{\ell m}
\Bigl\arrowvert^{r=R_{\infty}}_{r=0}
{\quad},\eqno(5.36)$$
where
$$P^{(-)}_{\ell m}{\;}{\,} 
={\;}{\,}\ell(\ell +1)
\Biggl(\biggl(\partial_{t}h^{(-)}_{1\ell m}\biggr)
-{\,}r^{2}{\,}
\partial_{r}\biggl({{h^{(-)}_{0\ell m}}\over{r^{2}}}\biggr)\Biggr)
{\;}.\eqno(5.37)$$ 
Thus, 
${\,}S^{(2)}_{\rm class}\bigl[(h^{(-)}_{ij})_{\ell m}\bigr]{\,}$ 
is gauge-invariant up to a boundary term.  
We shall return below to the question of boundary 
conditions for the odd-parity perturbations.

At first sight, the odd-parity action looks unwieldy.  
Ideally, we would like to work with a classical action 
(both for odd and for even parity) in the form 
${\int}dr{\,}\psi(\partial_{t}\psi){\,}$, 
of the same general kind as in the massless-scalar 
classical action in [33,34]. 
In the present gravitational case, $\psi$ would ideally 
also be gauge-invariant and would obey a wave equation 
with a real potential.
To achieve this form, first use Eqs.(5.21-24) 
for the RW functions, and substitute them 
into the field equation (5.26), to obtain
$$\eqalignno{\Bigl(\partial_{t}P^{(-)}_{\ell m}\Bigr){\,}
&={\,}\ell{\,}(\ell +1)F_{2\ell}(r)
\Biggl(h^{(-)}_{1\ell m}+{{1}\over{2}}{\,}
\partial_{r}
\biggl({{h^{(-)}_{2\ell m}}\over{r^{2}}}\biggr)\Biggr)
{\;},&(5.38)\cr
\Bigl(\partial_{r}P^{(-)}_{\ell m}\Bigr)
&=-{\,}{{2P^{(-)}_{\ell m}}\over{r}}
+\ell(\ell +1)
\biggl({{1}\over{2}}F_{1\ell}(r)
+{{1}\over{r^{2}}}\biggr)
\biggl(\Bigl(\partial_{t}h^{(-)}_{2\ell m}\Bigr)
+2h^{(-)}_{0\ell m}\biggr)
{\,}.&(5.39)\cr}$$
When we substitute into Eq.(5.35), 
using Eq.(5.38) for ${\,}h^{(-)}_{1\ell m}{\,}$, 
and then use Eqs.(5.29,39), 
the boundary term (5.36) vanishes.  
As a consequence, we find in the adiabatic approximation that
$$S^{(2)}_{\rm class}
\Bigl[\Bigl(h^{(-)}_{ij}\Bigr)_{\ell m}\Bigr]
=-{\,}{{1}\over{32\pi}}
\sum^{\infty}_{\ell =2}\sum^{\ell}_{m=-\ell}
{{(\ell -2)!}\over{(\ell +2)!}}
\int^{R_{\infty}}_{0}dr{\,}
e^{a}{\,}\xi^{(-)}_{2\ell m}
\Bigl(\partial_{t}\xi^{(-)*}_{2\ell m}\Bigr)
\Biggl\arrowvert_{t=T}
{\,},\eqno(5.40)$$
where 
${\,}\xi^{(-)}_{2\ell m}{\,}$ 
is defined by
$$\xi^{(-)}_{2\ell m}{\;}{\,}={\;}{\,}r{\,}P^{(-)}_{\ell m}
{\;}.\eqno(5.41)$$

Eqs.(5.21-23) show that 
${\,}\xi^{(-)}_{2\ell m}{\,}$ is gauge-invariant.
Indeed, ${\,}\xi^{(-)}_{2\ell m}{\,}$ 
is related to Moncrief's [68] gauge-invariant 
generalisation of the Zerilli function, 
${\,}Q^{(-)}_{\ell m}{\,}$ [64,70], 
defined as
$$Q^{(-)}_{\ell m}{\;}{\,} 
={\;}{\,}{{e^{-a}}\over{r}}
\Biggl(h^{(-)}_{1\ell m}+{{1}\over{2}}{\,}
r^{2}{\,}\partial_{r}
\biggl({{h^{(-)}_{2\ell m}}\over{r^{2}}}\biggr)\Biggr)
{\;},\eqno(5.42)$$
by
$$Q^{(-)}_{\ell m}{\;}{\,} 
={\,}-{\,}{{(\ell -2)!}\over{(\ell +2)!}}{\,}
\Bigl(\partial_{t}\xi^{(-)}_{2\ell m}\Bigr)
{\;},\eqno(5.43)$$
which, in effect, replicates Eq.(5.38).  
Note the simplifying property of 
${\,}Q^{(-)}_{\ell m}{\,}$, 
namely, that it is written entirely in terms 
of perturbations of the 3-geometry 
(our chosen boundary data).  
Further, $Q^{(-)}_{\ell m}$ is automatically gauge-invariant, 
since it is independent of the perturbed lapse and shift.

In the adiabatic approximation, 
the function $\xi^{(-)}_{\ell m}$
obeys the wave equation, of RW type:
$$e^{-a}{\,}\partial_{r}\biggl(e^{-a}
\Bigl(\partial_{r}\xi^{(-)}_{2\ell m}\Bigr)\biggr)
-\bigl(\partial_{t}\bigr)^{2}\xi^{(-)}_{2\ell m} 
-V^{(-)}_{\ell}(r){\,}\xi^{(-)}_{\ell m}{\;}{\,}
={\;}{\,}0
{\;},\eqno(5.44)$$
where
$$V^{(-)}_{\ell}(r){\;}{\,} 
={\;}{\,}e^{-a}\biggl({{\ell(\ell +1)}\over{r^{2}}}{\,}
-{{6m(r)}\over{r^{3}}}\biggr){\quad}>{\quad}0
{\;}.\eqno(5.45)$$
Further, in the adiabatic approximation, 
the function 
${\,}Q^{(-)}_{\ell m}{\,}$ 
obeys the same equation (5.44). 
In the RW gauge, one would solve 
for ${\,}Q^{(-)RW}_{\ell m}{\,}$, 
then determine ${\,}h^{(-)RW}_{1\ell m}$ from Eq.(5.42), 
and then determine ${\,}h^{(-)RW}_{0\ell m}$ 
with the help of Eq.(5.27).
\end{section}

\begin{section}{Boundary conditions and classical action 
-- Maxwell case}
In Sec.4, for the Maxwell field 
(whether odd- or even-parity), we derived an expression, 
Eq.(4.45), for the classical Maxwell action 
${\,}S^{\rm EM}_{\rm class}{\,}$, 
given {\it via} certain operations in terms 
of the magnetic field on the final boundary 
${\,}{\Sigma}_{F}{\,}$.  
In Sec.5, for odd-parity gravitational perturbations, 
we derived Eq.(5.40), a somewhat analogous expression
for the second-variation classical action 
${\,}S^{(2)}_{\rm class}{\,}$, 
depending on the odd-parity metric perturbations 
${\,}h^{(-)}_{ij}{\,}$ over the final surface 
${\Sigma}_{F}{\,}$.  In the present Section 6, 
concerning Maxwell theory, we relate 
${\,}S^{\rm EM}_{\rm class}{\,}$ 
more explicitly to the final boundary data.  
The following Sec.7 will similarly re-express 
${\,}S^{(2)}_{\rm class}{\,}$ 
for odd gravitational perturbations.  
A corresponding treatment for the more complicated
even-parity gravitational case will be given in Secs.8,9.  
The resulting expressions for the classical action 
will greatly simplify our understanding of the black-hole 
quantum amplitudes.

Physically, our gauge-invariant odd- and even-parity 
Maxwell variables
${\,}\psi^{(o)}_{1\ell m}{\,}$,
${\,}\psi^{(e)}_{1\ell m}{\,}$ 
are effectively the radial components of the magnetic 
and electric field strengths, respectively:
$$\eqalignno{B^{(1)r}_{\ell m}(x){\;}{\,}& 
={\;}{\,}{{e^{-a/2}}\over{r^{2}}}{\;}\psi^{(o)}_{1\ell m}(t{\,},r)
{\,}Y_{\ell m}(\Omega)
{\quad},&(6.1)\cr
E^{(1)r}_{\ell m}(x){\;}{\,}&
={\,}-{\,}{{e^{-a/2}}\over{r^{2}}}{\;}
\psi^{(e)}_{1\ell m}(t{\,},r){\,}Y_{\ell m}(\Omega)
{\quad},&(6.2)\cr}$$
where
$$B^{(1)i}_{\ell m}{\;}{\,} 
={\;}{\,}\bigl(^{(3)}\gamma\bigr)^{-{{1}\over{2}}}{\;}
{\cal B}^{(1)i}_{\ell m}
{\quad}.\eqno(6.3)$$
The remaining, transverse, magnetic field components are
$$\eqalignno{B^{(1)\theta}_{\ell m}(x){\;}
&={\;}{{e^{-a/2}}\over{r^{2}}}
\Biggl[a_{1\ell m}(t,r)
{{\bigl(\partial_{\phi}Y_{\ell m}\bigr)}\over{\bigl(\sin\theta\bigr)}} 
+{{\bigl(\partial_{r}\psi^{(o)}_{1\ell m}\bigr)}
\over{\ell(\ell+1)}}
\bigl(\partial_{\theta}Y_{\ell m}\bigr)\Biggr]
{\;}.&(6.4)\cr
B^{(1)\phi}_{\ell m}(x){\;}&
={\;}{{e^{-a/2}}\over{r^{2}\bigl(\sin\theta\bigr)}}
\Biggl[-{\,}a_{1\ell m}(t,r)
\bigl(\partial_{\theta}Y_{\ell m}\bigr)
+{{\bigl(\partial_{r}\psi^{(o)}_{1\ell m}\bigr)}
\over{\ell(\ell+1)}}{{\bigl(\partial_{\phi}Y_{\ell m}\bigr)}
\over{\bigl(\sin\theta\bigr)}}\Biggr]
{\;}.&(6.5)\cr}$$

The main aim of this paper is to calculate quantum 
amplitudes for weak spin-1 (Maxwell) perturbative data 
on the late-time final surface $\Sigma_{F}{\,}$, 
by evaluating the classical action 
$S^{EM}_{\rm class}$ 
and hence the semi-classical wave function 
${\rm (const.)}{\,}{\times}{\,}\exp\bigl(iS^{EM}_{\rm class}\bigr)$, 
{\it as functionals of the spin-1 final boundary data}.  
In Eq.(4.45), ${\,}S^{EM}_{\rm class}$ 
was expressed as an integral over the boundary, 
involving various perturbative quantities 
used in the description above of the dynamical 
perturbations.  The present task is to determine 
'optimal' or 'natural' boundary data, 
both for the odd-parity case and separately 
for the even-parity case, such that 
(i.) the classical boundary-value problem can readily 
be solved, given these data, and 
(ii.) the classical Maxwell action 
$S^{EM}_{\rm class}$ 
can be (re-)expressed in terms of the appropriate 
boundary data.  Under those conditions, 
we will then have a description of the spin-1 
radiation, associated with gravitational collapse 
to a black hole, analogous to that for the spin-0 
(massless-scalar) radiation, developed in [33,34].

Following the discussion of Secs.2 and 4, 
the relevant field components to be fixed 
on $\Sigma_{I}$ and $\Sigma_{F}$, 
in the odd-parity case, 
are ${\,}\psi^{(o)}_{1\ell m}{\,}$, 
as may be seen from Eq.(6.1).
In the even-parity case, the appropriate boundary 
conditions involve fixing $a_{1\ell m}{\,}$, 
as may be seen from Eqs.(6.4,5).  In the even case, 
this is equivalent, from Eqs.(4.40,42), to specifying  
${\,}\partial_{t}\psi^{(e)}_{1\ell m}$ on the boundary.  
Thus, we choose the boundary data to consist 
of ${\,}\psi^{(o)}_{1\ell m}{\,}$ 
in the odd-parity case, 
and ${\,}\partial_{t}\psi^{(e)}_{1\ell m}{\,}$ 
in the even-parity case.  
Hence, though ${\,}\psi^{(o)}_{1\ell m}{\,}$ 
and ${\,}\psi^{(e)}_{1\ell m}{\,}$ 
both obey the same dynamical field 
equations (4.36,43) and (4.47), 
the natural boundary conditions for them 
on $\Sigma_{I,F}{\,}$ are quite different 
-- Dirichlet for the odd-parity case 
${\,}\psi^{(o)}_{1\ell m}{\,}$, 
but Neumann in the even-parity case 
$\psi^{(e)}_{1\ell m}{\,}$.  
This is reminiscent of the situation obtaining 
when spin-2 gravity is coupled to all lower spins, 
fermionic as well as bosonic; that is, to spins 
$s={{3}\over{2}}{\,},{\,}1{\,},{\,}{{1}\over{2}}{\,}$ 
and $0{\,}$, especially in locally-supersymmetric 
models [22], such as models of $N=1$ supergravity 
with gauged supermatter [71-73].  
As found also, in this paper, for spin-2 (graviton) 
perturbed data on the final surface $\Sigma_{F}{\,}$, 
the natural boundary conditions are contrasting, 
for odd-parity ${\it vis-\grave a-vis}$ even-parity modes.  
For the remaining $s=0$ (scalar) bosonic case, 
if one requires, as in the Introduction, 
that the full theory be locally supersymmetric, 
so that quantum amplitudes are meaningful, 
one finds that all scalar fields must be 
${\it complex}{\,}$, whether as part 
of a multiplet or as a single complex scalar [22].  
The treatment of the real $s=0$ case in [33,34] 
can be replicated in the case of a complex scalar 
field $\phi{\,}$, except that the natural boundary 
conditions, consistent with local supersymmetry, 
require ${\rm Re}(\phi)$ to be fixed at a surface 
such as $\Sigma_{F}$ (Dirichlet), 
whereas the normal derivative
${\,}\partial[{\rm Im}(\phi)]/\partial n{\,}$ 
must also be fixed at a bounding surface (Neumann).  
Of course, this treatment extends to fermionic data 
$(s={{1}\over{2}}{\rm{\;}and}{\;}{{3}\over{2}}){\,}$, 
as described in [35,36].

In the gravitational-collapse model, 
by analogy with the simplifying choice 
${\,}\phi^{(1)}{\mid}_{\Sigma_{I}}=0{\,}$ 
for the initial perturbative scalar-field data, 
taken in [33,34], we take (for the purposes of exposition) 
the simplest Maxwell initial data at
$\Sigma_{I}{\;}(t=0){\,}$.  
That is, we consider a negligibly weak magnetic field 
outside the 'star':  
the boundary conditions are
$$\eqalignno{\psi^{(o)}_{1\ell m}(0{\,},r){\;}{\,}&
={\;}{\,}0
{\quad},&(6.6)\cr
\partial_{t}\psi^{(e)}_{1\ell m}(0{\,},r){\;}{\,}&
={\;}{\,}0
{\quad}.&(6.7)\cr}$$
Condition (6.6) is a Dirichlet condition 
on the initial odd-parity magnetic field 
-- see Eqs.(6.1,4,5).  
Condition (6.7) implies that we have an initially 
static even-parity multipole [65].

We now follow the analysis of the spin-0 field, 
and separate the radial-and time-dependence.  
In neighbourhoods of $\Sigma_{I}$ and $\Sigma_{F}{\,}$, 
where an adiabatic approximation is valid, 
we can 'Fourier-expand' the variables 
${\,}\psi^{(o)}_{1\ell m}(t{\,},r){\,}$ 
and 
${\,}\psi^{(e)}_{1\ell m}(t{\,},r){\,}$, 
subject to the initial conditions (6.6) and (6.7).  
By analogy with the scalar case [33,34], let us write
$$\eqalignno{\psi^{(o)}_{1\ell m}(t{\,},r){\quad}&
={\;}{\,}\int^{\infty}_{-\infty}dk{\;} 
a^{(o)}_{1k\ell m}{\,}\psi^{(o)}_{1k\ell}(r){\,}
{{\sin(kt)}\over{\sin(kT)}}
{\quad},&(6.8)\cr
\psi^{(e)}_{1\ell m}(t,r){\quad}&
={\;}{\,}\int^{\infty}_{-\infty}dk{\;}
a_{1k\ell m}^{(e)}{\,}\psi^{(e)}_{1k\ell}(r){\,}
{{\cos(kt)}\over{\sin(kT)}}
{\quad},&(6.9)\cr}$$
where the radial functions 
$\{\psi^{(o)}_{1k\ell}(r)\}$ 
and
$\{\psi^{(e)}_{1k\ell}(r)\}$ 
are independent of $m{\,}$, 
given the spherical symmetry of the background 
space-time.  Here, the position-independent 
quantities $\{a^{(o)}_{1k\ell m}\}$ 
and
$\{a^{(e)}_{1k\ell m}\}$ are some coefficients, 
with smooth dependence on the continuous 
variable $k{\,}$, which label the configuration 
of the electromagnetic field on the final surface 
$\Sigma_{F}{\,}$.

The radial functions $\{\psi^{(o)}_{1k\ell}(r)\}$ 
and
$\{\psi^{(e)}_{1k\ell}(r)\}$ 
each obey a regularity condition at the centre 
of spherical symmetry $\{r=0\}$ on the final surface 
$\Sigma_{F}{\,}$; 
this requires that the corresponding (spatial) electric 
or magnetic field, defined {\it via} Eqs.(6.1-5), 
should be smooth in a neighbourhood of $\{r=0\}{\,}$.  
As a consequence, the radial functions must be real:
$${\psi^{(o)}_{1k\ell}}^{*}(r){\;}{\,} 
={\;}{\,}\psi^{(o)}_{1,-k\ell}(r){\quad},
{\qquad}{\quad}{\psi^{(e)}_{1k\ell}}^{*}(r){\;}{\,} 
={\;}{\,}\psi^{(e)}_{1,-k\ell}(r)
{\quad}.\eqno(6.10)$$
For small $r{\,}$, 
the radial functions should be asymptotically
proportional to a spherical Bessel function [74]:
$$\eqalignno{\psi^{(o)}_{1k\ell}(r){\quad}&
\sim{\quad}{(\rm const.)}{\;}{\times}{\;}{\;}r{\,}j_{\ell}(kr)
{\quad},&(6.11)\cr
\psi^{(e)}_{1k\ell}(r){\quad}&
\sim{\quad}({\rm const.})'{\;}{\times}{\;}{\;}r{\,}j_{\ell}(kr)
{\quad},&(6.12)\cr}$$
as ${\,}r{\,}\rightarrow{\,}0_{+}{\;}$.  
Also, the reality of the radial electric and magnetic 
fields implies that
$$\psi^{(o)}_{1\ell m}(t{\,},r){\;} 
={\;}(-1)^{m}{\;}\psi^{(o)*}_{1\ell,-m}(t{\,},r){\;},
{\qquad}\psi^{(e)}_{1\ell m}(t{\,},r){\;} 
={\;}(-1)^{m}{\;}\psi^{(e)*}_{1\ell,-m}(t{\,},r)
{\;}.\eqno(6.13)$$
This in turn implies that
$$a^{(o)}_{1k\ell m}{\;} 
={\;}(-1)^{m}{\;}a^{(o)*}_{1,-k\ell,-m}{\;}{\,},
{\qquad}a^{(e)}_{1k\ell m}{\;}
={\;}(-1)^{m+1}{\;}a^{(e)*}_{1,-k\ell,-m}
{\;}{\,}.\eqno(6.14)$$
Since the potential (4.37), 
appearing in the $(t{\,},r)$ wave equation (4.36), 
tends sufficiently rapidly to zero as 
${\,}r{\,}\rightarrow{\,}\infty{\,}$, 
in the region where the space-time is almost 
Schwarzschild, one has asymptotic (large-$r$) behaviour 
of ${\,}\psi^{(o)}_{1k\ell}(r)$ 
and 
${\,}\psi^{(e)}_{1k\ell}(r)$ 
which is analogous to that in the scalar case:
$$\eqalignno{\psi^{(o)}_{1k\ell}(r){\quad}
&\sim{\quad}\Bigl(z^{(o)}_{k\ell}{\,}\exp(ik{\,}r^{*}_{s}) 
+z^{(o)*}_{k\ell}{\,}
\exp\bigl(-ik{\,}r^{*}_{s}\bigr)\Bigr)
{\;}.&(6.15)\cr
\psi^{(e)}_{1k\ell}(r){\quad}
&\sim{\quad}\Bigl(z^{(e)}_{k\ell}{\,}
\exp(ik{\,}r^{*}_{s})
+z^{(e)*}_{k\ell}{\,}\exp(-ik{\,}r^{*}_{s})\Bigr)
{\;}.&(6.16)\cr}$$
Here, $\{z^{(o)}_{k\ell}\}$ 
and $\{z^{(e)}_{k\ell}\}$ 
are complex coefficients, depending smoothly 
on the continuous variable $k{\,}$.
Also, $r^{*}_{s}$ again denotes the Regge-Wheeler 
'tortoise' coordinate of Eq.(4.38) 
for the Schwarzschild geometry [49,63].  
As in the scalar case [33,34], 
the inner product (normalisation) 
for the radial functions follows in the limit 
${\,}R_{\infty}{\,}\rightarrow{\,}\infty{\,}$:
$$\eqalignno{\int^{R_\infty}_{0}dr{\;}e^{a}{\,}
\psi^{(o)}_{1k\ell}(r){\,}\psi^{(o)}_{1k'\ell}(r){\;}{\,}&
={\;}{\,}2\pi{\,}{\mid}z^{(o)}_{k\ell}{\mid}^{2}{\,} 
\Bigl[\delta(k{\,},k')+\delta(k{\,},-k')\Bigr]
{\;},&(6.17)\cr
\int^{R_\infty}_{0}dr{\;}e^{a}{\,}
\psi^{(e)}_{1k\ell}(r){\,}\psi^{(e)}_{1k'\ell}(r){\;}{\,}&
={\;}{\,}2\pi{\,}{\mid}z^{(e)}_{k\ell}{\mid}^{2}{\,}
\Bigl[\delta(k{\,},k')+\delta(k{\,},-k')\Bigr]
{\;}.&(6.18)\cr}$$

Finally, we are in a position to compute 
the classical Maxwell action
$S^{EM}_{\rm class}{\,}$ 
as a functional of the spin-1 boundary data 
on the final surface $\Sigma_{F}{\,}$.  
This gives straightforwardly the semi-classical 
wave function for the complexified time-interval 
$T{\,}$, leading to the Lorentzian quantum 
amplitude or wave function.  
Our boundary conditions (6.6,7) 
above on the initial hypersurface $\Sigma_{I}{\,}$, 
at time $t=0{\,}$ say, were designed so as to give 
zero contribution from $\Sigma_{I}$ 
to the expression (4.45) for the classical action 
${\,}S^{EM}_{\rm class}{\,}$.  
The contribution to (4.45) from $\Sigma_{F}$ is found, 
using Eqs.(6.8,9,17,18), to be
$$\eqalign{&S^{EM}_{\rm class}
\bigl[\{a^{(o)}_{1k\ell m}{\,},{\,}
a^{(e)}_{1k\ell m}\}\bigr]\cr
=-&{{1}\over{2}}{\,}\sum_{\ell m}
{{(\ell -1)!}\over{(\ell +1)!}}
\int^{\infty}_{0}dk{\;}k{\,}
\biggl[{\mid}z^{(o)}_{k\ell}{\mid}^{2}{\,}
\biggl({\mid}a^{(o)}_{1k\ell m}{\mid}^{2}
+{\rm Re}\Bigl(a^{(o)}_{1k\ell m}{\,}
a^{(o)*}_{1,-k\ell m}\Bigr)\biggr)\cr
&+{\,}{\mid}z^{(e)}_{k\ell}{\mid}^{2}{\,}
\biggl({\mid}a^{(e)}_{1k\ell m}{\mid}^{2}
+{\rm Re}\Bigl(a^{(e)}_{1k\ell m}{\,}
a^{(e)*}_{1,-k\ell m}\Bigr)\biggr)\biggr]{\,}
{\cot(kT)}{\mid}_{\Sigma_{F}}
{\;}.\cr}\eqno(6.19)$$
As promised, this does now express the classical 
Maxwell part  
${\,}S^{EM}_{\rm class}{\,}$ 
of the action as an explicit functional 
of suitably chosen boundary data, namely, 
$\{a^{(o)}_{1k\ell m}\}$ 
and $\{a^{(e)}_{1k\ell m}\}{\,}$.
\end{section}

\begin{section}{Boundary conditions and asymptotically-flat gauge 
-- odd-parity gravitational perturbations}
In classical Lorentzian general relativity, 
one would expect to choose regular Cauchy data 
on an initial space-like hypersurface $\Sigma_{I}{\,}$, 
which would then evolve smoothly into $\{x^{0}>0\}{\,}$, 
subject to the linear hyperbolic equation (5.44).  
A natural initial condition, for given quantum numbers 
$(\ell{\,},m)$ [65], would be to assume an initially 
stationary odd-parity multipole:
$$\Bigl(\partial_{t}\xi^{(-)}_{2\ell m}\Bigr)
\Bigl\arrowvert_{t=0}
{\;}{\,}={\;}{\,}0
{\quad}.\eqno(7.1)$$

The combined Einstein/massless-scalar 
boundary-value problem, originally posed in [33,34], 
for complex time-separation 
${\,}T={\mid}T{\mid}{\,}\exp(-{\,}i\theta){\;},
{\;}0<\theta\leq\pi/2{\,}$,
involved specifying the intrinsic 3-metric 
$(h_{ij})_{I,F}$
and the value of the scalar field $(\phi)_{I,F}$ 
on the initial and final space-like hypersurfaces 
$\Sigma_{I}{\,},\Sigma_F{\,}$.  
By Eq.(5.43), the above Eq.(7.1) reads
$$Q^{(-)}_{\ell m}(0{\,},r){\;}{\,}
={\;}{\,}0
\eqno(7.2)$$
or, equivalently,
$$\eqalignno{h^{(-)}_{1\ell m}(0{\,},r){\;}{\,} 
&={\;}{\,}0
{\quad},&(7.3)\cr
h^{(-)}_{2\ell m}(0{\,},r){\;}{\,} 
&={\;}{\,}0
{\quad},&(7.4)\cr}$$
We therefore take these as our (Dirichlet) boundary 
conditions on the odd-parity gravitational perturbations, 
on the initial surface $\Sigma_{I}{\,}$, 
even though they may have arisen from consideration 
of the Cauchy problem.

In [11,16,33,34] for the $s=0$ case, 
we made use of the adiabatic approximation in order 
to separate the perturbation problem with respect 
to the variables $t{\,}$ and $r{\,}$.  
Here, for $s=2{\,}$, we first separate the odd-parity 
Eqs.(5.25,26) in the RW gauge, 
and then use Eqs.(5.21-23) to determine 
the time-dependence (in particular) in any gauge.

As in the massless-scalar $(s=0)$ case, 
we introduce a 'Fourier-type' expansion: 
$$h^{(-)RW}_{0,1,2\ell m}(t{\,},r){\;}{\,} 
={\;}{\,}\int^{\infty}_{-\infty}
dk{\;}a^{(-)}_{k\ell m}
{\;}h^{(-)RW}_{0,1,2k\ell m}(t{\,},r)
{\quad},\eqno(7.5)$$
where the $\{a^{(-)}_{k\ell m }\}$ 
are certain odd-parity 'Fourier'coefficients.  
With suitable treatment of any arbitrary phase factors
involved, in order to separate the odd-parity field 
equations (5.22,23) in the adiabatic approximation, 
one must have 
$$\eqalignno{h^{(-)RW}_{0\ell m}(t{\,},r){\quad}&
\propto{\quad}\cos(kt)
{\quad},&(7.6)\cr
h^{(-)RW}_{1\ell m}(t{\,},r){\quad}&
\propto{\quad}\sin(kt)
{\quad},&(7.7)\cr}$$
(Of course, a normal-mode $e^{-ikt}$ time dependence 
for the functions $h^{(-)RW}_{0{\,},1{\,},\ell m}$ 
would also satisfy the field equations.)  
In Eq.(5.23), if ${\,}h^{(-)RW}_{0\ell m}$, 
which is related to the odd-parity shift 
and can thus be freely specified, has $\cos(kt)$ 
time dependence, then $h^{(-)}_{0\ell m}$ 
must have the same time-dependence, 
while $\Lambda_{\ell m}(t{\,},r)$ 
must have $\sin(kt)$ time dependence.  
But, by Eq.(5.21), $h^{(-)}_{2\ell m}$ 
must then have $\sin(kt)$ time-dependence.  
Similarly, from Eq.(5.22), 
given that $h^{(-)RW}_{1\ell m}$ 
has $\sin(kt)$ time-dependence, 
$h^{(-)}_{1\ell m}$ 
must also have $\sin(kt)$ time-dependence, 
as must $\Lambda_{\ell m}(t{\,},r){\,}$.  
These conclusions are consistent with our choice 
of boundary conditions (7.3,4).  
Noting Eqs.(5.37,42,43), 
the Dirichlet conditions (7.3,4) are equivalent 
to the boundary condition (7.1), which is analogous 
to a specification of momenta 
in a 
$\bigl(\xi^{(-)}_{2\ell m}{\,},
{\,}\partial_{t}\xi^{(-)}_{2\ell m}{\,}\bigr)$ 
representation.  
This also accounts for the minus sign in Eq.(5.40).

For large $r{\,}$, the potential term in Eq.(5.44) 
vanishes sufficiently rapidly that 
$\xi^{(-)}_{2\ell m}$ becomes a superposition 
of outgoing and ingoing waves at radial infinity.  
Note that $Q^{(-)}_{\ell m}$ also obeys Eq.(5.44); 
thus, Eq.(5.42) in the RW gauge tells us that 
${\,}h^{(-)RW}_{1\ell m}
={\,}r{\,}Q^{(-)RW}_{\ell m}e^{a}
={\,}O(r){\;}$ 
at large $r{\,}$.  Now, the field equation (5.27) 
implies that 
$$\Bigl(\partial_{t}h^{(-)RW}_{o\ell m}\Bigr){\;}{\,} 
={\;}{\,}e^{-a}{\,}\partial_{r}
\Bigl(r{\,}Q^{(-)RW}_{\ell m}\Bigr){\quad}.
\eqno(7.8)$$
That is, odd-parity metric perturbations diverge 
at large $r{\,}$, in the RW gauge.  
This is only a coordinate effect, 
as the Riemann-curvature invariants decay 
at a rate $O(r^{-1})$ at large $r$ [42,43].  
(A similar phenomenon occurs for the even-parity 
perturbations in the RW gauge.)  
Here, in the odd-parity case, we construct 
a gauge transformation to an asymptotically-flat (AF) 
gauge, in which the radiative behaviour 
of the metric perturbations becomes manifest.

Our odd-parity AF gauge is chosen such that
$$h^{(-)AF}_{0\ell m}(t{\,},r){\;}{\,}
={\;}{\,}0{\quad}.
\eqno(7.9)$$
Thus, in terms of the preceding RW gauge:
$$\eqalignno{h^{(-)AF}_{0\ell m}{\;}{\,}
&={\;}{\,}0{\;}{\,} 
={\;}{\,}h^{(-)RW}_{0\ell m}
-\Bigl(\partial_{t}\Lambda_{\ell m}\Bigr){\quad},
&(7.10)\cr
h^{(-)AF}_{1\ell m}{\;}{\,} 
&={\;}{\,}h^{(-)RW}_{1\ell m}
-\Bigl(\partial_{r}\Lambda_{\ell m}\Bigr)
+{{2\Lambda_{\ell m}}\over{r}}{\quad},
&(7.11)\cr
h^{(-)AF}_{2\ell m}{\;}{\,} 
&={\;}{\,}2{\,}\Lambda_{\ell m}{\quad}.
&(7.12)\cr}$$
Given ${\,}h^{(-)RW}_{0\ell m}$ 
and ${\,}h^{(-)RW}_{1\ell m}$  
as a starting-point, one can, from the above, 
determine ${\,}\Lambda_{\ell m}(t{\,},r)$ 
and hence 
${\,}h^{(-)AF}_{1\ell m}$ 
and 
${\,}h^{(-)AF}_{2\ell m}$.
On substituting for $h^{(-)RW}_{1\ell m}$ 
from Eq.(7.11) into Eq.(5.24), one finds
$$\bigl(\partial_{t}\bigr)^{2}h^{(-)AF}_{1\ell m}{\;}{\,} 
={\,}-{\,}{{2\lambda e^{-a}}\over{r^{2}}}{\,} 
h^{(-)RW}_{1\ell m}{\quad}.
\eqno(7.13)$$
Now, following the approach used throughout 
when studying boundary conditions at the final 
surface ${\,}\Sigma_{F}{\;}{\,}(t=T){\,}$, set:
$$h^{(-)AF}_{1\ell m}(t{\,},r){\;}{\,} 
={\;}{\,}\int^{\infty}_{-\infty}dk{\;}
a^{(-)}_{k\ell m}{\;}h^{(o)AF}_{1k\ell}(r){\,}
{{\sin(kt)}\over{\sin(kT)}}{\quad},
\eqno(7.14)$$
where the ${\;}\{h^{(-)AF}_{1k\ell}(r)\}{\;}$ 
are real radial functions. 

The property 
$a^{(-)*}_{k\ell m}
=(-1)^{m}a^{(-)}_{-k\ell ,-m}$
holds for the coefficients.
Similarly, one can construct a corresponding expansion for 
${\,}Q^{(-)RW}_{\ell m}(t{\,},r){\,}$.  
Then, from Eq.(7.13), one has
$$h^{(-)AF}_{1\ell m}(t{\,},r){\;}{\,} 
={\;}{\,}{{2\lambda}\over{r}}
\int^{\infty}_{-\infty}{\,}dk{\;}
{{a^{(-)}_{k\ell m}}\over{k^{2}}}{\;}
Q^{(-)RW}_{k\ell}(r){\;}
{{\sin (kt)}\over{\sin(kT)}}{\quad},
\eqno(7.15)$$
which is ${\,}O(r^{-1})$ at large $r{\,}$, as required.  
On using Eq.(5.37), one further finds
$$\eqalign{\xi^{(-)AF}_{2\ell m}(t{\,},r){\;}{\,}
&={\;}{\,}r\ell(\ell +1)
\Bigl(\partial_{t}h^{(-)AF}_{1\ell m}\Bigr)\cr
&={\;}{\,}\int^{\infty}_{-\infty}
dk{\;}{\hat a}^{(-)}_{2k\ell m}{\;}
\xi^{(-)AF}_{2k\ell}(r){\;}
{{\cos(kt)}\over{\sin(kT)}}{\quad},\cr}
\eqno(7.16)$$
where
$${\hat a}^{(-)}_{2k\ell m}{\;}{\,} 
={\;}{\,}k\ell(\ell +1)a^{(-)}_{k\ell m}{\quad},
\eqno(7.17)$$ 
and where
$$\xi^{(-)AF}_{2k\ell}(r){\;}{\,} 
={\;}{\,}r{\,}h^{(-)AF}_{1k\ell}(r)
\eqno(7.18)$$
satisfies
$$e^{-a}\Bigl(e^{-a}{\,}\xi^{(-)AF'}_{2k\ell}\Bigr)' 
+\Bigl(k^{2}-V^{(-)}_{\ell}(r)\Bigr)
\xi^{(-)AF}_{2k\ell}{\;}{\,}
={\;}{\,}0{\quad}.
\eqno(7.19)$$

As in the spin-0 case [33,34] 
and in the spin-1 case above, we have, for ${\,}k>0{\;}$:
$$\eqalignno{\xi^{AF}_{2k\ell -}(r){\,} 
&\sim{\,}r{\,}j_{\ell}(kr){\;}{\,},
{\qquad}{\quad}{\rm as}{\quad}r{\,}\rightarrow{\,}0{\quad}, 
&(7.20)\cr
\xi^{AF}_{2k\ell -}(r){\,} 
&\sim
\Bigl(\Bigl(z_{2k\ell-}\Bigr)
\exp\bigl(ikr^{*}_{s}\bigr) 
+\Bigl(z^{*}_{2k\ell -}\Bigr)
\exp\bigl(-ikr^{*}_{s}\bigr)\Bigr),
{\:}{\rm as}{\;}r^{*}_{s}
\rightarrow\infty ,
&(7.21)\cr}$$
where the $j_{\ell}(z)$ are spherical Bessel 
functions, and where $r^{*}_{s}{\,}$ is the Schwarzschild 
Regge-Wheeler coordinate [49,63] of Eq.(4.38).
Thence, one deduces the normalisation property
$$\int^{R\infty}_{0}dr{\,}e^{a}{\,}
\xi^{AF}_{2k\ell -}(r){\;}\xi^{AF}_{2k'\ell -}(r)
\Bigl\arrowvert_{\Sigma_{F}}{\;} 
={\;}2\pi{\,}{\mid}z_{2k\ell -}{\mid}^{2}{\,}
\Bigl(\delta(k{\,},k')+\delta(k{\,},-k')\Bigr){\;}.
\eqno(7.22)$$
The resulting form of the classical action 
for odd-parity (spin-2) gravitational perturbations 
can then be expressed as a functional of the complex 
quantities $\{a_{2k\ell m-}\}$ which encode the boundary
data on $\Sigma_{F}$ for the odd-parity 
gravitational perturbations.  Here,
$$\eqalign{&S^{(2)}_{\rm class}
\Bigl[\{a_{2k\ell m-}\}\Bigr]\cr
&={\,}{{1}\over{16}}
\sum^{\infty}_{\ell =2}\sum^{\ell}_{m=-\ell}
{{(\ell -2)!}\over{(\ell +2)!}}
\int^{\infty}_{0}dk{\,}k{\,}
{\mid}z_{2k\ell-}{\mid}^{2}{\;}
{\mid}(a_{2k\ell m-})-(a_{2,-k\ell m-}){\mid}^{2}
\cot(kT){\,}.\cr}
\eqno(7.23)$$
From this expression, one proceeds as in [33,34] 
(for spin-0) and as above (for spin-1) to study 
the semi-classical quantum amplitude or wave function, 
proportional to 
${\,}\exp\bigl(iS^{(2)}_{\rm class}\bigr){\,}$, 
as a function of the complexified time-interval 
${\,}T={\mid}T{\mid}\exp(-{\,}i\theta){\,}$,
for ${\,}0<\theta\leq\pi/2{\,}$.  
Just as in the spin-0 and spin-1 case,
one straightforwardly recovers the complex Lorentzian 
amplitude for odd-parity gravitational perturbations, 
on taking the limit 
${\,}\theta\rightarrow 0_{+}{\;}$.
\end{section}

\begin{section}{Regge-Wheeler formalism 
-- even-parity gravitational perturbations}
Working with even-parity gravitational perturbations 
in the RW formalism is notoriously more difficult 
than working with those of odd parity.  
Yet, Chandrasekhar [57] showed that solutions 
to Zerilli's even-parity equation [64,70] 
[Eq.(9.10) below] can be expressed in terms 
of the odd-parity solutions.  
One might therefore expect that our results 
for the even-parity action should mirror 
those for the odd-parity action.

We expand the even-parity perturbations as
$$\eqalign{\Bigl(h^{(+)}_{ij}\Bigr)_{\ell m}(x){\,}
=&{\,}h^{(+)}_{1\ell m}(t{\,},r){\,}
\Bigl[(f_{1})_{ij}\Bigr]_{\ell m}
+H_{2\ell m}(t{\,},r){\,}e^{(a-b)/2}{\,}
\Bigl[(f_{2})_{ij}\Bigr]_{\ell m}\cr
&+{\,}r^{2}{\,}K_{\ell m}(t{\,},r)
\Bigl[(f_{3})_{ij}\Bigr]_{\ell m} 
+r^{2}{\,}G_{\ell m}(t{\,},r)
\Bigl[(f_{4})_{ij}\Bigr]_{\ell m}{\quad},\cr}
\eqno(8.1)$$
Here, the non-zero components 
of the (un-normalised) basis of the symmetric 
tensor spherical harmonics 
$\bigl[(f_{1,2,3,4})_{ij}\bigr]_{\ell m}$ 
are defined by 
$$\bigl[(f_1)_{r\theta}\bigr]_{\ell m}{\;}{\,}
={\;}{\,}\bigl({\partial}_{\theta}Y_{\ell m}\bigr){\quad},
\eqno(8.2)$$
$$\bigl[(f_1)_{r\phi}\bigr)]_{\ell m}{\;}{\,}
={\;}{\,}\bigl({\partial}_{\phi}Y_{\ell m}\bigr){\quad},
\eqno(8.3)$$
$$\bigl[(f_2)_{rr}\bigr]_{\ell m}{\;}{\,}
={\;}{\,}Y_{\ell m}{\quad},
\eqno(8.4)$$
$$\bigl[(f_3)_{\theta\theta}\bigr]_{\ell m}{\;}{\,}
={\;}{\,}Y_{\ell m}{\quad},
\eqno(8.5)$$
$$\bigl[(f_3)_{\phi\phi}\bigr]_{\ell m}{\;}{\,}
={\;}{\,}\bigl({\sin}^{2}\theta\bigr){\,}Y_{\ell m}{\quad},
\eqno(8.6)$$
$$\bigl[(f_4)_{\theta\theta}\bigr]_{\ell m}{\;}{\,}
={\;}{\,}\bigl({\partial}_{\theta}\bigr)^{2}{\,}Y_{\ell m}{\quad},
\eqno(8.7)$$
$$\bigl[(f_4)_{\theta\phi}\bigr]_{\ell m}{\;}{\,}
={\;}{\,}\bigl[{\partial}_{\theta}{\partial}_{\phi}
-(\cot\theta){\partial}_{\phi}\bigr]Y_{\ell m}{\quad},
\eqno(8.8)$$
$$\bigl[(f_4)_{\phi\phi}\bigr]_{\ell m}{\;}{\,}
={\;}{\,}\bigl[({\partial}_{\phi})^{2}
+(\sin\theta{\,}\cos\theta){\partial}_{\theta}\bigr]Y_{\ell m}
{\quad},
\eqno(8.9)$$

The non-zero inner products are given by
$$\eqalignno{\int d\Omega{\,}
\bigl[(f_{1})^{ij}\bigr]_{\ell m}{\,}
\bigl[(f_{1})_{ij}\bigr]^{*}_{\ell'm'}{\;}
&={\;}{{2e^{-a}}\over{r^{2}}}{\,}\ell(\ell +1)
\delta_{\ell\ell'}{\,}\delta_{mm'}
{\quad},&(8.10)\cr
\int d\Omega{\,}\bigl[(f_{2})^{ij}\bigr]_{\ell m}{\,}
\bigl[(f_{2})_{ij}\bigr]^{*}_{\ell'm'}{\;}
&={\;}e^{-2a}{\,}\delta_{\ell\ell'}{\,}\delta_{mm'}
{\quad},&(8.11)\cr
\int d\Omega{\,}\bigl[(f_{3})^{ij}\bigr]_{\ell m}{\,}
\bigl[(f_{3})_{ij}\bigr]^{*}_{\ell'm'}{\;}
&={\;}{{2}\over{r^{4}}}{\,}
\delta_{\ell\ell'}{\,}\delta_{mm'}
{\quad},&(8.12)\cr
\int d\Omega{\,}\bigl[(f_{4})^{ij}\bigr]_{\ell m}{\,}
\bigl[(f_{4})_{ij}\bigr]^{*}_{\ell'm'}{\;}
&={\;}{{{\Lambda}_{\ell}
({\Lambda}_{\ell}-1)}\over{r^{4}}}{\,}
\delta_{\ell\ell'}{\,}\delta_{mm'}
{\quad},&(8.13)\cr
\int d\Omega{\,}\bigl[(f_{3})_{ij}\bigr]_{\ell m}{\,}
\bigl[(f_{4})^{ij}\bigr]^{*}_{\ell'm'}{\;}
&={\,}-{\,}{{\ell(\ell+1)}\over{r^{4}}}{\,}
\delta_{\ell\ell'}{\,}\delta_{mm'}
{\quad},&(8.14)\cr}$$
where we define 
${\Lambda}_{\ell}{\,}={\ell}({\ell}+1){\,}$.  
The even-parity basis is also orthogonal 
to the odd-parity basis of Sec.4.

Further, for the even-parity perturbed shift, 
one can write
$$\Bigl[N^{(+)}_{i}\Bigr]_{\ell m}{\,} 
={\,}\Bigl[H_{1\ell m}(t{\,},r){\,}Y_{\ell m}{\,},
{\,}h^{(+)}_{0\ell m}(t{\,},r)
\bigl(\partial_{\theta}Y_{\ell m}\bigr){\,},
{\,}h^{(+)}_{0\ell m}(t{\,},r)
\bigl(\partial_{\phi}Y_{\ell m}\bigr)\Bigr]
{\;}.\eqno(8.15)$$
For the perturbed lapse,
$$\bigl[N^{(1)(+)}\bigr]_{\ell m}{\,} 
={\,}-{\,}{{1}\over{2}}{\,}
H_{0\ell m}(t{\,},r){\,}e^{-a/2}{\,}Y_{\ell m}
{\quad}.\eqno(8.16)$$
Again, 
${\,}H^{*}_{0\ell m}=(-1)^{m}{\,}H_{0\ell,-m}{\;}$, etc.  
Hence, for the linear-order perturbation 
$h^{(1)}_{\mu\nu}$ 
of the 4-dimensional metric, 
the quantities $h^{(1)}_{tt},{\,}h^{(1)}_{rr}$ 
and $h^{(1)}_{tr}$ 
behave as scalars under rotations 
(their odd-parity part vanishes), 
while 
${\,}h^{(1)}_{t\theta}{\,},
{\,}h^{(1)}_{t\phi}{\,},{\,}h^{(1)}_{r\theta}{\,}$ 
and $h^{(1)}_{r\phi}$ behave as vectors.  
For 
${\,}a{\,},b{\,}={\,}\theta,\phi{\;}$, 
the $2{\times}2$ angular block $h^{(1)}_{ab}$ 
is a tensor under rotations.
The even-parity gravitational momentum components can,
correspondingly, be written in the form
$$\eqalign{\bigl(\pi^{(+)}_{ij}\bigr)_{\ell m}{\,} 
=&{\,}(^{3}\gamma)^{{1}\over{2}}{\,}
\Bigl(P_{h_{1}\ell m}(t{\,},r)
\bigl[(f_{1})_{ij}\bigr]_{\ell m}
+P_{H_{2}\ell m}(t{\,},r)
\bigl[(f_{2})_{ij}\bigr]_{\ell m}\cr
&+{\,}r^{2}{\,}P_{K\ell m}(t{\,},r)
\bigl[(f_{3})_{ij}\bigr]_{\ell m}
+r^{2}{\,}P_{G\ell m}(t{\,},r)
\bigl[(f_{4})_{ij}\bigr]_{\ell m}\Bigr){\;}.\cr}
\eqno(8.17)$$
Again, one can easily show that the $P$'s in Eq.(8.17) 
are related to
${\,}h_{1}{\,},H_{2}{\,},K{\,}$ 
and ${\,}G$ of the corresponding Eq.(8.1) by:
$$\eqalignno{P_{h1\ell m}(t{\,},r){\,}
&={\,}{{1}\over{2}}{\,}e^{a/2}
\Biggl(\Bigl(\partial_{t}h^{(+)}_{1\ell m}\Bigr)
-r^{2}{\,}\partial_{r}
\Biggl({{h^{(+)}_{0\ell m}}\over{r^{2}}}\Biggr)\Biggr)
{\;},
&(8.18)\cr
P_{G\ell m}(t{\,},r){\,}
&={\,}{{1}\over{2}}{\,}e^{a/2}
\Biggl(\Bigl(\partial_{t}G_{\ell m}\Bigr)
-\Biggl({{2h^{(+)}_{0\ell m}}
\over{r^{2}}}\Biggr)\Biggr){\;},
&(8.19)\cr
P_{K\ell m}(t{\,},r){\,}
=&{\,}-{\,}{{1}\over{2}}{\,}e^{a/2}
\Biggl(\Bigl(\partial_{t}H_{2\ell m}\Bigr)
+\Bigl(\partial_{t}K_{\ell m}\Bigr)\cr
&+\biggl(a'-{{2}\over{r}}\biggr)
e^{-a}H_{1\ell m}
-2{\,}e^{-a}\Bigl(\partial_{r}H_{1\ell m}\Bigr)\cr
&+\Biggl({{2\ell(\ell +1){\,}h^{(+)}_{0\ell m}}
\over{r^{2}}}\Biggr)
-\ell(\ell +1)
\Bigl(\partial_{t}G_{\ell m}\Bigr)\Biggr){\;},
&(8.20)\cr
P_{H2\ell m}(t{\,},r){\,}
=&{\,}-{\,}e^{a/2}\Bigl(\partial_{t}K_{\ell m}\Bigr)
+\biggl({{2}\over{r}}\biggr)e^{-a/2}H_{1\ell m}
-\Biggl({{\ell(\ell +1){\,}h^{(+)}_{0\ell m}}
\over{r^{2}}}\Biggr)e^{a/2}\cr
&+{\,}{{1}\over{2}}{\,}\ell(\ell+1)e^{a/2}
\Bigl(\partial_{t}G_{\ell m}\Bigr){\;}.
&(8.21)\cr}$$

For even-parity gravitational perturbations, 
gauge transformations are induced by even-parity 
gauge vector fields 
${\,}(\xi^{(+)\mu})_{\ell m}{\;}$,
of the form:
$$\eqalign{\Bigl(\xi^{(+)t}\Bigr)_{\ell m}{\;}
&={\;}X^{(+)}_{0\ell m}(t{\,},r){\,}Y_{\ell m}{\;},
{\quad}\Bigl(\xi^{(+)r}\Bigr)_{\ell m}{\;}
={\;}X^{(+)}_{1\ell m}(t{\,},r){\,}Y_{\ell m}{\;},\cr
\Bigl(\xi^{(+)\theta}\Bigr)_{\ell m}
&=X^{(+)}_{2\ell m}(t{\,},r)
\bigl(\partial_{\theta}Y_{\ell m}\bigr),
{\quad}\Bigl(\xi^{(+)\phi}\Bigr)_{\ell m}
=\Biggl({{X^{(+)}_{2\ell m}(t{\,},r)}
\over{\sin^{2}\theta}}\Biggr)
\bigl(\partial_{\phi}Y_{\ell m}\bigr){\,}.\cr}
\eqno(8.22)$$
Within the adiabatic approximation, 
these induce the following even-parity 
gauge transformations:
$$\eqalignno{H'_{0\ell m}{\;}{\,}
&={\;}{\,}H_{0\ell m}-a' X^{(+)}_{1\ell m}
+2\Bigl(\partial_{t}X^{(+)}_{0\ell m}\Bigr){\quad},
&(8.23)\cr
H'_{1\ell m}{\;}{\,}
&={\;}{\,}H_{1\ell m}{\,}
-{\,}e^{-a}\Bigl(\partial_{r}X^{(+)}_{0\ell m}\Bigr)
-{\,}e^{a}\Bigl(\partial_{t}X^{(+)}_{1\ell m}\Bigr){\quad},
&(8.24)\cr
H'_{2\ell m}{\;}{\,}
&={\;}{\,}H_{2\ell m}-{\,}a'{\,}X^{(+)}_{1\ell m}
-2\Bigl(\partial_{r}X^{(+)}_{1\ell m}\Bigr){\quad},
&(8.25)\cr
K'_{\ell m}{\;}{\,}
&={\;}{\,}K_{\ell m}-
\biggl({{2X^{(+)}_{1\ell m}}\over{r}}\biggr){\quad},
&(8.26)\cr
G'_{\ell m}{\;}{\,}
&={\;}{\,}G_{\ell m}-2X^{(+)}_{2\ell m}{\quad},
&(8.27)\cr
h^{(e)'}_{0\ell m}{\;}{\,}
&={\;}{\,}h^{(+)}_{0\ell m}
+e^{-a}X^{(+)}_{0\ell m}
-r^{2}\Bigl(\partial_{t}X^{(+)}_{2\ell m}\Bigr){\quad},
&(8.28)\cr
h^{(+)'}_{1\ell m}{\;}{\,}
&={\;}{\,}h^{(+)}_{1\ell m}
-e^{a}X^{(+)}_{1\ell m}
-{\,}r^{2}\Bigl(\partial_{r}X^{(+)}_{2\ell m}\Bigr){\quad}.
&(8.29)\cr}$$

As in the odd-parity case, we would like to construct 
gauge-invariant combinations of components 
of the perturbed 3-geometry.  
Following [68], we define
$$\eqalignno{k_{1\ell m}{\;}{\,}
&={\;}{\,}K_{\ell m}
+e^{-a}\Biggl(r\Bigl(\partial_{r}G_{\ell m}\Bigr)
-\Biggl({{2h^{(+)}_{1\ell m}}\over{r}}\Biggr)\Biggr){\quad},
&(8.30)\cr
k_{2\ell m}{\;}{\,}
&={\;}{{1}\over{2}}\biggl(e^{a}{\,}H_{2\ell m}
-e^{a/2}{\,}\partial_{r}
\Bigl(r{\,}e^{a/2}{\,}K_{\ell m}\Bigr)\biggr){\quad}.
&(8.31)\cr}$$
It can be shown that both the functions 
${\,}k_{1\ell m}{\,}$ 
and 
${\,}k_{2\ell m}{\,}$ are gauge-invariant.  
For future use, in the calculation of the even-parity 
classical action, we define [73] the linear 
combination of ${\,}k_{1\ell m}{\,}$ 
and ${\,}k_{2\ell m}{\;}$:
$$q_{1\ell m}{\;}{\,}
={\;}{\,}r\ell(\ell +1)k_{1\ell m}
+4re^{-2a}{\,}k_{2\ell m}{\quad}.
\eqno(8.32)$$

At this stage, as with the odd-parity case, 
we again make use of the property of the uniqueness 
of the (even-parity) RW gauge.  
In the RW gauge, one has
$$\eqalign{H^{RW}_{0\ell m}{\;}
=&{\;}H_{0\ell m} 
-{{1}\over{2}}{\,}r^{2}{\,}a'{\,}e^{-a}
\Biggl({{2h^{(+)}_{1\ell m}}\over{r^{2}}}
-\Bigl(\partial_{r}G_{\ell m}\Bigr)\Biggr)
+r^{2}{\,}e^{a}\bigl(\partial_{t}\bigr)^{2}G_{\ell m}\cr
&-2e^{a}\Bigl(\partial_{t}h^{(+)}_{0\ell m}\Bigr){\quad},\cr}
\eqno(8.33)$$
Then,
$$\eqalign{H^{RW}_{1\ell m}{\;}{\,}
=&{\;}{\,}H_{1\ell m}
+r^{2}\Bigl(\partial_{r}\partial_{t}G_{\ell m}\Bigr)
\Bigl(\partial_{r}h^{(+)}_{0\ell m}\Bigr)
-\Bigl(\partial_{t}h^{(+)}_{1\ell m}\Bigr)\cr
&+{\,}r\biggl(1+{{1}\over{2}}{\,}ra'\biggr)
\Bigl(\partial_{t}G_{\ell m}\Bigr)
-a'{\,}h^{(+)}_{0\ell m}{\quad}.\cr}
\eqno(8.34)$$

Next,
$$\eqalign{H^{RW}_{2\ell m}{\;}{\,}
&={\;}{\,}H_{2\ell m}
+\biggl(a'-{{4}\over{r}}\biggr)e^{-a}
\biggl(h^{(+)}_{1\ell m}
-{{1}\over{2}}r^{2}
\Bigl(\partial_{r}G_{\ell m}\Bigr)\biggr)\cr
&+r^{2}{\,}e^{-a}
\Biggl(\bigl(\partial_{r}\bigr)^{2}G_{\ell m}
-2{\,}\partial_{r}
\Biggl({{h^{(+)}_{1\ell m}}\over{r^{2}}}\Biggr)\Biggr)
{\quad}.\cr}
\eqno(8.35)$$
Further,
$$K^{RW}_{\ell m}{\;}{\,}
={\;}{\,}K_{\ell m}
-\biggl({{2e^{-a}}\over{r}}\biggr)
\biggl(h^{(+)}_{1\ell m} 
-{{1}\over{2}}r^{2}
\Bigl(\partial_{r}G_{\ell m}\Bigr)\biggr){\quad};
\eqno(8.36)$$
with
$$G^{RW}_{\ell m}{\;}{\,}
={\;}{\,}0{\;}{\,}
={\;}{\,}G_{\ell m}-2X^{(+)}_{2\ell m}{\quad},
\eqno(8.37)$$
together with
$$h^{(+)RW}_{0\ell m}{\;}{\,}
={\;}{\,}0{\quad},
\eqno(8.38)$$
and
$$h^{(+)RW}_{1\ell m}{\;}{\,}
={\;}{\,}0{\quad},
\eqno(8.39)$$
where (in the RW gauge)
$$\eqalignno{X^{(+)}_{0\ell m}{\;}{\,} 
&={\;}{\,}e^{-b}\biggl({{1}\over{2}}{\,}r^{2}
\Bigl(\partial_{t}G_{\ell m}\Bigr)
-h^{(+)}_{0\ell m}\biggr){\quad},
&(8.40)\cr
X^{(+)}_{1\ell m}{\;}{\,} 
&={\;}{\,}e^{-a}\biggl(h^{(+)}_{1\ell m}
-{{1}\over{2}}r^{2}\Bigl(\partial_{r}G_{\ell m}\Bigr)\biggr)
{\quad},&(8.41)\cr
X^{(+)}_{2\ell m}{\;}{\,} 
&={\;}{\,}{{1}\over{2}}{\,}G_{\ell m}{\quad}.
&(8.42)\cr}$$

At late times, following gravitational collapse 
to a black hole, in the absence of background matter 
and in the adiabatic approximation, 
the even-parity RW field equations are seven coupled
equations for the four unknowns 
${\,}(H^{RW}_{0\ell m}{\,},H^{RW}_{1\ell m}{\,},
H^{RW}_{2\ell m}{\,},K^{RW}_{\ell m}){\,}$.  
Assuming that ${\,}\ell\geq{\,}2{\;}{\,}$ 
--- that is, that we are studying dynamical modes 
--- we give here those RW field equations 
which are of first order in $r$ and $t{\,}$ [70].  
These are, respectively, 
the ${\,}(t\theta){\,},{\,}(tr){\,}$ 
and 
${\,}(r\theta){\,}$ 
components of the linearised field equations:
$$\eqalignno{\Bigl(\partial_{r}H^{RW}_{1\ell m}\Bigr)
&+{\,}{{2m}\over{r^{2}}}{\,}e^{a}{\,}H^{RW}_{1\ell m}{\;}{\,}
={\;}{\,}e^{a}{\,}\partial_{t}
\Bigl(K^{RW}_{\ell m}+H^{RW}_{2\ell m}\Bigr){\quad},
&(8.43)\cr
{{1}\over{2}}{\,}\ell(\ell +1)H^{RW}_{1\ell m}{\;}
=&-{\,}r^{2}
\Bigl(\partial_{t}\partial_{r}K^{RW}_{\ell m}\Bigr)
+r^{2}\Bigl(\partial_{t}H^{RW}_{0\ell m}\Bigr)\cr
&-{\,}re^{a}\biggl(1-{{3m}\over{r}}\biggr)
\Bigl(\partial_{t}K^{RW}_{\ell m}\Bigr){\quad},
&(8.44)\cr
\Bigl(\partial_{t}H^{RW}_{1\ell m}\Bigr){\,}
&={\,}e^{-a}\Bigl(\partial_{r}H^{RW}_{0\ell m}\Bigr)
-e^{-a}\Bigl(\partial_{r}K^{RW}_{\ell m}\Bigr)
+{{2m}\over{r^{2}}}H^{RW}_{0\ell m}{\,},
&(8.45)\cr}$$
and the ${\,}(\theta\phi){\,}$ component
$$H^{RW}_{0\ell m}{\;}{\,}
={\;}{\,}H^{RW}_{2\ell m}{\;}{\,}
\equiv{\;}{\,}H^{RW}_{\ell m}{\quad}.
\eqno(8.46)$$
We also give one second-order equation, 
namely, the ${\,}(rr){\,}$ component:
$$\eqalign{e^{2a}
\bigl(\partial{_t}\bigr)^{2}{\,}K^{RW}_{\ell m}{\;}{\,}
=&{\;}{{2}\over{r}}{\,}e^{a} 
\Bigl(\partial_{t}H^{RW}_{1\ell m}\Bigr)
-{{1}\over{r}}\Bigl(\partial_{r}H^{RW}_{2\ell m}\Bigr)\cr
&+{\,}{{e^{a}}\over{r}}\biggl(1-{{m}\over{r}}\biggr)
\Bigl(\partial_{r}K^{RW}_{\ell m}\Bigr)\cr
&-{\,}{{e^{a}}\over{2r^{2}}}{\,}(\ell +2)(\ell -1)
\Bigl(K^{RW}_{\ell m}
-H^{RW}_{2\ell m}\Bigr){\quad}.\cr}
\eqno(8.47)$$

Following Eq.(8.44), we find, 
for the gauge-invariant component defined in Eq.(8.32): 
$$\bigl(\partial_{t}q_{1\ell m}\bigr){\;}{\,} 
={\;}{\,}\ell(\ell +1)
\biggl(r\Bigl(\partial_{t}K^{RW}_{\ell m}\Bigr)
-e^{-a}{\,}H^{RW}_{1\ell m}\biggr){\quad}.
\eqno(8.48)$$
We also find
$$\Bigl(\partial_{r}q_{1\ell m}\Bigr){\;}{\,}
={\,}-{\,}\ell(\ell +1)
\Biggl(2e^{-a}{\,}k_{2\ell m}
+\biggl(1+{{1}\over{2}}
ra'\biggr)k_{1\ell m}\Biggr){\quad},
\eqno(8.49)$$
where, in the RW gauge,
$$q_{1\ell m}{\quad}
\equiv{\quad}2re^{-a}{\,}H^{RW}_{\ell m}
-2r^{2}{\,}e^{-a}
\Bigl(\partial_{r}K^{RW}_{\ell m}\Bigr)
+2\bigl(\lambda r+3m\bigr)K^{RW}_{\ell m}{\quad},
\eqno(8.50)$$
with 
${\;}\lambda{\,}={\,}{{1}\over{2}}(\ell +2)(\ell -1){\;}$.  
We can now solve for 
${\,}k_{1\ell m}{\,},{\,}k_{2\ell m}{\,}$ 
in terms of 
${\,}q_{1\ell m}{\,}$ 
and its radial derivative 
${\,}(\partial_{r}q_{1\ell m}){\,}$, giving
$$\eqalignno{k_{1\ell m}{\;}{\,} 
&={\;}{\,}{{1}\over{2\bigl(\lambda r+3m\bigr)}}
\Biggl(\Biggl({{re^{-a}
\bigl(\partial_{r}q_{1\ell m}\bigr)}
\over{\bigl(\lambda +1\bigr)}}\Biggr)
+q_{1\ell m}\Biggr){\quad},
&(8.51)\cr
k_{2\ell m}{\;}{\,}
&={\,}-{\,}{{e^{a}}\over{4\bigl(\lambda r+3m\bigr)}}
\Biggl(r\bigl(\partial_{r}q_{1\ell m}\bigr)
+e^{a}\biggl(1-{{3m}\over{r}}\biggr)q_{1\ell m}\Biggr){\;}.
&(8.52)\cr}$$
Further, $H^{RW}_{\ell m}$ and $K^{RW}_{\ell m}$ 
can also be written in terms of 
${\,}k_{1\ell m}{\,}$ and ${\,}k_{2\ell m}{\,}$.
\end{section}

\begin{section}{ Classical action and boundary conditions 
-- even-parity gravitational perturbations}
As in the case of odd-parity gravitational 
perturbations, we can exploit the uniqueness 
of the RW gauge to simplify the even-parity action 
and to obtain a general gauge-invariant form 
for the even-parity classical action 
${\,}S^{(2)}_{\rm class}{\,}$.  
In the RW gauge, this is
$$\eqalign{S^{(2)}_{\rm class}
\Bigl[(h^{(+)}_{ij})_{\ell m}\Bigr]{\,}
&={\,}{{1}\over{32\pi}}
\int_{\Sigma_F}d^{3}x
\sum_{\ell \ell' mm'}
\Bigl(\pi^{(+)ij}\Bigr)_{\ell m}
\Bigl(h^{(+)}_{ij}\Bigr)^{*}_{\ell' m'}\cr
&+{{1}\over{32\pi}}\sum_{\ell m}
\int^{R_{\infty}}_{0}dr{\,}e^{a/2}
\biggl(H^{RW*}_{\ell m}{\,}P^{RW}_{H 2\ell m}
+2K^{RW*}_{\ell m}{\,}P^{RW}_{K\ell m}\biggr)
\Bigl\arrowvert_T\cr}
{\,}.
\eqno(9.1)$$
Again, we would like to put the action into the form
$\int{\,}dr{\,}\psi{\,}(\partial_{t}\psi){\,}$,  
where $\psi$ is gauge-invariant and obeys a decoupled 
wave equation.  
Since $q_{1\ell m}$ 
is the only unconstrained gauge-invariant even-parity 
quantity which involves only perturbations 
of the intrinsic 3-geometry, 
one might expect that Eq.(9.1) should reduce 
to the form
$$S^{(2)}_{\rm class}\Bigl[\{q_{1\ell m}\}\Bigr]{\;} 
={\;}{{1}\over{32\pi}} 
\sum^{\infty}_{\ell=2}\sum^{\ell}_{m=-\ell}
\int^{R_{\infty}}_{0}dr
\Bigl(\pi_{1\ell m}{\,}q^{*}_{1\ell m}
+\bigl(\partial_{r}Z_{\ell m}\bigr)\Bigr)
\Bigl\arrowvert_{t=T}{\;}{\,},
\eqno(9.2)$$
for some variable ${\,}Z_{\ell m}{\,}$, 
where $\pi_{1\ell m}$ is the gauge-invariant momentum 
conjugate to $q_{1\ell m}{\,}$.  
This is in fact the case.  
First, make in Eq.(9.1) the substitutions 
(as mentioned at the end of Sec.8) 
for $H^{RW}_{2\ell m}$ 
and 
$K^{RW}_{\ell m}$ in terms of ${\,}k_{1\ell m}{\,}$ 
and 
${\,}k_{2\ell m}{\;}$; 
then substitute the expressions (8.51,52) 
for ${\,}k_{1\ell m}{\,}$ 
and 
${\,}k_{2\ell m}{\,}$ 
in terms of 
${\,}q_{1\ell m}{\;}$.  
After several integrations by parts, 
we arrive at an action of the form (9.2), with
$$\eqalignno{\pi_{1\ell m}{\;}
=&{\;}{{r^{2}}\over{2\bigl(\lambda r+3m\bigr)}} 
\Biggl({\hat P}^{RW}_{\ell m}
-\biggl(1-{{3m}\over{r}}\biggr)
P^{RW}_{H_{2}\ell m}{\,}e^{3a/2}\Biggr)\cr
&-{\,}{{1}\over{2}}{\,}\partial_{r}
\Biggl({{r^{3}}\over{\bigl(\lambda r+3m\bigr)}}
\Biggl({{{\hat P}^{RW}_{\ell m}e^{-a}}
\over{\bigl(\lambda +1\bigr)}}
-P^{RW}_{H_{2}\ell m}{\,}e^{a/2}\Biggr)\Biggr){\quad},
&(9.3)\cr
Z_{\ell m}{\,}
&={\,}r^{3}{\,}e^{a/2}{\,}
P^{RW}_{H_{2}\ell m}{\,}K^{RW}_{\ell m}
+{{r^{3}{\,}q_{1\ell m}}\over{2\bigl(\lambda r+3m\bigr)}}
\Biggl({{\hat P^{RW}_{\ell m}e^{-a}}
\over{\bigl(\lambda +1\bigr)}}
-P^{RW}_{H_{2}\ell m}{\,}e^{a/2}\Biggr){\,},
&(9.4)\cr 
{\hat P}^{RW}_{\ell m}{\;}
&={\;}e^{a/2}\biggl(2P^{RW}_{K\ell m}
-2P^{RW}_{H_{2}\ell m}
-r\Bigl(\partial_{r}P^{RW}_{H_{2}\ell m}\Bigl)\biggr){\quad}.
&(9.5)\cr}$$ 

This expression for the even-parity classical action 
simplifies yet further, since the linearised field 
equations imply that
$${\hat P}^{RW}_{\ell m}{\;}{\,} 
={\;}{\,}{{(\lambda +1)}\over{r}}{\,}
e^{a}{\,}H^{RW}_{1\ell m}{\quad}.
\eqno(9.6)$$
Further, Eqs.(8.43,44) show, with the help of Eq.(8.48), 
that
$$\pi_{1\ell m}{\;}{\,}
={\;}{\,}{{\lambda r e^{a}}
\over{2\bigl(\lambda r+3m\bigr)}} 
\bigl(\partial_{t}q_{1\ell m}\bigr){\quad}, 
\eqno(9.7)$$
Eq.(9.2) for the even-parity ${\,}S^{(2)}_{\rm class}{\,}$ 
then reduces to an expression of the desired form:
$$\eqalign{S^{(2)}_{\rm class}&
\Bigl[\{\bigl(h^{(+)}_{ij}\bigr)_{\ell m}\}\Bigr]\cr 
&={\;}{{1}\over{32\pi}}
\sum^{\infty}_{\ell =2}
\sum^{\ell}_{m=-\ell}
{{(\ell -2)!}\over{(\ell +2)!}}
\int^{R_{\infty}}_{0}dr{\,}
e^{a}{\,}\xi^{(+)}_{2\ell m}{\,}
\bigl(\partial_{t}\xi^{(+)*}_{2\ell m}\bigr)
\Bigl\arrowvert_{t=T}{\;},\cr}
\eqno(9.8)$$ 
where ${\,}\xi^{(+)}_{2\ell m}{\,}$ is defined as
$$\xi^{(+)}_{2\ell m}{\;}{\,}
={\;}{\,}{{\lambda r{\,}q_{1\ell m}}
\over{\bigl(\lambda r+3m\bigr)}}
{\quad}.
\eqno(9.9)$$

We have made use of the assumption above 
that the specified perturbations
$h^{(1)}_{ij}{\mid}_{\Sigma_{F}}$ 
of the spatial 3-metric on the final boundary 
$\Sigma_{F}$ have been taken to be real.  
Of course, for the Dirichlet boundary-value problem 
with $T$ rotated into the complex,
the classical solution for the metric and scalar field 
will have both an imaginary part and a real part.

Given the uniqueness of the RW gauge 
for even-parity modes, one can see that Eq.(9.8) 
for $S^{(2)}_{\rm class}$ is in fact valid 
in any gauge, with a vanishing contribution 
from the total divergence since $q_{1\ell m}{\,}$, 
as given by Eq.(8.28), and therefore also
$\xi^{(+)}_{2\ell m}{\,}$, 
are gauge-invariant.  
There are obvious similarities between Eq.(9.8) 
and the classical massless-scalar action of [33,34], 
with ${\,}\xi^{(+)}_{2\ell m}{\,}$ 
and ${\,}\xi_{0\ell m+}{\,}$
differing only by an $\ell$-dependent normalisation factor.  
This should not be surprising, 
as scalar spherical harmonics have even parity.

Again, one can show that the gauge-invariant quantity
${\,}\xi^{(+)}_{2\ell m}{\,}$ satifies Zerilli's equation [70]
$$e^{-a}{\,}\partial_{r}\Bigl(e^{-a}
\bigl(\partial_{r}\xi^{(+)}_{2\ell m}\bigr)\Bigr)
-\bigl(\partial_{t}\bigr)^{2}\xi^{(+)}_{2\ell m}
-V^{(+)}_{\ell}\xi^{(+)}_{2\ell m}{\quad} 
={\quad}0{\quad},
\eqno(9.10)$$
$$V^{(+)}_{\ell}{\,} 
={\,}\biggl(1-{{2m}\over r}\biggr)
{2\Bigl({{\lambda^{2}(\lambda +1)r^{3}+3\lambda^{2}mr^{2}
+9\lambda m^{2}r+9m^{3}}}\Bigr)
\over{r^{3}
\bigl(\lambda r+3m\bigr)^{2}}}{\quad}>{\quad}0{\;}.
\eqno(9.11)$$
Now, both for odd and even parity, 
the field equations for the metric perturbations 
have been reduced to the two wave equations (5.44) and (9.10).
  
In contrast to the odd-parity case, 
where we assumed an initially stationary multipole, 
here for even parity we treat 
${\,}\xi^{(+)}_{2\ell m}{\,}$ 
by analogy with the massless-scalar-field quantity 
${\,}\xi_{0\ell m+}{\;}$, 
and impose the Dirichlet boundary condition 
$$\xi^{(+)}_{2\ell m}(0{\,},r){\;}{\,}
={\;}{\,}0
\eqno(9.12)$$
at the initial surface ${\,}\Sigma_{I}{\;}{\,}(t=0)$.  
Proceeding now by analogy 
with the separation-of-variables analysis of Sec.7 
for the odd-parity case, we find that, 
if ${\,}K^{RW}_{\ell m}{\,}$ 
has ${\,}\sin(kt){\,}$ time-dependence, 
then so must $H^{RW}_{\ell m}$ also, 
whereas
$H^{RW}_{1\ell m}$ must have 
${\,}\cos(kt)$ time-dependence.  
Consistency with the gauge transformations (8.33-39) 
implies that these time dependences are valid 
in an arbitrary gauge, and further that 
$G_{\ell m}$ and $h^{(+)}_{1\ell m}$ 
have ${\,}\sin(kt){\,}$ 
time dependence, 
whereas $h^{(+)}_{0\ell m}$ 
has ${\,}\cos(kt){\,}$ 
time dependence.  
Consequently, ${\,}q_{1\ell m}$ must have 
${\,}\sin(kt){\,}$ time dependence, 
whence the boundary condition (9.12) is justified 
through Eq.(9.9).  
(Alternatively, one could instead have studied 
normal-mode time dependence.)

Following the scalar-field analysis of [33,34], 
we can write
$$\xi^{(+)}_{2 \ell m}(t{\,},r){\;}{\,}
={\;}{\,}\int^{\infty}_{-\infty}
dk{\,}a^{(+)}_{2k\ell m}{\,}
\xi^{(+)}_{2k\ell}(r){\,}
{{\sin(kt)}\over{\sin(kT)}}{\quad},
\eqno(9.13)$$
where the $\{a^{(+)}_{2k\ell m}\}$ 
are suitable even-parity 'Fourier coefficients', 
and where  $\{\xi^{(+)}_{2k\ell m}(r)\}$ 
are real radial functions.  These functions satisfy
$$e^{-a}{\,}{d\over{dr}}
\Biggl(e^{-a}{\,}{{d\xi^{(+)}_{2k\ell}}\over{dr}}\Biggr)
+\biggl(k^{2}
-V^{(+)}_{\ell}(r)\biggr)
\xi^{(+)}_{2k\ell}{\quad}
={\quad}0{\quad}.
\eqno(9.14)$$
Regularity at the origin implies that
$$\xi^{(+)}_{2k\ell}(r)\quad 
\sim\quad{\rm (const.)}{\;}\times{\;}{\;}r{\,}j_{\ell}(kr)
\eqno(9.15)$$ 
for small $r{\,}$.  
Again, at large $r{\,}$, the potential vanishes
sufficiently rapidly that 
${\,}\xi^{(+)}_{2k\ell}(r){\,}$ 
has the asymptotic form
$$\xi^{(+)}_{2k\ell}(r)\quad 
\sim\quad\biggl(\Bigl(z^{(+)}_{2k\ell}\Bigr)
\exp(ikr^{*}_{s})
+\Bigl(z^{(+)*}_{2k\ell}\Bigr)
\exp(-ikr^{*}_{s})\biggr){\quad},
\eqno(9.16)$$
where $\{z^{(+)}_{2k\ell m}\}$ are complex constants.  
Then, the classical action $S^{(2)}_{\rm class}$ 
for even-parity gravitational perturbations reads
$$\eqalign{S^{(2)}_{\rm class}&
\Bigl[\{a^{(+)}_{2k\ell m}\}\Bigr]\cr
=&{\,}{{1}\over{16}}\sum^{\infty}_{\ell =2}
\sum^{\ell}_{m=-\ell}{{(\ell -2)!}\over{(\ell +2)!}}
\int^{\infty}_{0}dk{\,}k{\,}
{\mid}z_{2k\ell +}{\mid}^{2}{\,}
{\mid}\bigl(a_{2k\ell m+}\bigr)
+\bigl(a_{2,-k\ell m+}\bigr){\mid}^{2}
\cot(kT){\,},\cr}
\eqno(9.17)$$ 
where the notation is in line with that for spin-0 
and for odd-parity fields.  
The coordinates  $\{a_{2k\ell m+}\}$ 
label the configuration in $k$-space 
of the even-parity part of the metric perturbations 
on the final surface $\Sigma_{F}{\,}$.

Let us now re-assemble both the odd-and even-parity 
metric perturbations.  As above, we consider 
for simplicity odd-parity metric perturbations 
which are initially static (Neumann problem) 
and even-parity metric perturbations 
which vanish initially (Dirichlet problem), 
on the space-like hypersurface $\Sigma_{I}{\,}$.  
The total classical spin-2 action is then
$$\eqalign{&S^{(2)}_{\rm class}{\,}
={\,}{{1}\over{32\pi}}
\sum_{\ell mP}{{(\ell -2)!}\over{(\ell +2)}}{\,}
P\int^{R\infty}_{0}dr{\,}
e^{(a-b)/2}{\,}\xi_{2\ell mP}
\Bigl(\partial_{t}\xi^{*}_{2\ell mP}\Bigr)
\Bigl\arrowvert_{\Sigma_{F}}\cr
&={{1}\over{16}}
\sum^{\infty}_{\ell =2}\sum^{\ell}_{m=-\ell}
\sum_{P=\pm}{{(\ell -2)!}\over{(\ell +2)!}}
\int^{\infty}_{0}dk{\,}k
{\mid}z_{2k\ell P}{\mid}^{2}
{\mid}(a_{2k\ell mP})
+(Pa_{2,-k\ell mP}){\mid}^{2}\cot(kT){\,},\cr}
\eqno(9.18)$$
where the complex coefficients $\{a_{2k\ell mP}\}$ obey
$$a_{2k\ell mP}{\quad} 
={\quad}P{\,}(-1)^{m}{\;}a^{*}_{2,-k\ell,-mP}{\quad}.
\eqno(9.19)$$
Here, $P$ takes the value $\pm 1$ 
according as the parity is even or odd.

As in the case of odd-parity metric 
perturbations (Sec.7), the even-parity metric 
perturbations also diverge at large $r{\,}$, 
except in a special gauge, 
the asymptotically-flat (AF) gauge.  
In the AF gauge for even parity, as for odd parity, 
all physical components
${\,}h^{(1)}_{(\mu)(\nu)}
={\,}{\mid}\gamma^{\mu\mu}\gamma^{\nu\nu}{\mid}^{{1}\over{2}}
{\,}h^{(1)}_{\mu\nu}{\,}$ 
(that is, all components of ${\,}h^{(1)}_{\mu\nu}{\,}$ 
projected onto the legs of a pseudo-orthonormal tetrad 
oriented along the unperturbed 
$(t{\,},r{\,},\theta{\,},\phi)$ directions) 
fall off in the wave zone more rapidly than $r^{-1}{\,}$, 
except for the transverse (angular) components, 
which carry information about the gravitational radiation.  
In the new (AF) gauge, for even parity, one has
$$h^{(+)AF}_{0\ell m}{\quad} 
={\quad}H^{AF}_{0\ell m}{\quad} 
={\quad}H^{AF}_{1\ell m}{\quad}
={\quad}0{\quad}.
\eqno(9.20)$$
Further, from the even-parity gauge 
transformations (8.23-29), one has
$$\eqalignno{0{\;}{\,}
&={\;}{\,}H^{RW}_{0\ell m}
-a'{\hat X}^{(+)}_{1\ell m}
+2\Bigl(\partial_{t}{\hat X}^{(+)}_{0\ell m}\Bigr){\quad},
&(9.21)\cr
0{\;}{\,}
&={\;}{\,}H^{RW}_{1\ell m}
+e^{-a}
\Bigl(\partial_{r}{\hat X}^{(+)}_{0\ell m}\Bigr)
-e^{a}\Bigl(\partial_{t}{\hat X}^{(+)}_{1\ell m}\Bigr){\quad},
&(9.22)\cr
H^{AF}_{2\ell m}{\;}{\,}
&={\;}{\,}H^{RW}_{2\ell m}
-a'{\,}{\hat X}^{(+)}_{1\ell m}
-2\Bigl(\partial_{r}{\hat X}^{(+)}_{1\ell m}\Bigr){\quad},
&(9.23)\cr
K^{AF}_{\ell m}{\;}{\,}
&={\;}{\,}K^{RW}_{\ell m}
-\biggl({{2{\hat X}^{(+)}_{1\ell m}}\over{r}}\biggr){\quad},
&(9.24)\cr
G^{AF}_{\ell m}{\;}{\,}
&={\,}-2{\hat X}^{(+)}_{2\ell m}{\quad},
&(9.25)\cr
0{\;}{\,}
&={\;}{\,}e^{-a}{\,}{\hat X}^{(+)}_{0\ell m}
-r^{2}\Bigl(\partial_{t}{\hat X}^{(+)}_{2\ell m}\Bigr){\quad},
&(9.26)\cr
h^{(+)AF}_{1\ell m}{\;}{\,} 
&={\,}-{\,}e^{a}{\hat X}^{(+)}_{1\ell m}
-r^{2}\Bigl(\partial_{r}{\hat X}^{(+)}_{2\ell m}\Bigr){\quad},
&(9.27)\cr}$$
where a hat denotes a gauge function 
in the AF gauge.  
Therefore, once given 
${\,}H^{RW}_{\ell m}{\,},{\;}H^{RW}_{1\ell m}{\,}$ 
and 
$K^{RW}_{\ell m}{\,}$, 
then Eqs.(9.21,22) can be solved for 
$\hat X^{(+)}_{0\ell m}{\,}$ 
and 
$\hat X^{(+)}_{1\ell m}{\;}$.  
Thence, Eq.(9.26) can be used, in order to solve 
for 
$\hat X^{(+)}_{2\ell m}{\;}$.  
In solving these equations, 
one chooses the arbitrary functions 
which arise such that asymptotic flatness is still satisfied.  
Thus, the AF gauge is consistent.
\end{section}

\begin{section}{Conclusion}
In this paper, we have taken over the scalar (spin-0) 
calculations of [33,34], with the help of the angular 
harmonics of Regge and Wheeler [63], 
to include the more complicated Maxwell (spin-1) 
and linearised graviton (spin-2) cases.  
For spin-1, the linearised Maxwell field splits 
into a part with even parity and a part with odd parity; 
a different treatment is needed for each of these two cases.  
In both cases the relevant boundary conditions involve 
fixing the magnetic field on the initial space-like boundary 
$\Sigma_{I}$ and final boundary $\Sigma_{F}{\;}$.  
The main result is an explicit expression (6.19) 
for the classical (linearised) Maxwell action, 
as a functional of the final magnetic field, 
subject to the simplifying assumption that the magnetic 
field on the initial surface $\Sigma_{I}$ is zero.  
From this, the Lorentzian quantum amplitude for photon 
final data can be derived, as in [34] for spin-0 
perturbative final data, by taking the limit
$\theta\rightarrow 0_{+}$ 
of
$\exp\bigl(iS_{\rm class}\bigr){\,}$, 
where $S_{\rm class}$ is the action of the classical 
solution of the boundary-value problem with prescribed 
initial and final data, with complexified time-interval 
$T={\mid}T{\mid}\exp(-{\,}i\theta){\,}$, 
where $0<\theta\leq\pi/2{\,}$. 

Linearised gravitational-wave $(s=2)$ perturbations 
about a spherically-symmetric Einstein/massless-scalar 
collapse to a black hole have also been studied here. 
As for Maxwell $(s=1)$ perturbations, 
the principal aims for $s=2$ also are 
(1) to specify suitable perturbative boundary data 
on the final space-like hypersurface $\Sigma_{F}$ 
at a late time $T{\,}$, subject (for simplicity) 
to the initial boundary data on $\Sigma_{I}{\,}$ 
(time ${\,}t=0{\,}$) being spherically symmetric;  
(2) to express the spin-2 Lorentzian classical action 
$S_{\rm class}$ as an explicit functional 
of the 'suitable' boundary data above, 
and of the proper-time interval $T{\,}$, 
once $T$ has been rotated into the complex: 
$T\rightarrow{\mid}T{\mid}\exp(-{\,}i\theta){\,}$, 
for ${\,}0<\theta\leq\pi/2{\,}$;  
(3) given $S_{\rm class}{\;}$, to compute, following Feynman, 
the quantum amplitude for the weak-field final data, 
by taking the limit of the semi-classical amplitude 
${\rm (const.)}\times\exp\bigl(iS_{\rm class}\bigr)$ 
as $\theta\rightarrow 0_{+}{\,}$.

As in the $s=1$ case, it is also necessary for $s=2$ 
to decompose the metric perturbations into parts 
with odd and even parity.  The main difference on moving 
from the $s=1$ to the $s=2$ case is a considerable increase 
in algebraic or analytic complexity, 
to be expected since one deals with tensor fields 
rather than vector fields.

Some indications towards unification of these ideas 
for perturbative fields of different spin $s$ 
appear already in Secs. 2,3.  
For $s=1{\,}$, the quantity most naturally specified 
as an argument of the quantum wave-functional, 
on a bounding hypersurface such as $\Sigma_{F}{\,}$, 
is the (spatial) magnetic field $B_{i}{\;}$, 
subject to the condition 
${\,}^{3}\nabla_{k}B^{k}{\,}={\,}0{\;}$.  
Correspondingly, for linearised gravitational waves $(s=2)$, 
the natural boundary data were found 
to be the (symmetric, trace-free) magnetic part $H_{ik}$ 
of the Weyl tensor [52,53], 
subject to ${\,}^{3}\nabla_{k}H^{ik}{\,}={\,}0{\;}$.  
In 2-component spinor language, these correspond $(s=1)$ 
to a particular 'projection' of the (complex) symmetric 
Maxwell field-strength spinor 
$\phi_{AB}{\,}={\,}\phi_{(AB)}{\,}$, 
and $(s=2)$ to a corresponding projection 
of the totally-symmetric (complex) Weyl spinor  
$\Psi_{ABCD}{\,}={\,}\Psi_{(ABCD)}{\,}$. 
Of course, as treated in [35], 
these boundary conditions constructed from $\phi_{AB}{\,}$ 
and $\Psi_{ABCD}{\,}$ are special cases 
of the natural boundary conditions for gauged 
supergravity [71-73].  In our work on quantum amplitudes 
for spin-${{1}\over{2}}$ [35], the natural boundary 
conditions also involved a corresponding projection 
of the spin-${{1}\over{2}}$ field.  
Although 2-component spinor language might 
(to some people) seem a luxury in treating bosonic fields 
describing photons or gravitons, above, 
it is practically a necessity in treating 
the corresponding fermionic (massless) neutrino 
spin-${{1}\over{2}}$ field, as in [35], 
and (for supergravity) the gravitino 
spin-${{3}\over{2}}$ field, on which work is in progress [36].
\end{section}

\parindent = 1 pt

\begin{section}*{References} 
\everypar{\hangindent\parindent} 
        
\noindent [1] M.K.Parikh and F.Wilczek, 
Phys. Rev. Lett. {\bf 85}, 5042 (2000).

\noindent [2] M.Parikh, 
Gen. Relativ. Gravit., {\bf 36}, 2419 (2004).

\noindent [3] S.W.Hawking, 
Nature (London) {\bf 248}, 30 (1974);
Phys. Rev. D {\bf 14}, 2460 (1976);
Commun. Math. Phys. {\bf 43}, 199 (1975).

\noindent [4] P.H{\'a}j{\'i}{\v c}ek and W.Israel, 
Phys. Lett. A {\bf 80}, 9 (1980).

\noindent [5] J.Bardeen, 
Phys. Rev. Lett. {\bf 46}, 382 (1981).

\noindent [6] S.W.Hawking, 
Commun. Math. Phys. {\bf 87}, 395 (1982).

\noindent [7] S.W.Hawking, 
`Boundary Conditions of the Universe' 
in {\it Astrophysical Cosmology}, 
Proceedings of the Study Week on Cosmology and Fundamental Physics, 
eds. H.A.Br{\"u}ck, G.V.Coyne and M.S.Longair. 
Pontificia Academiae Scientarium: Vatican City,
{\bf 48}, 563 (1982).

\noindent [8] R.M.Wald, 
in {\it Quantum Theory of Gravity}, 
ed. S.Christensen, (Adam Hilger, Bristol) 160 (1984).

\noindent [9] A.N.St.J.Farley, 
'Quantum Amplitudes in Black-Hole Evaporation', 
Cambridge Ph.D. dissertation, approved 2002 (unpublished).

\noindent [10] S.W.Hawking, 
communication, GR17 Conference, Dublin, 18-24 July (2004). 

\noindent [11] A.N.St.J.Farley and P.D.D'Eath, 
Phys Lett. B {\bf 601}, 184 (2004); arXiv gr-qc/0407086.     
 
\noindent [12] E.P.S.Shellard, 
'The future of cosmology:  observational and computational prospects', 
in {\it The Future of Theoretical Physics and Cosmology}, 
eds. G.W.Gibbons, E.P.S.Shellard and S.J.Rankin
(Cambridge University Press, Cambridge) 755 (2003).

\noindent [13] R.P.Feynman and A.R.Hibbs, 
{\it Quantum Mechanics and Path Integrals}, 
(McGraw-Hill, New York) (1965).

\noindent [14] P.R.Garabedian, 
{\it Partial Differential Equations}, 
(Wiley, New York) (1964).

\noindent [15] P.D.D'Eath, 
{~}{\it Supersymmetric {~}Quantum {~}Cosmology}, 
{~}(Cambridge {~}University {~}Press, {~}Cambridge) (1996).

\noindent [16] A.N.St.J.Farley and P.D.D'Eath, 
Phys Lett. B {\bf 613}, 181 (2005); arXiv gr-qc/0510027.
          
\noindent [17] W.McLean, 
{\it Strongly Elliptic Systems and Boundary Integral Equations}, 
(Cambridge University Press, Cambridge) (2000); 
O.Reula, 
'A Configuration space for quantum gravity and solutions 
to the Euclidean Einstein equations in a slab region',
Max-Planck-Institut fur Astrophysik, {\bf MPA}, 275 (1987). 

\noindent [18] D.Christodoulou, 
Commun. Math. Phys. {\bf 105} 337 (1986); 
{\bf 106} 587 (1986); 
{\bf 109} 591, 613 (1987);
Commun. Pure Appl. Math. {\bf 44}, 339 (1991); 
{\bf 46}, 1131 (1993).

\noindent [19] P.D.D'Eath and A.Sornborger, 
Class. Quantum Grav. {\bf 15}, 3435 (1998).

\noindent [20] P.D.D'Eath, 
'Numerical and analytic estimates for the Einstein/scalar 
boundary-value problem', in progress. 

\noindent [21] A.Das, M.Fischler and M. Ro\v cek, 
Phys.Lett. B {\bf 69}, 186 (1977).

\noindent [22] J.Wess and J.Bagger, 
{\it Supersymmetry and Supergravity}, 2nd. edition, 
(Princeton University Press, Princeton) (1992).

\noindent [23]  J.B.Hartle and S.W. Hawking, 
Phys. Rev. D {\bf 28}, 2960 (1983).

\noindent [24]  P.A.M.Dirac, 
{\it Lectures on Quantum Mechanics}, 
(Academic Press, New York) (1965).

\noindent [25]  P.D.D'Eath, 
'Loop amplitudes in supergravity by canonical quantization', 
in {\it Fundamental Problems in Classical, Quantum and String Gravity}, 
ed. N.S{\'a}nchez (Observatoire de Paris) 
166 (1999); arXiv hep-th/9807028.

\noindent [26] P.D.D'Eath, 
'What local supersymmetry can do for quantum cosmology', 
in {\it The Future of Theoretical Physics and Cosmology}, 
eds. G.W.Gibbons, E.P.S.Shellard and S.J.Rankin 
(Cambridge University Press, Cambridge) 693 (2003); 
arXiv gr-qc/0511042.

\noindent [27] L.D.Faddeev, 
in {\it Methods in Field Theory}, 
ed. R.Balian and J.Zinn-Justin 
(North-Holland, Amsterdam) (1976).

\noindent [28] L.D.Faddeev and A.A.Slavnov, 
{\it Gauge Fields: Introduction to Quantum Theory} 
(Benjamin/Cummings, Reading, Mass.) (1980).

\noindent [29] C.Itzykson and J.-B.Zuber, 
{\it Quantum Field Theory},
(McGraw-Hill, New York) (1980).

\noindent [30] P.D.D'Eath, 
{\it Black Holes: Gravitational Interactions}, 
(Oxford University Press, Oxford) (1996).

\noindent [31] G. 't Hooft, 
Phys. Lett. B {\bf 198}, 61 (1987).

\noindent [32] S. Giddings, 
'Black holes at accelerators',
in {\it The Future of Theoretical Physics and Cosmology}, 
eds. G.W.Gibbons, E.P.S.Shellard and S.J.Rankin 
(Cambridge University Press, Cambridge) 278 (2003).

\noindent [33] A.N.St.J.Farley and P.D.D'Eath, 
'Quantum Amplitudes in Black-Hole Evaporation: I. Complex Approach', 
submitted for publication (2006); arXiv gr-qc/0510028.

\noindent [34] A.N.St.J.Farley and P.D.D'Eath, 
'Quantum Amplitudes in Black-Hole Evaporation: II. Spin-0 Amplitude', 
submitted for publication (2006); arXiv gr-qc/0510029.

\noindent [35] A.N.St.J.Farley and P.D.D'Eath, 
Class. Quantum Grav. {\bf 22}, 3001 (2005); arXiv gr-qc/0510036.

\noindent [36] A.N.St.J.Farley and P.D.D'Eath, 
'Spin-3/2 Amplitudes in Black-Hole Evaporation', in progress.

\noindent [37] J.Mathews, 
J. Soc. Ind. Appl. Math. {\bf 10}, 768 (1962).

\noindent [38] J.N.Goldberg, A.J.MacFarlane, E.T.Newman, F.Rohrlich 
and E.C.G.Sudarshan, 
J. Math. Phys. {\bf 8}, 2155 (1967).

\noindent [39] J.A.H.Futterman, F.A.Handler and R.A.Matzner,   
{\it Scattering from Black Holes} 
(Cambridge University Press, Cambridge) (1988).

\noindent [40]  H.Stephani et al., 
{\it Exact Solutions to Einstein's Field Equations}, 2nd. ed, 
(Cambridge University Press, Cambridge) (2003).

\noindent [41]  P.C.Vaidya, 
Proc. Indian Acad. Sci. {\bf A33}, 264 (1951).

\noindent [42] R.Penrose and W.Rindler, 
{\it Spinors and Space-Time}, vol. 1
(Cambridge University Press, Cambridge) (1984)

\noindent [43] R.Penrose and W.Rindler, 
{\it Spinors and Space-Time}, vol. 2
(Cambridge University Press, Cambridge) (1986)

\noindent [44] T.Eguchi, P.B.Gilkey and A.J.Hanson, 
Phys. Rep. {\bf 66}, 214 (1980).

\noindent [45] A.N.St.J.Farley and P.D.D'Eath,
'Bogoliubov Transformations in Black-Hole Evaporation', 
submitted for publication (2006).

\noindent [46] R.S.Ward and R.O.Wells, 
{\it Twistor Geometry and Field Theory}
(Cambridge University Press, Cambridge) (1990).

\noindent [47] M.Nakahara, 
{\it Geometry, Topology and Physics}, 2nd. edition
(Institute of Physics, Bristol) (2003).

\noindent [48] S.W.Hawking and G.F.R.Ellis, 
{\it The large scale structure of space-time}, 
(Cambridge University Press, Cambridge) (1973).

\noindent [49] C.W.Misner, K.S.Thorne and J.A.Wheeler, 
{\it Gravitation},
(Freeman, San Francisco) (1973).

\noindent [50] J.A.Wheeler, 
'Geometrodynamics and the Issue of the Final State', p. 317, 
in {\it Relativity, Groups and Topology}, 
eds. C.DeWitt and B.DeWitt (Blackie and Son, London) (1964).

\noindent [51] J.A.Wheeler, 
'Superspace and the Nature of Quantum Geometrodynamics', p.242, 
in {\it Battelle Rencontres},
eds. C.M.DeWitt and J.A.Wheeler (Benjamin, New York) (1968).

\noindent [52]  P.D.D'Eath, 
Phys. Rev. D {\bf 11}, 1387 (1975).

\noindent [53] K.S.Thorne, R.H.Price and D.A.Macdonald, 
{\it Black holes. The membrane paradigm.} 
(Yale University Press, New Haven, Ct.) (1986).

\noindent [54] M.F.Atiyah, N.J.Hitchin and I.M.Singer,
Proc. R. Soc. Lond. A. {\bf 362}, 425 (1978).

\noindent [55] E.T.Newman and R.Penrose, 
J. Math. Phys. {\bf 3}, 566 (1962).

\noindent [56] S.A.Teukolsky, 
Astrophys. J. {\bf 185}, 635 (1973).

\noindent [57] S.Chandrasekhar, 
{\it The Mathematical Theory of Black Holes}
(Oxford University Press, Oxford) (1992).

\noindent [58] R.Geroch, A.Held and R.Penrose, 
J. Math. Phys. {\bf 14}, 874 (1973).

\noindent [59] W.Kinnersley, 
J. Math. Phys {\bf 10}, 1195 (1969).

\noindent [60]J.D.Jackson, 
{\it Classical Electrodynamics}, 
(Wiley, New York) (1975).

\noindent [61] P.L.Chrzanowski, 
Phys. Rev D {\bf 11}, 2042 (1975).

\noindent [62] R.M.Wald, 
Phys. Rev. Lett. {\bf 41}, 203 (1978).

\noindent [63] T.Regge and J.A.Wheeler, 
Phys. Rev. {\bf 108}, 1063 (1957).

\noindent [64] F.J.Zerilli, 
Phys. Rev. D {\bf 2}, 2141 (1970).

\noindent [65] C.T.Cunningham, R.H.Price and V.Moncrief, 
Astrophys. J. {\bf 224}, 643 (1978); 
ibid. {\bf 230}, 870 (1979).

\noindent [66] J.A.Wheeler, 
{\it Geometrodynamics} 
(Academic Press, New York) (1962).

\noindent [67] K.S.Thorne and A.Campolattaro, 
Astrophys. J. {\bf 149}, 591 (1967).

\noindent [68] V.Moncrief, 
Ann. Phys. (N.Y.) {\bf 88}, 323 (1974).

\noindent [69]  A.N.St.J.Farley and P.D.D'Eath, 
'Vaidya Space-Time in Black-Hole Evaporation', 
to appear in General Relativity and Gravitation (2006); 
arXiv gr-qc/0510040.

\noindent [70] F.J.Zerilli, 
Phys. Rev. D {\bf 9}, 860 (1974).

\noindent [71]  P.Breitenlohner and D.Z.Freedman, 
Phys. Lett. B {\bf 115}, 197 (1982).

\noindent [72] P.Breitenlohner and D.Z.Freedman, 
Ann. Phys. (N.Y.) {\bf 144}, 249 (1982).

\noindent [73] S.W.Hawking, 
Phys. Lett. B {\bf 126}, 175 (1983).

\noindent [74] M.Abramowitz and I.A.Stegun, 
{\it Handbook of Mathematical Functions}, 
(Dover, New York) (1964).

\end{section}

\end{document}